\documentclass[traditabstract]{aa}

\usepackage{graphicx}
\usepackage{enumerate}
\usepackage{epsfig}
\usepackage{color}
\usepackage{txfonts}
\usepackage{multirow}
\usepackage{float}
\usepackage[export]{adjustbox}

\newcommand\gaia{\textit{Gaia}}
\newcommand{\kms}{km~s$^{-1}$}

\begin{document}
\title{The Milky Way has no in-situ halo other than the heated thick disc}
\subtitle{Composition of the stellar halo and age-dating the last significant merger with \gaia~DR2 and APOGEE}
\titlerunning{Composition of the stellar halo and age-dating the last significant merger}

\author{P.~Di Matteo\inst{1, 2},  M.~Haywood\inst{1, 2}, M.~D.~Lehnert\inst{2},  D.~Katz\inst{1}, S.~Khoperskov\inst{3,4}, O.~N.~Snaith\inst{5,1}, A.~G$\rm \acute{o}$mez\inst{1}, N.~Robichon\inst{1}}

\authorrunning{P. Di Matteo et al.}

\institute{GEPI, Observatoire de Paris, PSL Research University, CNRS,
Place Jules Janssen, 92190 Meudon, France
\email{paola.dimatteo@obspm.fr}
\and Sorbonne Universit\'{e}, CNRS UMR 7095, Institut d'Astrophysique de Paris, 98bis bd Arago, 75014 Paris, France
\and Max Planck Institute for Extraterrestrial Physics, 85741 Garching, Germany
\and Institute of Astronomy, Russian Academy of Sciences, Pyatnitskaya st., 48, 119017 Moscow, Russia
\and School of Physics, Korea Institute for Advanced Study, 85 Hoegiro, 
Dongdaemun-gu, Seoul 02455, Republic of Korea
}

\date{Accepted, Received}

\abstract{Previous studies based on the analysis of \gaia~DR2 data have revealed that accreted stars, possibly originating from a single progenitor satellite, are a significant component of the halo of our Galaxy, potentially constituting most of the halo stars at $\rm [Fe/H] < -1$ within a few kpc from the Sun and beyond. In this paper, 
we couple astrometric data from \gaia~DR2 with
elemental abundances from APOGEE~DR14 to characterize the kinematics
and chemistry of in-situ and accreted populations up to [Fe/H]$\sim-2$. 
Accreted stars appear to significantly impact the Galactic chemo-kinematic relations, not only at $\rm [Fe/H]< -1$, but also at metallicities typical of the thick and metal-poor thin discs. They constitute about 60\% of all stars at $\rm [Fe/H] < -1$, the remaining 40\% being made of (metal-weak) thick disc stars.
We find that
the stellar kinematic fossil record shows the imprint left by this accretion 
event which heated the old Galactic disc. We are able to age-date this kinematic imprint, showing that the accretion occurred between 9 and
11~Gyr ago, and that it led to the last significant heating of the Galactic disc. 
An important fraction of stars with abundances typical of the (metal-rich) thick disc, and heated by this interaction, is now found in the Galactic halo. Indeed about half  of the kinematically defined halo at few kpc from the Sun is composed of metal-rich thick disc stars. Moreover, we suggest that this metal-rich thick disc component dominates the stellar halo of the inner Galaxy.
The new picture that emerges from this study is one where
the standard  non-rotating in-situ halo population, the collapsed halo, seems to be more elusive than ever.}


\maketitle

\section{Introduction}\label{intro}

The presence of accreted stars in galaxy halos is a natural expectation of $\Lambda CDM$ Cosmology \citep{cole91, white91}.
Stars and stellar systems deposited in our galactic halo by past accretion events have been known and postulated for decades \citep{searle78, zinn93, majewski96, zinn96, helmi99, chiba00, venn04, bullock04, font06, carollo07, bell08, delucia08, johnston08, forbes10, xue11, leaman13, pillepich15}.  Chemical abundances of stars in the solar vicinity and on a larger scale of a few kpc have provided evidence that two halo components co-exist: an $\alpha-$enhanced, metal-poor population, which possibly set the initial conditions for the formation of the Galactic disc, and a low $\alpha-$abundance, metal-poor population, possibly accreted early in the evolution of our Galaxy \citep{nissen10, nissen11, navarro11, ramirez12, schuster12, hawkins15a, hayes18}.

Together with the discovery of a number of thin stellar streams, possibly associated to disrupted globular clusters \citep{ibata18, malhan18, price-whelan18}, the analysis of the first and second \gaia~releases \citep{gaia16, gaia17, gaia18} have revealed the presence of tidal debris from an ancient massive accretion  \citep{belokurov18, haywood18, myeong18, helmi18, mackereth18, fattahi18}. As shown by \citet{haywood18}, and later confirmed by \citet{helmi18}, stars belonging to this massive accretion event mostly redistribute along the bluer of the two sequences discovered in the \gaia~DR2 HR diagram of kinematically selected halo stars \citep{gaiababusiaux18}. Their chemical abundances overlap with the low-$\alpha$ sequence discovered by \citet[][see \citealp{haywood18}]{nissen10} and extensively studied by \citet[][see \citealp{haywood18, helmi18}]{hayes18}. The dynamics and orbits of these low-$\alpha$ stars, e.g., their positions in a ``Toomre diagram'', are such that they dominate regions with no or retrograde rotation and high total orbital energy \citep{koppelman18, haywood18}. This is true even if the overlap with the red HR diagram sequence in  \gaia~DR2 is significant \citep{haywood18}, as expected in a scenario where the stellar halo is made of both accreted and in-situ disc stars\footnote{By ``in-situ disc stars'' we mean stars that formed in the early Milky Way disc and that were then heated by interactions \citep{purcell10, zolotov10, font11, qu11b, mccarthy12}} \citep{jeanbaptiste17}. 
Accreted stars have been suggested to be the dominant metal-poor (i.e. $\mathrm{[Fe/H]}<-1$)  halo component at few kpc from the Sun and beyond \citep{haywood18, belokurov18, iorio18} and may extend to
relatively high metallicities \citep[i.e. at $\mathrm{[Fe/H]}>-1$; see][]{nissen10, hayes18}.

In this paper, by coupling kinematics $and$ chemical abundances, we aim to: \textit{(1)} discuss the imprint of the accreted stars on the Galactic chemo-kinematic relations, at metallicities typical of the halo, the metal-poor thin and thick discs; \textit{(2)} estimate each components relative fraction in different regions of the [Fe/H]--[Mg/Fe] plane, especially at $\mathrm{[Fe/H]}<-1$, where a discrimination on the basis of the chemical abundances alone is likely not possible; and \textit{(3)} age-date the last significant accretion event experienced by the Galaxy, by making use of kinematics to investigate signatures of the heating of the early Galactic disc, and using the abundances in the early Galactic disc to constraint the time when this heating event occurred. Finally, the accretion of relatively massive satellites is expected to generate a significant fraction of stars with halo kinematics, but abundances typical of the early disc from which they were kicked out. Stars heated by merging events should have kinematics which overlap with the accreted population \citep{jeanbaptiste17}. We conclude our analysis by investigating the presence of in-situ stars with disc chemistry but with halo kinematics to estimate their relative fractions of their original population and of the total halo population within few kpc from the Sun.

The paper is organized as follows: in Sect.~\ref{sample}, how we selected the sample of stars used in this analysis; in Sect.~\ref{results} we present our results, which include the mean Galactic chemo-kinematic relations (Sect.~\ref{mean}), the relative fractions of in-situ and accreted stars in the [Fe/H]-[Mg/Fe] plane (Sect.~\ref{disentangling}), and how we age-dated the last significant accretion event experienced by our Galaxy (Sect.~\ref{dating}). In Sect.~\ref{discussion}, we discuss the implications of our findings and finally in Sect.~\ref{conclusions}, we summarize our conclusions.
 
\section{Data}\label{sample}
\begin{figure}
\centering
\includegraphics[clip=true, trim = 0mm 0mm 0mm 0mm, width=\linewidth]{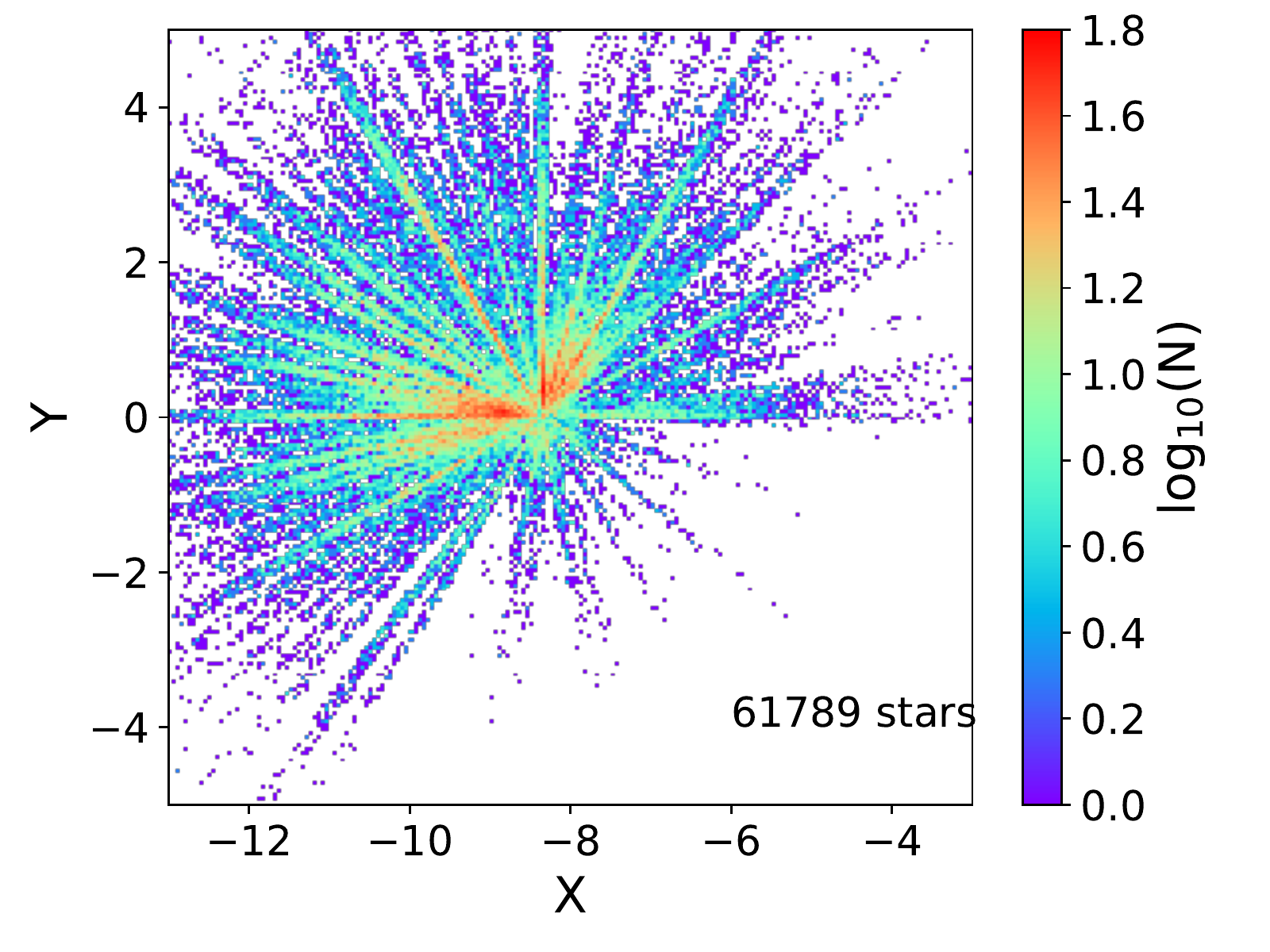}
\includegraphics[clip=true, trim = 0mm 0mm 0mm 0mm, width=\linewidth]{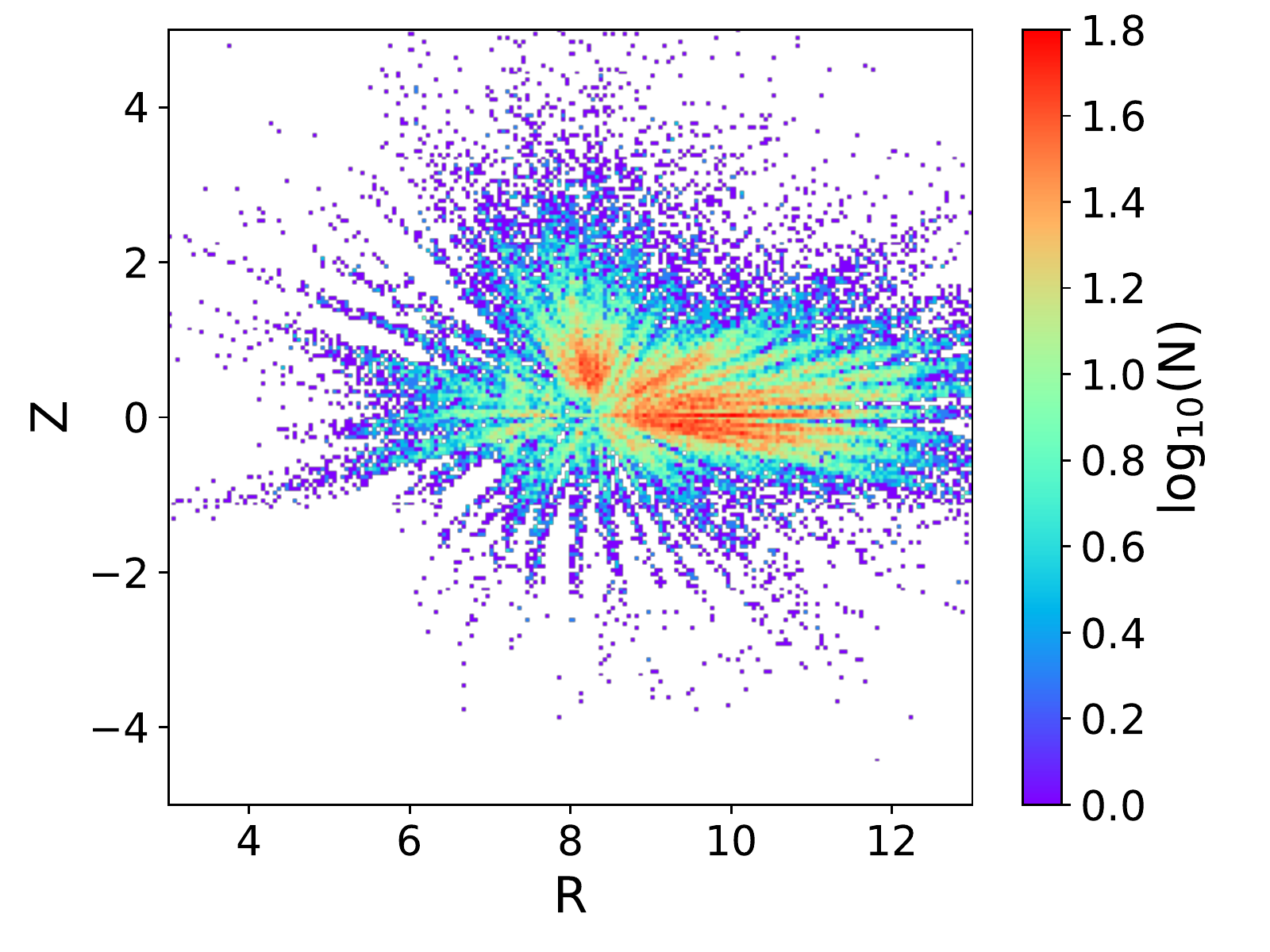}
\caption{Spatial distribution of stars in the sample we are analyzing in the x-y (\emph{top panel}) and R-z (\emph{bottom panel}) planes. The total number of stars in the sample is provided in the lower left-hand corner of the top panel. In both maps, number of stars, in logarithmic scale, per pixel, the pixel size being $50 \times 50 $ pc$^2$ is color coded as given in the bar on the right-hand side of each panel. The Sun lies at  $x = -8.34$~kpc, $y =0$ and $z = 27$~pc.}
\label{XYRZ}
\end{figure}

\begin{figure*}
\flushleft
\includegraphics[clip=true, trim = 0mm  0mm 0mm 0mm, width=0.3\linewidth]{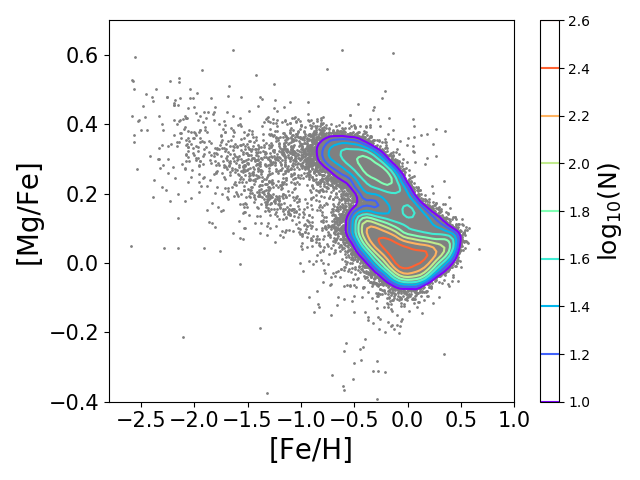}
\includegraphics[clip=true, trim = 0mm  0mm 0mm 0mm, width=0.3\linewidth]{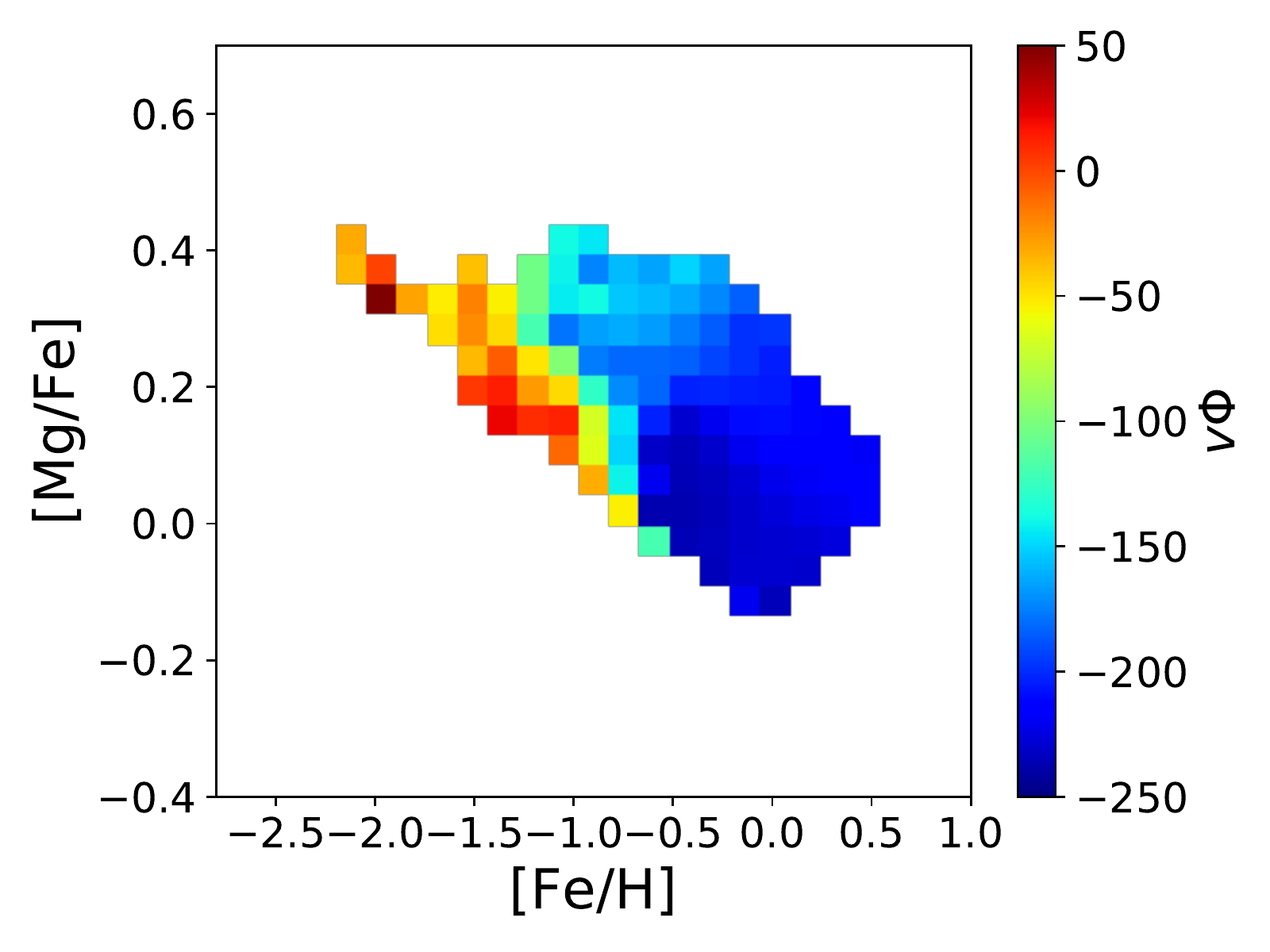} \\
\includegraphics[clip=true, trim = 0mm  0mm 0mm 0mm, width=0.3\linewidth]{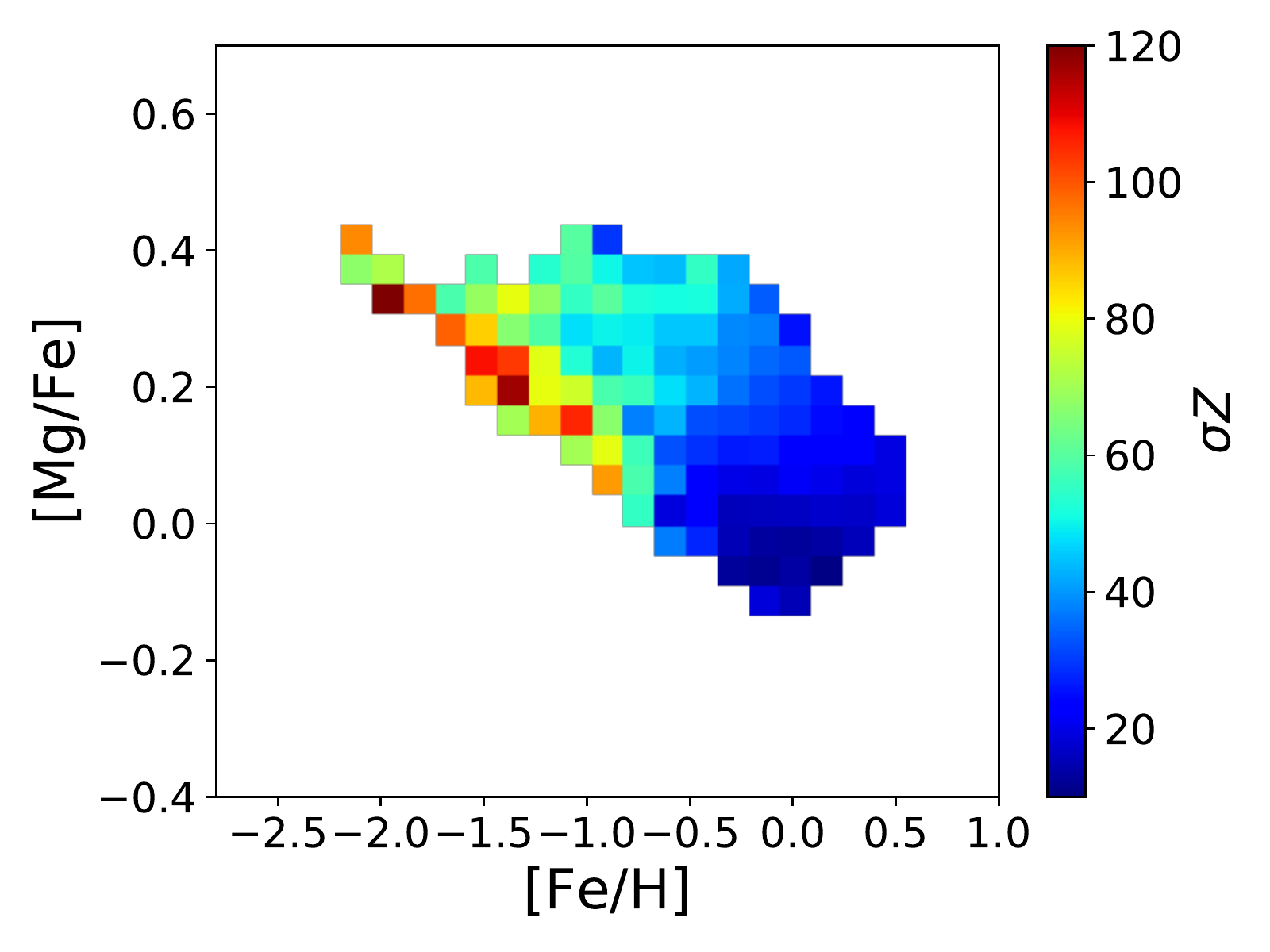}
\includegraphics[clip=true, trim = 0mm 0mm 0mm 0mm, width=0.3\linewidth]{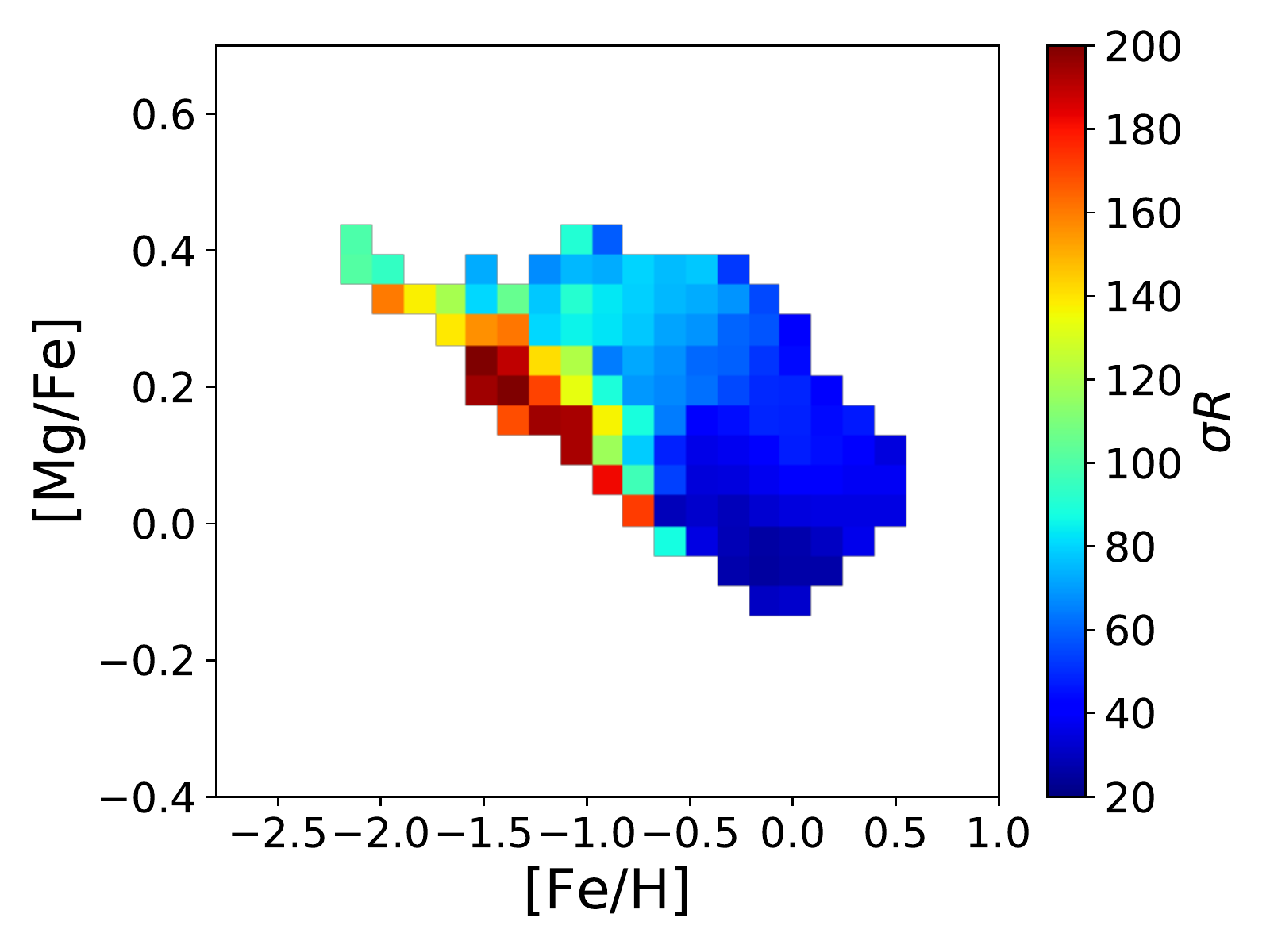}
\includegraphics[clip=true, trim = 0mm 0mm 0mm 0mm, width=0.3\linewidth]{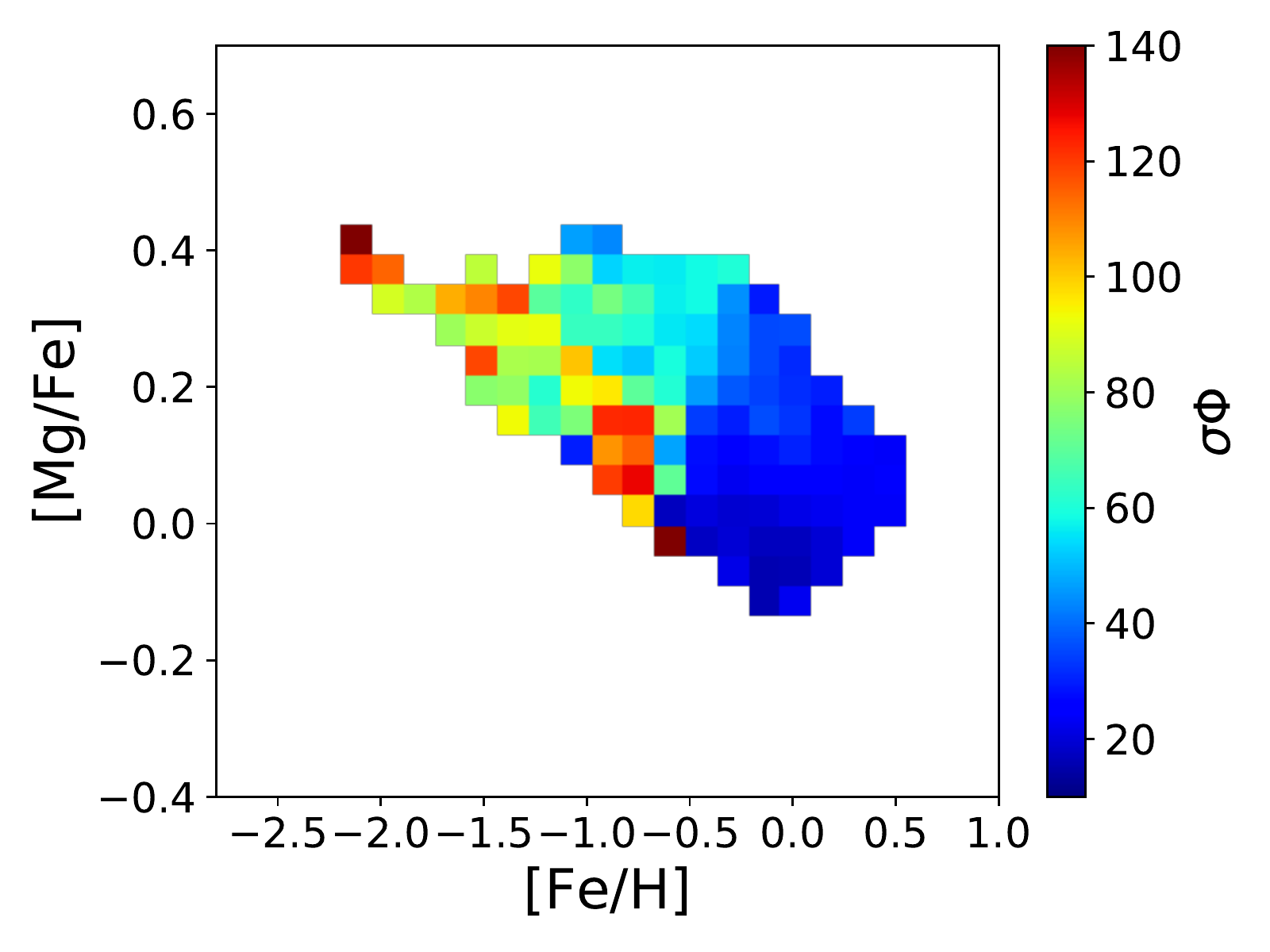}
\caption{(\emph{Top-left panel}): Distribution of stars in our sample in the  [Fe/H]--[Mg/Fe] plane. Colored lines correspond to isodensity contours, in logarithmic scale, as indicated in the color bar on the right-hand side of the panel. (\emph{Top-right panel}): Mean azimuthal velocity of stars in the [Fe/H]--[Mg/Fe]  plane. In this, as well as in all following panels, the pixel size is 0.152$\times$0.044, and only pixels containing more than 10 stars are shown. (\emph{Bottom-left panel}):  Mean vertical velocity dispersions of stars in the [Fe/H]--[Mg/Fe]  plane. (\emph{Bottom-middle panel}):  Mean radial velocity dispersions of stars in the [Fe/H]--[Mg/Fe]  plane. (\emph{Bottom-right panel}):  Mean azimuthal velocity dispersions of stars in the [Fe/H]--[Mg/Fe]  plane.}
\label{FeHMgFe_map}
\end{figure*}

The sample that we analyse in this paper is the result of cross-matching the \gaia~DR2 \citep{gaia18} with APOGEE data from DR14 \citep{majewski17}, using the CDS X-match service\footnote{http://cdsxmatch.u-strasbg.fr/xmatch}. To construct this sample, we have selected stars in the two catalogues with a position mismatch tolerance of of 0.5~arcsec, and retained only those with positive parallaxes $\pi$, relative error on parallaxes $\sigma_\pi/\pi < 0.2$,  and a signal-to-noise ratio in the APOGEE spectra, 
SNR$>100$. All line-of-sight velocities used in this paper are from APOGEE.
Following the study of \citet{fernandez18}, we have applied addition selection criteria only retaining stars with effective temperatures, $T_{eff} > 4000$, and gravities, $1 < log(g) < 3.5$. Finally, we have also removed all APOGEE stars with ASCAPFLAG and STARFLAG warning of any problems with the determinations of the atmospheric parameters (specifically those with warning about the reliability of the effective temperature, log(g), rotation and having a very bright neighbour). 
After applying all these selection criteria, our final sample consists of 61789 stars, whose density distribution, projected onto the Galactic and meridional planes, is shown in Fig.~\ref{XYRZ}. As expected, 
most of the stars in the analyzed sample are at a distance of 2-3~kpc from the Sun, with a dearth of stars in the forth quadrant, due to the lack of coverage of this area in the APOGEE footprint. For calculating positions and velocities in the galactocentric rest-frame, we have assumed an in-plane distance of the Sun from the Galactic centre, R$_\odot$ = 8.34~kpc \citep{reid14}, a height of the Sun above the Galactic plane, z$_\odot$  = 27~pc \citep{chen01}, a velocity for the Local Standard of Rest, $V_{LSR}$= 240 \kms  \citep{reid14}, and a peculiar velocity of the Sun with respect to the LSR, U$_\odot$  = 11.1~\kms, V$_\odot$=12.24~\kms, W$_\odot$=7.25~\kms\ \citep{schonrich10}.
Individual uncertainties in the velocities of stars in the sample, due to the propagation of the uncertainties on the observables (parallaxes, proper motions and radial velocities) are discussed in  Appendix~\ref{uncertainties}.
Parallaxes have been corrected by the zero-point offset of  $-0.03 \rm mas$  \citep{arenou18, gaia18, lindegren18}, and distances have been derived by inverting parallaxes. While the correction of the zero-point offset affects, as expected, the absolute values of our derived relations, and mean/median kinematics, all trends and conclusions presented in the following of the paper are maintained also when no correction of the parallax zero-point is made. 

In the following sections of this paper, we discuss fractions of in situ (thick disc) stars relative to 
accreted stars. The APOGEE selection function is not expected to introduce any bias against some specific population at a given metallicity, $\alpha-$abundance, or kinematics, because the selection criteria in color and magnitude of the survey (see Majewski et al. 2017) are not related to these parameters. The  color cut of the survey introduces a selection on metal poor stars, but there is no reason that this would introduce a bias against a specific population, in the metallicity range studied in this paper. So we do expect that the fractions discussed here are representative of a local sample of the Galaxy.  However, it goes without saying that the relative fractions of the different populations are affected by the distance limit of the APOGEE sample, and may not be representative of the entire Milky Way. Fig.~\ref{XYRZ} shows the complex distribution of the APOGEE stars resulting from the adopted footprint of the survey. It is seen in particular that the APOGEE footprint favors anticentre directions, where the thick disc population is known to be less well represented because of its short scale length. It has been shown recently that the radial distributions of these two populations are widely different \citep{Sahlholdt19}, the accreted stars having a much more uniform distribution than thick disc stars within a few kpc from the Sun. It is therefore possible that these fractions are biased in the sense that the thick disc is relatively less represented locally. 

In Fig.~\ref{FeHMgFe_map}, we show the distribution of our sample in the [Mg/Fe]-[Fe/H] plane. While the vast majority of the stars have [Fe/H]$\geq -1$, and occupies the two chemically defined sequences of the Galactic thick and thin discs, about $1.5\%$ of the sample has [Fe/H]$\leq -1$. These relatively low Fe abundance stars lie along two sequences: a high [Mg/Fe]-abundance sequence, which joins to the thick disc sequence at high metallicities, but with a possible dip in the density of stars at  [Fe/H]$\sim -1$ \citep{hayes18, gaiababusiaux18} and a sequence extending from high [Mg/Fe] abundances, at metallicities [Fe/H]$\sim -2$, to [Mg/Fe] abundances about 0.2 lower, at [Fe/H]$\sim -1$. The continuation of the two sequences, at low ([Fe/H]$\sim -2$) and high ([Fe/H]$\sim -1$) metallicities is still uncertain, and difficult to constrain presently. However, we show subsequently that, by coupling chemical abundances with kinematics, it is possible to relate the low-[Mg/Fe] sequence to the metal-poor tail of the thin disc, namely, thin disc stars with [Fe/H]$\sim -0.5$ and solar and sub-solar [Mg/Fe].

\section{Results}\label{results}

Before presenting the results, we wish to introduce the  nomenclature and conventions adopted in this paper as this often leads to confusion in this type
of study. The definition and the borders of the Galactic stellar populations are not always trivial to set, because of the overlap that all populations, from those of the bulge, to the halo, passing through the thin and thick discs show.

\emph{The Galactic halo:}  In this paper, the halo is sometimes defined on the basis of its chemistry, that is as stars with  $\rm [Fe/H] < -1$. Sometimes we define the halo kinematically, that is, made of stars with absolute velocities, relative to the LSR, greater than 180~\kms.  We endeavour to be specific which definition we employ and, in Sect.~\ref{discussion}, we summarize all our findings in terms of what the ``Galactic halo'' consists of within a few kpc from the Sun. 

\emph{Convention adopted for the azimuthal velocities:} In our choice of the Galactocentric coordinate system, the Sun lies on the x-axis with a negative value of $x=-8.34$~kpc, and the $V_\odot$ is positive, that is parallel to the  $y$ axis. This implies that the disc rotates clockwise, and -- as a consequence -- the $z-$component of the disc angular momentum and the disc azimuthal velocity $v_\Phi$ are negative. Thus, negative $v_\Phi$ correspond to prograde velocity rotation, and positive $v_\Phi$ to retrograde velocity rotation. 

Finally, a note on  the results that will be presented in the following of this work. As shown in Fig.~\ref{XYRZ}, the sample studied in this paper is still relatively local, and most of it is restricted to distances of 2-3~kpc from the Sun. It is thus natural to discuss how general the results are, and to what extent they could be generalized to other regions of the Galactic disc. We have started investigating this issue  in the simulations presented in \citet[][see Figs. 5 and 7, and Tables 3, 4 and 5 in that paper]{jeanbaptiste17}. This work contains 3 simulations of a Milky Way-type galaxy accreting one or several satellites. We showed that, 1~Gyr after the accretion of a satellite, accreted stars are already very well mixed, and their fraction, as well as their kinematic properties, do not significantly depend on the "solar volume" chosen, for solar volumes at the same Galactocentric distance. These simulations thus suggest that the results presented in this paper should be generalized to different regions of the Galactic disc, at similar distances from the Galactic centre. This similarity can be understood because the dynamical times at the solar radius are relatively short: with a typical rotational period around the Galactic centre of about 200 Myr, after 1 or few Gyrs, accreted stars can redistribute rather homogeneously in the disc. Therefore, for mergers which took place 1~Gyr ago or earlier, we expect that the mixing of accreted material at the solar radius is now completed.  

\subsection{Kinematics versus abundances: mean relations}\label{mean}

\begin{figure}
\centering
\includegraphics[clip=true, trim = 0mm 0mm 0mm 0mm, width=0.8\linewidth]{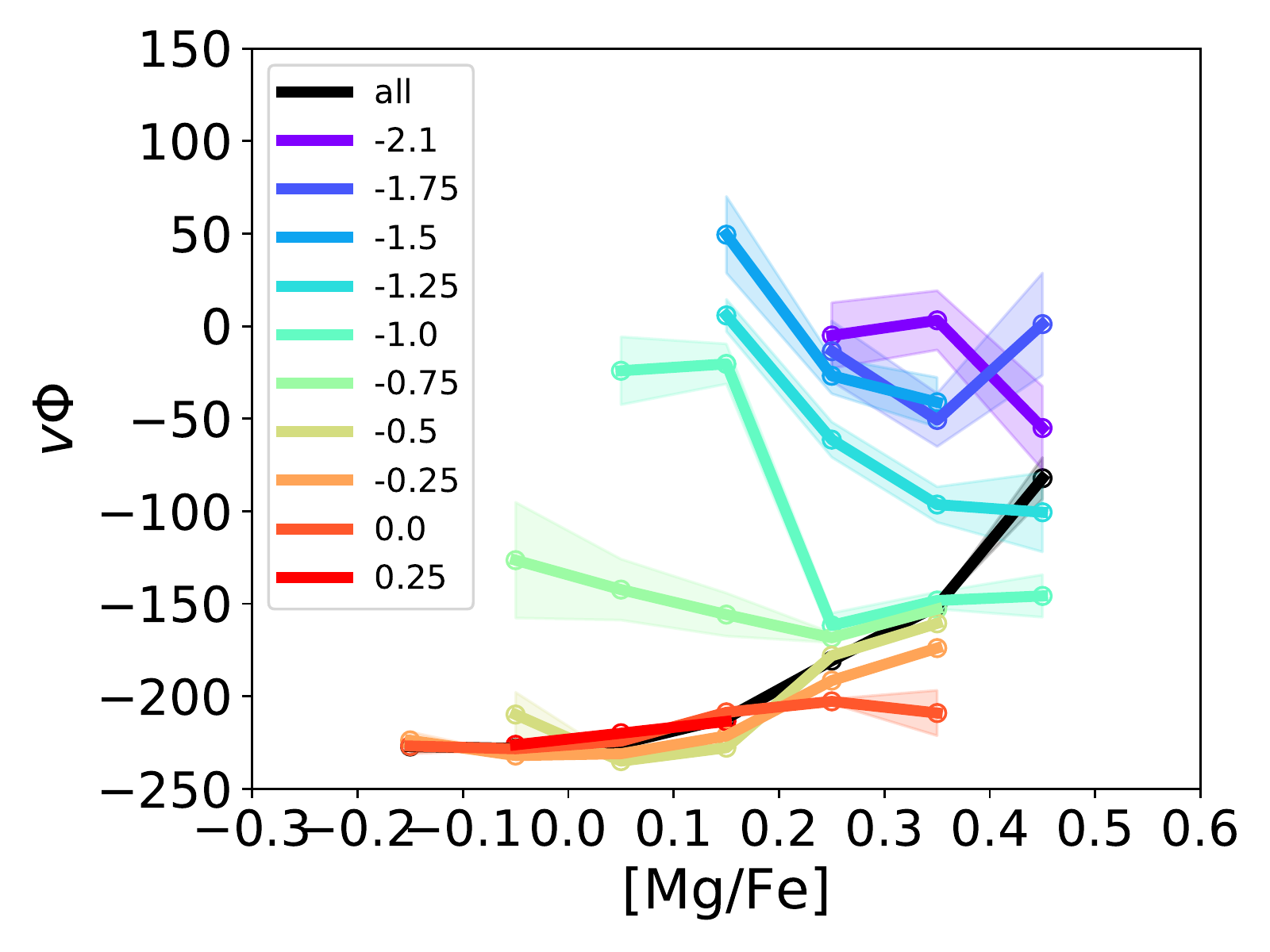}
\includegraphics[clip=true, trim = 0mm 0mm 0mm 0mm, width=0.8\linewidth]{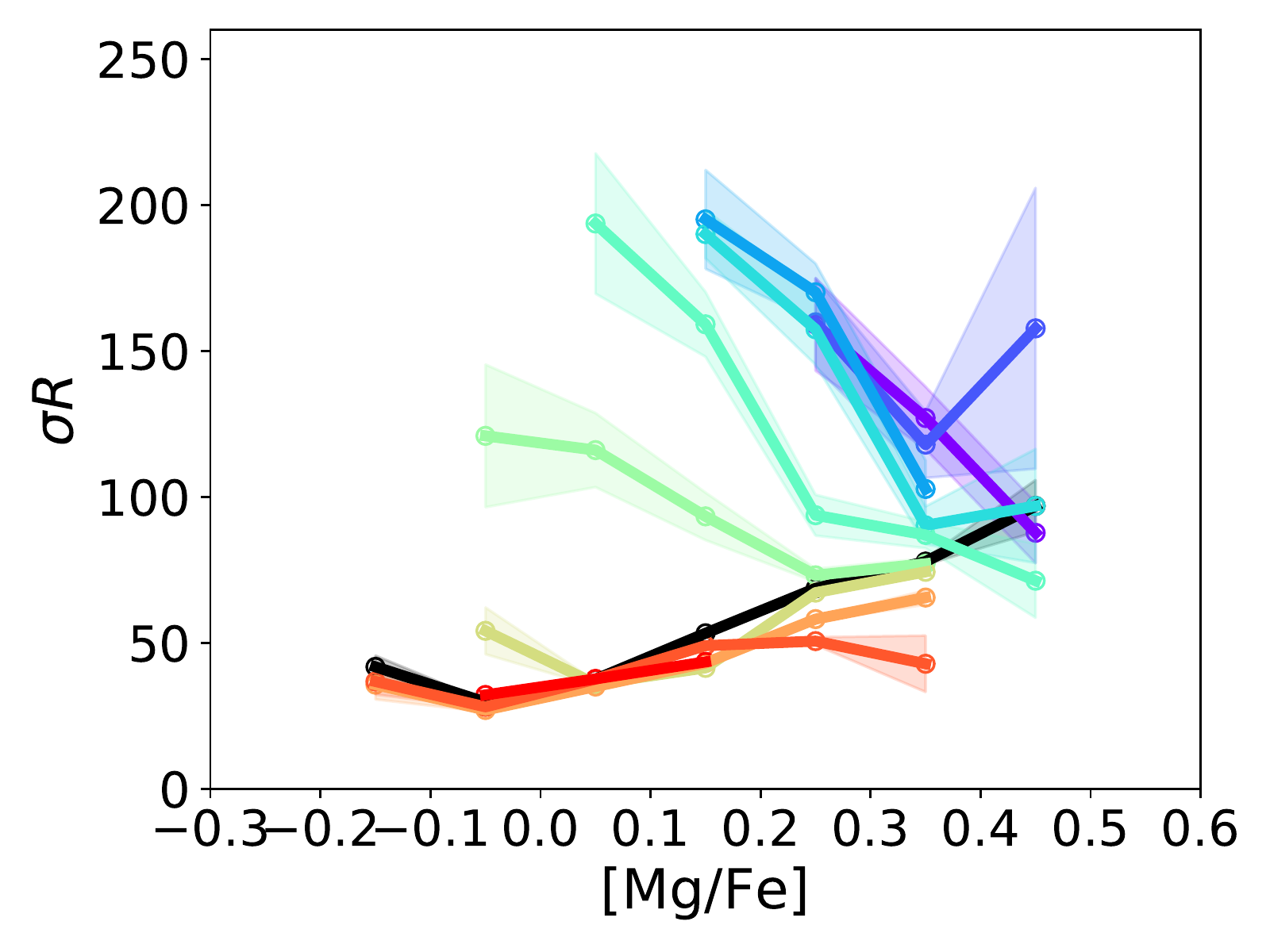}
\includegraphics[clip=true, trim = 0mm 0mm 0mm 0mm, width=0.8\linewidth]{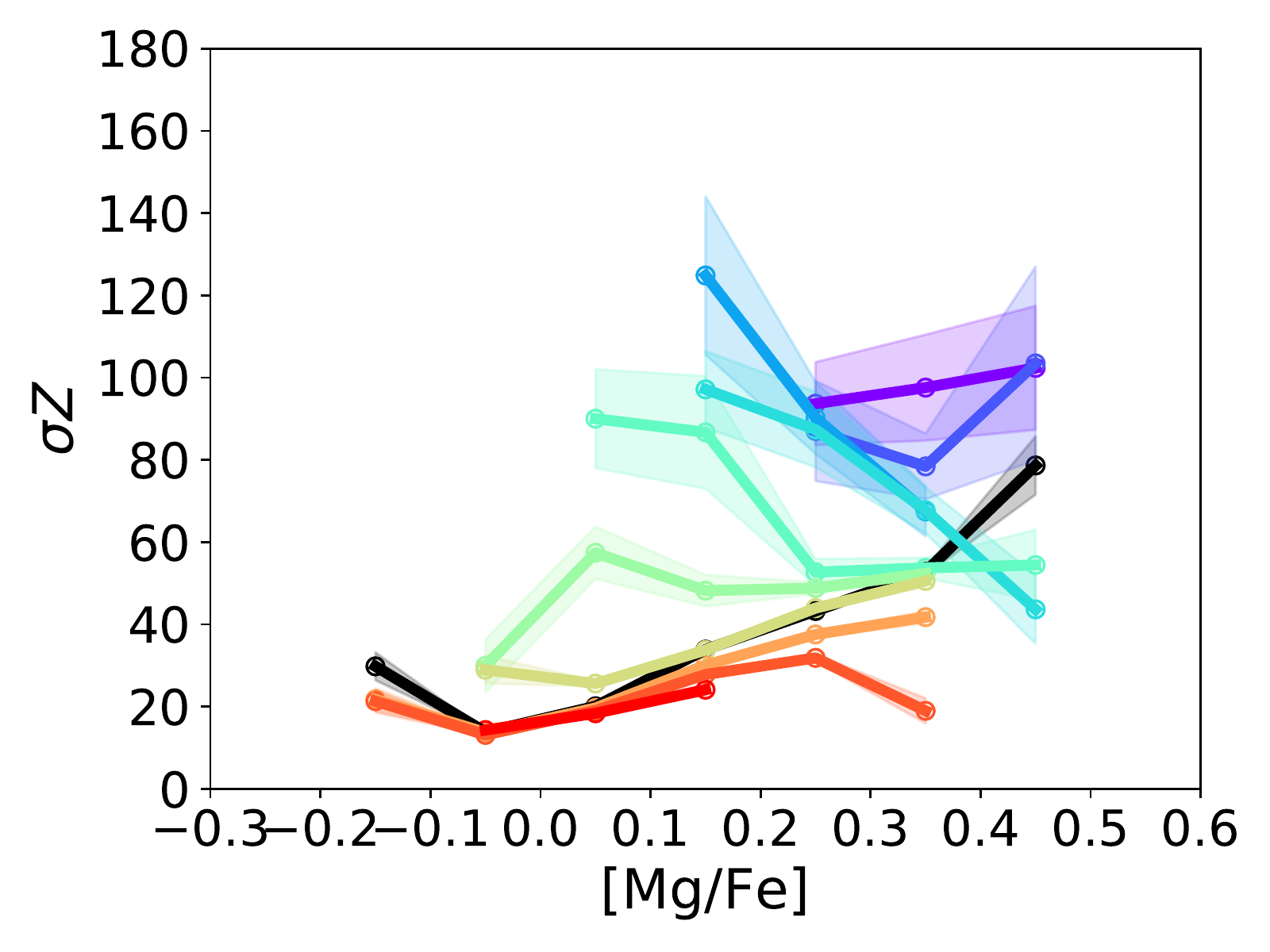}
\includegraphics[clip=true, trim = 0mm 0mm 0mm 0mm, width=0.8\linewidth]{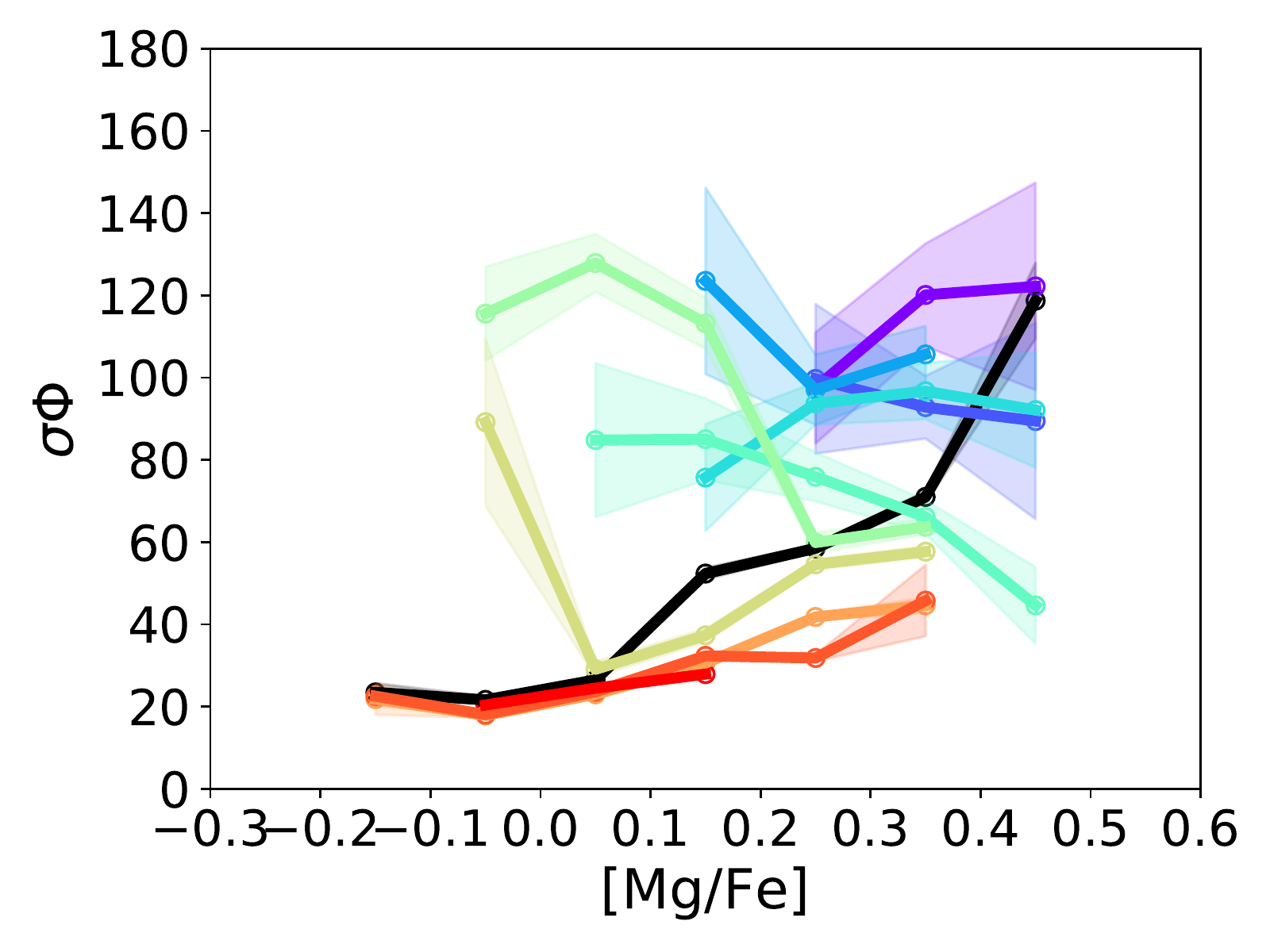}
\caption{(\emph{From top to bottom}): Mean azimuthal velocity, radial, vertical and azimuthal velocity dispersion of stars, as a function of their [Mg/Fe] ratio. In each panel, the relations are given for bins in [Fe/H], as indicated in the legend in the top-left panel. The black curves show the corresponding relation, for the total sample not binned in [Fe/H]. The $1\sigma$ uncertainty in each relation (colored shaded regions) is estimated through 1000 bootstrapped realizations. In all panels, only bins containing more than 10 stars are shown.}
\label{kins_vs_MgFe}
\end{figure}

\begin{figure}
\centering
\includegraphics[clip=true, trim = 0mm 0mm 0mm 0mm, width=0.8\linewidth]{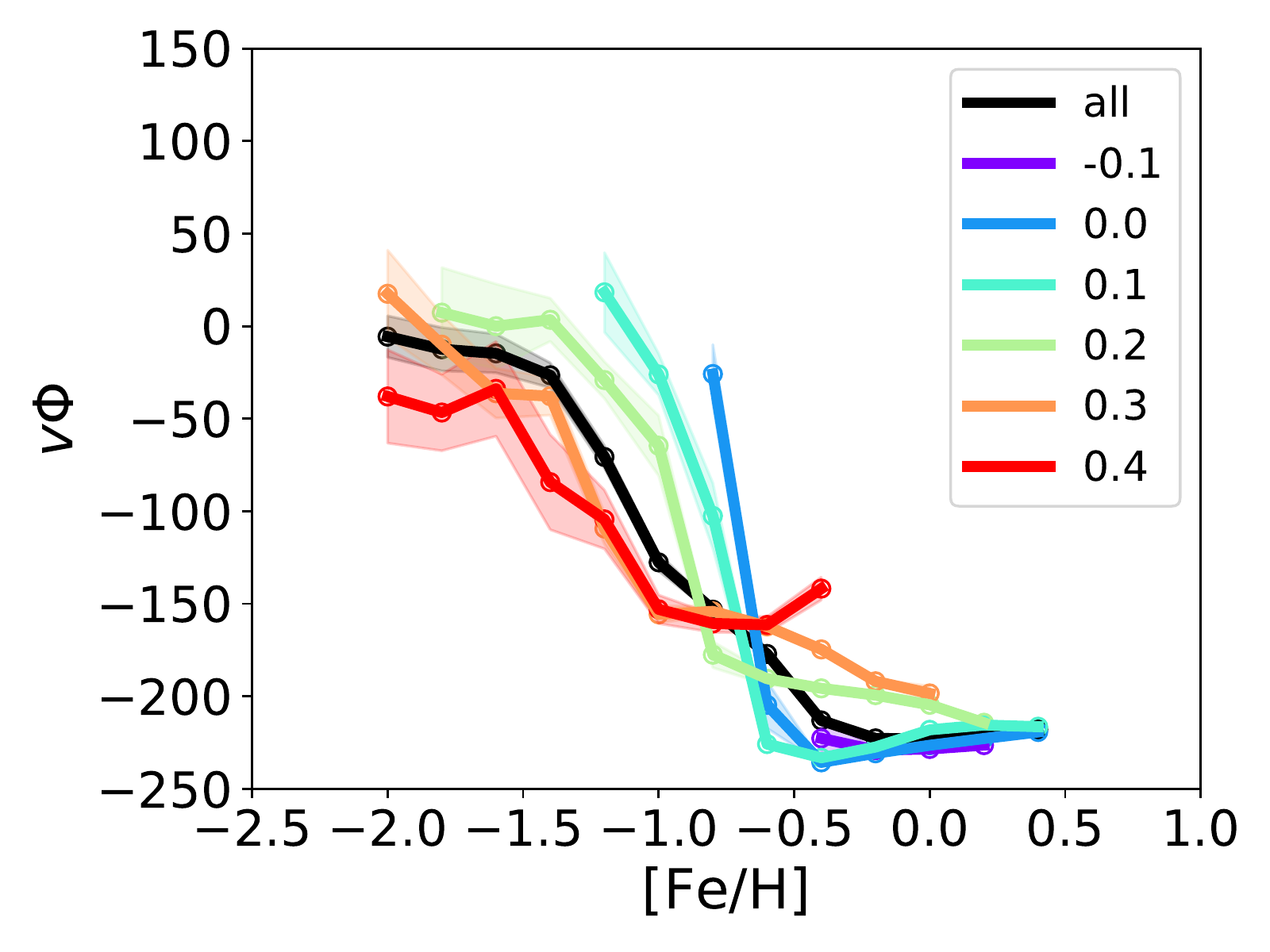}
\includegraphics[clip=true, trim = 0mm 0mm 0mm 0mm, width=0.8\linewidth]{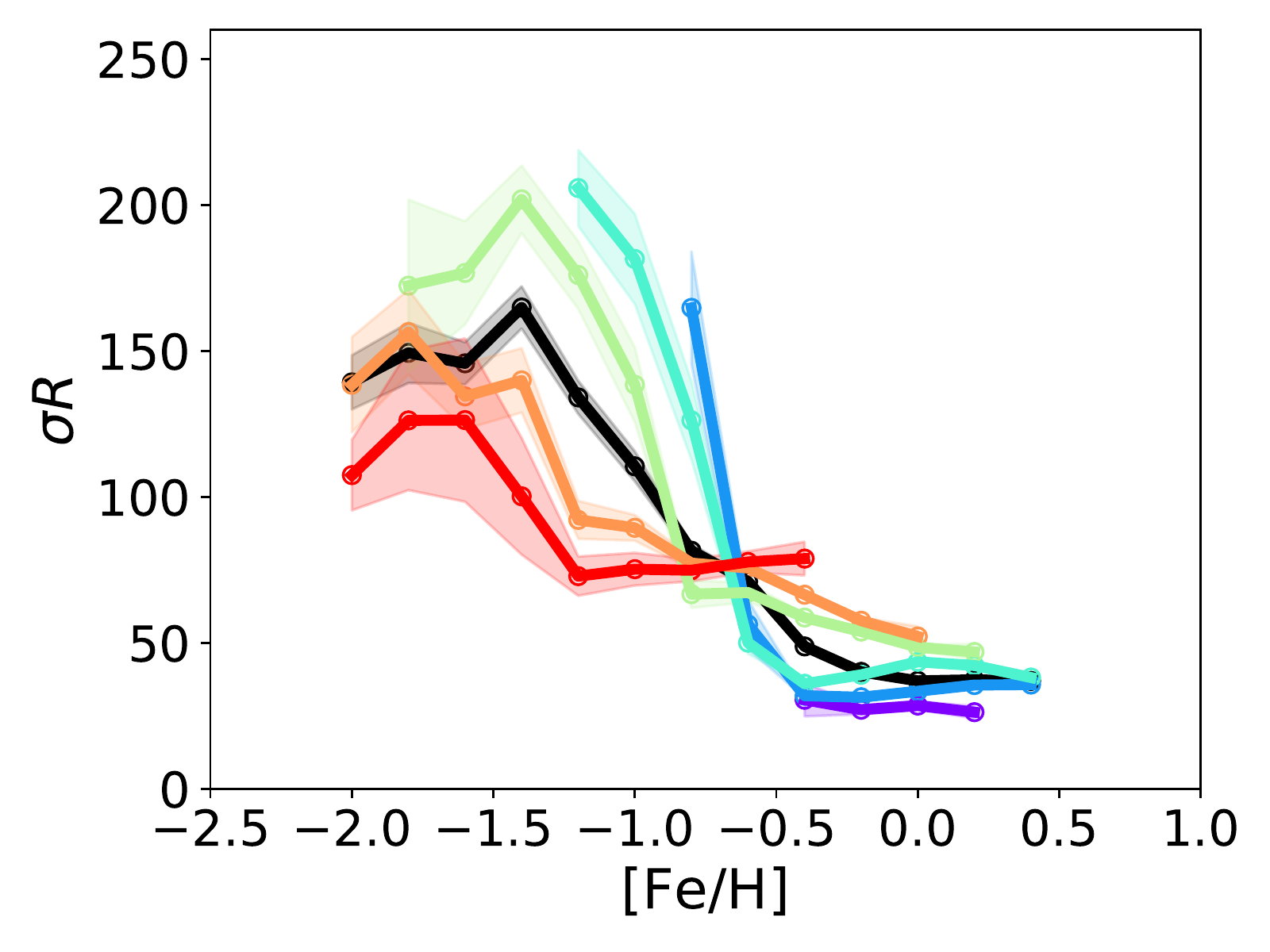}
\includegraphics[clip=true, trim = 0mm 0mm 0mm 0mm, width=0.8\linewidth]{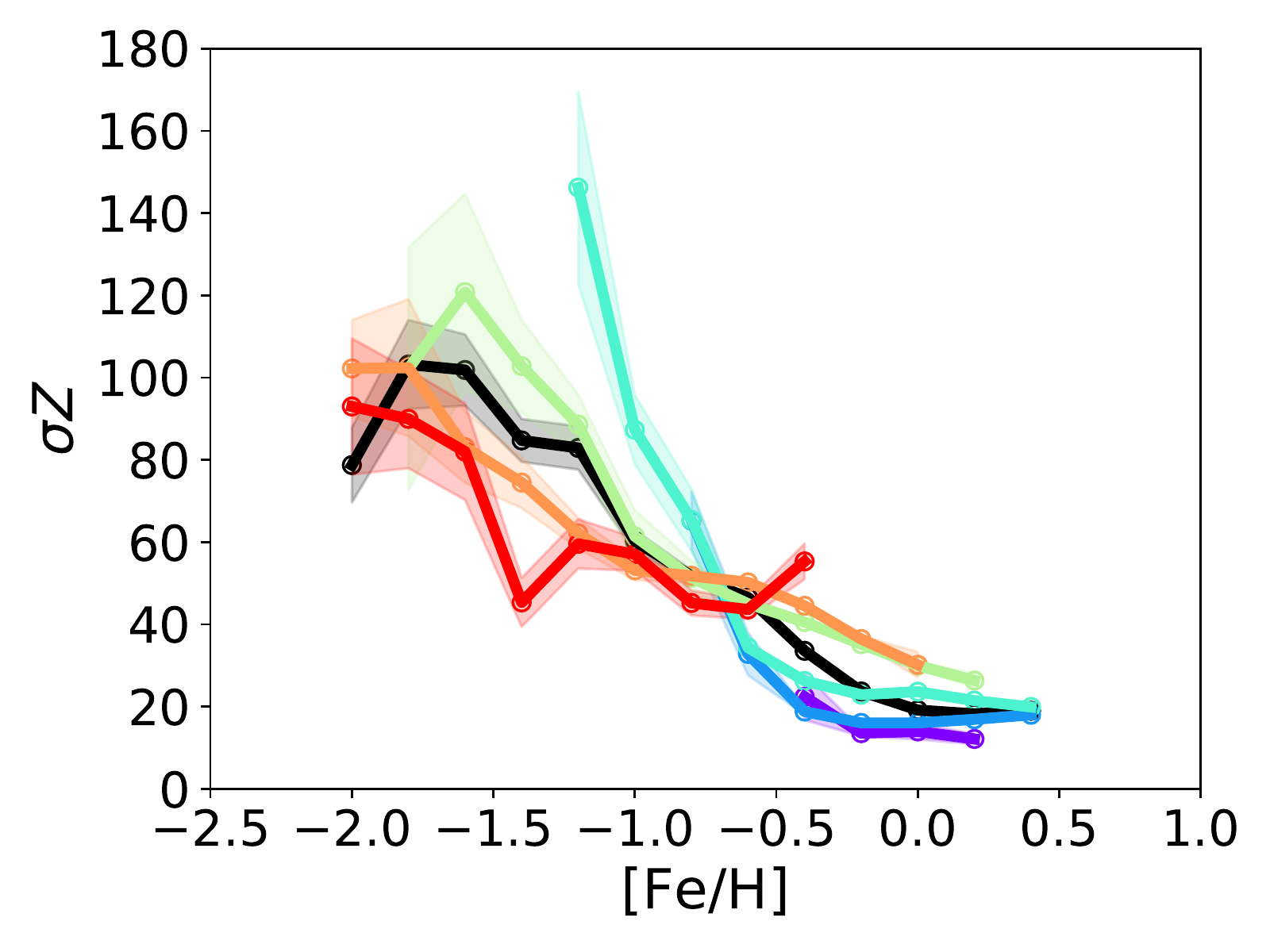}
\includegraphics[clip=true, trim = 0mm 0mm 0mm 0mm, width=0.8\linewidth]{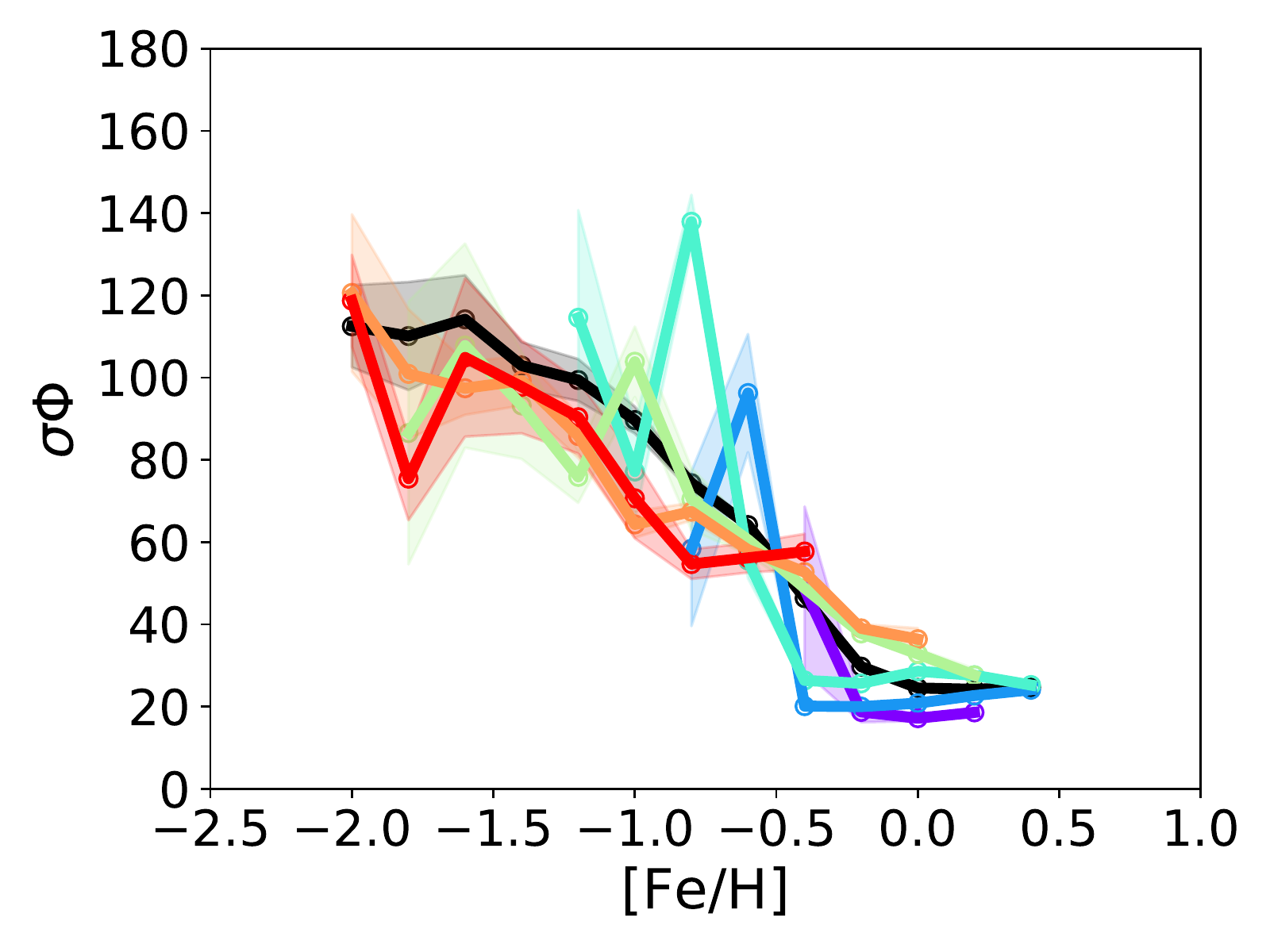}
\caption{(\emph{Panels from top to bottom}):  Mean azimuthal velocity, radial, vertical and azimuthal velocity dispersion of stars, as a function of their [Fe/H] abundance. In each panel, the relations are for various bins in [Mg/Fe], as indicated in the legend in the top-left panel. The black curves in each panel show the relation for the total sample. The $1\sigma$ uncertainty in each relation (colored region around each relation) is estimated using 1000 bootstrapped realizations. In all panels, only bins containing more than 10 stars are shown.}
\label{kins_vs_FeH}
\end{figure}

Fig.~\ref{FeHMgFe_map} (top-right panel and  bottom panels) shows the mean azimuthal velocity,  and the radial, vertical, and azimuthal velocity dispersions -- respectively $v_\Phi$,  $\sigma_R$,  $\sigma_Z$, $\sigma_\Phi$ -- of stars in the sample in different loci in the [Fe/H]-[Mg/Fe] plane. The [Fe/H] and [Mg/Fe] axis have been divided in $25$~bins. For each bin the mean azimuthal velocity and velocity dispersion of stars in that bin have been estimated and only bins containing more than 10 stars are shown (Fig.~\ref{FeHMgFe_map}).

The result is clear that, in the [Mg/Fe]-[Fe/H] plane, the low-$\alpha$ sequence of halo stars stands out as a distinct sequence in its kinematic properties with respect to both the high-$\alpha$ halo sequence and disc stars.
As an example, part of this sequence shows a mean retrograde motion as high as $v_\Phi=50$~\kms. At [Fe/H]$\sim-1$, the vertical velocity dispersion, $\sigma_Z$, is about $50$~\kms\ at the high-$\alpha$ end and increases to about $90$~\kms\ at the low-$\alpha$ end. Fig.~\ref{FeHMgFe_map} also shows that this oblique band of low $v_\Phi$ and high velocity dispersions in the [Fe/H]-[Mg/Fe] plane, appears to extend also to metallicities [Fe/H]$> -1$.  This implies that some accreted stars also have metallicities typical of disc stars. This is in agreement with the findings of \citet{mackereth18}, who found that some high eccentricity, low-$\alpha$ stars also have [Fe/H]$> -1$ (see Fig.~1 in their paper). It is also in agreement with the studies of \citet{nissen10, hayes18}, who show that the chemical pattern of the low-$\alpha$ sequence extends up to at least [Fe/H]$\sim -0.8$. In Sect.~\ref{disentangling}, we will show that accreted stars within this same sequence can have metallicities as high as [Fe/H]$\sim -0.5$. 

Fig.~\ref{FeHMgFe_map} shows that at all [Fe/H] where accreted, low-$\alpha$ stars are found, a rise in the velocity dispersions is expected. This result has naturally a direct impact on Galactic chemo-kinematic relations, as we show in  Fig.~\ref{kins_vs_MgFe}  by slicing the [Mg/Fe]-[Fe/H] in 10 metallicity bins, ranging from [Fe/H]$=-2.1$ up to [Fe/H]$=0.25$ (see Appendix~\ref{numbers_ck} for plots of the number of stars). For each bin, we consider all stars in the sample between [Fe/H]$_{\rm bin} \pm \Delta \rm[Fe/H]$,  [Fe/H]$_{\rm bin}$ with the central value of each metallicity bin given in the legend of Fig.~\ref{kins_vs_MgFe}, and $\Delta \rm[Fe/H]$=0.125 for all except the lowest metallicity bin, where $\Delta \rm[Fe/H]=0.225$ to increase its significance. We only show bins which contain more than 10 stars. 

We find that (Fig.~\ref{kins_vs_MgFe}): \textit{(1)} for the high metallicity bins, [Fe/H]$ > -0.5$, there is an overall increase in all velocity dispersions with  [Mg/Fe], as expected when moving from the cold kinematics of the thin disc, at low [Mg/Fe], to the hot kinematics of the $\alpha$-enhanced thick disc\footnote{The only exception to this trend is observed in the $\sigma_z$ versus [Mg/Fe] relation, for the bin centred at [Fe/H]$=0$ and [Mg/Fe]$=0.35$, where one observes a drop in the value of $\sigma_z$ comparable to those found for $\alpha$-poor bins}; \textit{(2)} for the metallicity bins, [Fe/H]=-0.5 and -0.75, the velocity dispersions are no longer increasing with [Mg/Fe] but become rather constant with [Mg/Fe]; \textit{(3)} at lower metallicities, the trend appears completely opposite to that observed at [Fe/H]$ > -0.5$, and all components of the velocity dispersions decrease with [Mg/Fe]; and \textit{(4)} finally, for [Fe/H]$< -1.75$, $\sigma_Z$ and $\sigma_\Phi$ do not show significant variations with [Mg/Fe].  It is also interesting that in the lowest metallicity bins the maximum of the velocity dispersion is displaced to lower [Mg/Fe], as [Fe/H] increases. This behaviour was noted already by  \citet{minchev14}, and here we demonstrate that it is a natural signature  of accreted stars whose abundances are characterized by an anti-correlation between [Mg/Fe] and [Fe/H].

Accreted stars also leave specific signatures in the $v_\Phi$--[Mg/Fe] relations, for different [Fe/H] (see Fig.~\ref{kins_vs_MgFe}). While the whole sample of stars shows a monotonic relation, with the mean rotation decreasing with [Mg/Fe], different trends are found when stars are grouped in bins of [Fe/H]: a monotonic relation of $v_\Phi$ with [Mg/Fe] is found for [Fe/H]$> -0.5$; at [Fe/H]$= -0.5$, an upturn is observed at lowest [Mg/Fe] end, with stars at  [Mg/Fe]$\sim -0.05$ rotating as slow as stars at [Mg/Fe]$\sim 0.2$; the mean $v_\Phi$ appears nearly flat and independent of [Mg/Fe] at [Fe/H]$= -0.75$, while the rotation increases with [Mg/Fe], for [Fe/H]$< -0.75$. Particularly remarkable are the trends observed at  [Fe/H]$=-1.$ and [Fe/H]$=-1.25$: here the high $\alpha-$bins have a mean azimuthal velocity lagging that of the LSR by 100-150 \kms, but have a mean prograde rotation, while the lowest $\alpha-$bins have a mean null or positive $v_\Phi$, indicative of a null or retrograde rotation. As for the velocity dispersion relations, we interpret the upturn observed at [Fe/H]$= -0.5$, and the features found at lower metallicities and low [Mg/Fe], as the consequence of the presence of the accreted population, which is characterized by a mean null or slightly retrograde rotation (see \citealt{nissen10, koppelman18, haywood18, helmi18}, and Sects.~\ref{disentangling} and ~\ref{dating}).

We conclude this section by presenting a similar analysis to that of Fig.~\ref{kins_vs_MgFe}, but this time slicing the [Fe/H]--[Mg/Fe] plane in bins of [Mg/Fe] (Fig.~\ref{kins_vs_FeH}).  Again, we investigate $v_\Phi$, $\sigma_R$, $\sigma_Z$ and $\sigma_\Phi$ relations as a function of [Fe/H].  For this, we have sliced the [Mg/Fe]-[Fe/H] in 6 bins of [Mg/Fe], ranging from [Mg/Fe]$=-0.1$ up to [Mg/Fe]$=0.4$. For each bin, we consider all stars in the sample between [Mg/Fe]$_{\rm bin} \pm \Delta \rm[Mg/Fe]$, [Mg/Fe]$_{\rm bin}$ is the central value of the bin (see legend of Fig.~\ref{kins_vs_FeH}). $\Delta \rm[Mg/Fe]$ is 0.05 for all bins. The trends of $v_\Phi$, $\sigma_R$, $\sigma_Z$ and $\sigma_\Phi$ as a function of [Fe/H] for the whole sample are as we expected -- increasing rotation and decreasing velocity dispersions with increasing [Fe/H]. However, for different bins of [Mg/Fe], these relations show specific characteristics. \textit{(1)} For [Fe/H]$> -0.5$ and [Mg/Fe]$< 0.1$, the rotation decreases as [Fe/H] increases. This trend, visible also in the $v_\Phi$ relation (Fig.~\ref{FeHMgFe_map}), has already been found in a number of studies \citep{haywood08,lee11} and, we confirm the existence of a positive gradient in the $v_\Phi-\rm [Fe/H]$ relation for thin disc stars with [Fe/H]$> -0.5$. \textit{(2)} For metallicities $\rm [Fe/H] \le -0.5$, there is a sharp increase in all the velocity dispersions--metallicity relations, and also in the $v_\Phi-\rm [Fe/H]$ relation, for all [Mg/Fe] bins. The steepness of the relations at $\rm [Fe/H] \le -0.5$ reaches a maximum for $\rm [Mg/Fe] = 0.0$, and decreases with increasing $\rm [Mg/Fe]$, and the metallicity where the increase becomes the most evident shifts towards lower metallicities as  $\rm [Mg/Fe]$ increases. In other words, for any given $\rm [Fe/H]$ below $-0.5$, the stellar kinematics become increasingly hotter as the relative $\alpha$-abundance ratio decrases. In particular at  $\rm [Fe/H] \le -1.5$, on the one hand, high $\rm [Mg/Fe]$ populations  ($\rm [Mg/Fe] =0.4$) have a mean prograde motion, while on the other hand, stars with $\rm [Mg/Fe] =0.2$ have a mean retrograde motion. This is a clear indication that at these metallicities, at least two different populations, with different mean kinematic properties, co-exist.

\begin{figure}[!ht]
\centering
\includegraphics[clip=true, trim = 0mm 0mm 0mm 0mm, width=0.8\linewidth]{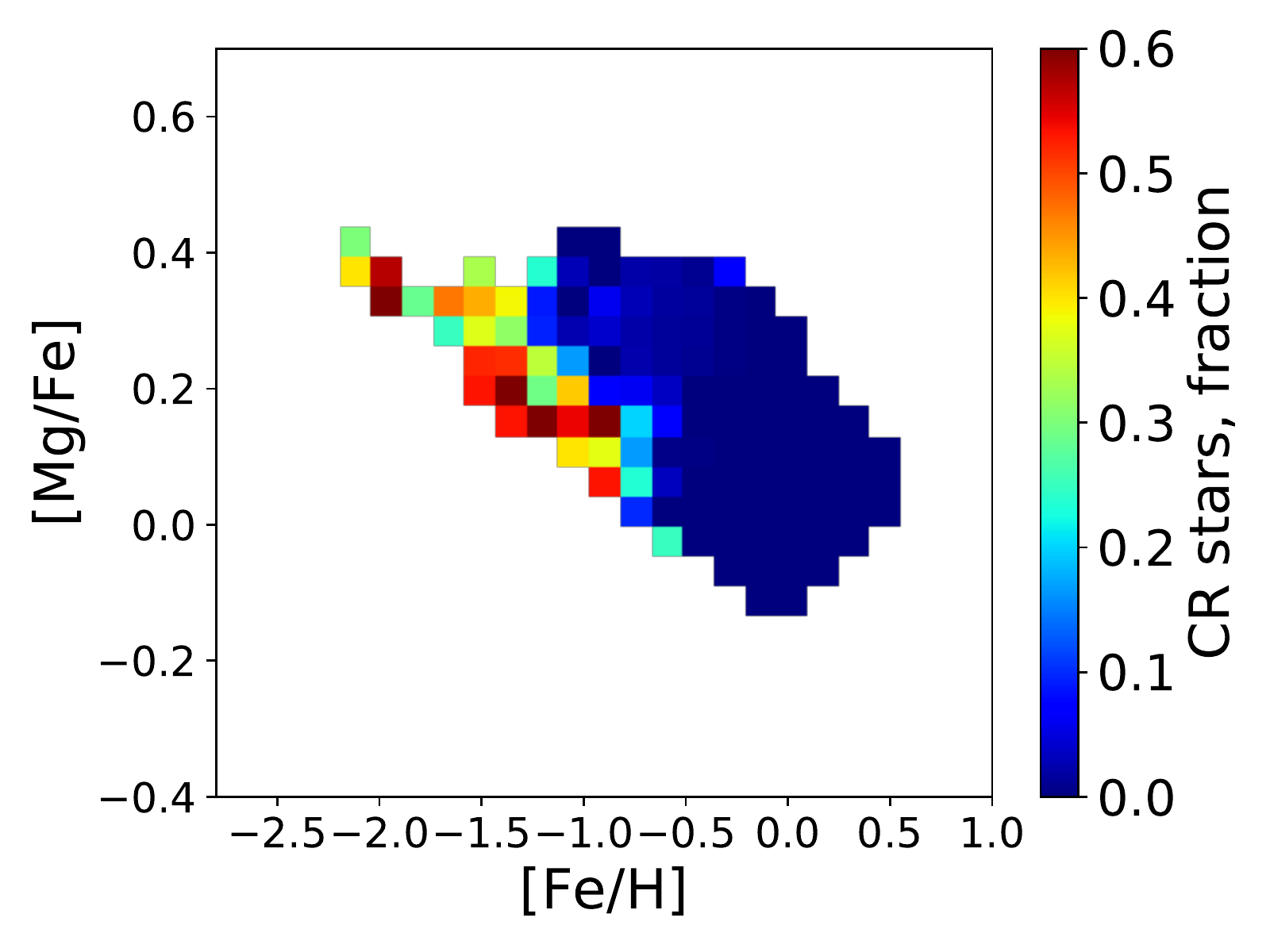}
\includegraphics[clip=true, trim = 0mm 0mm 0mm 0mm, width=0.8\linewidth]{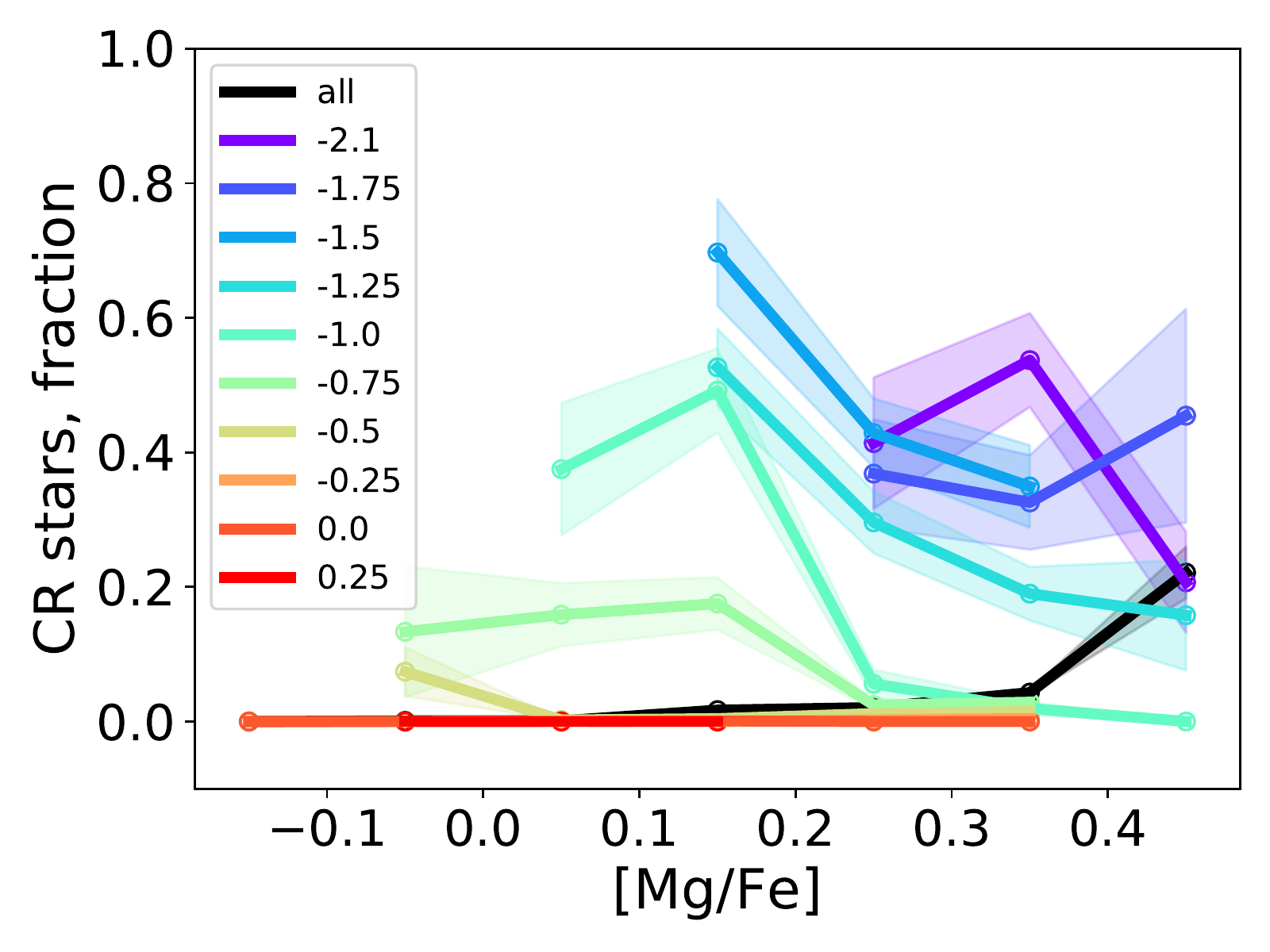}
\includegraphics[clip=true, trim = 0mm 0mm 0mm 0mm, width=0.8\linewidth]{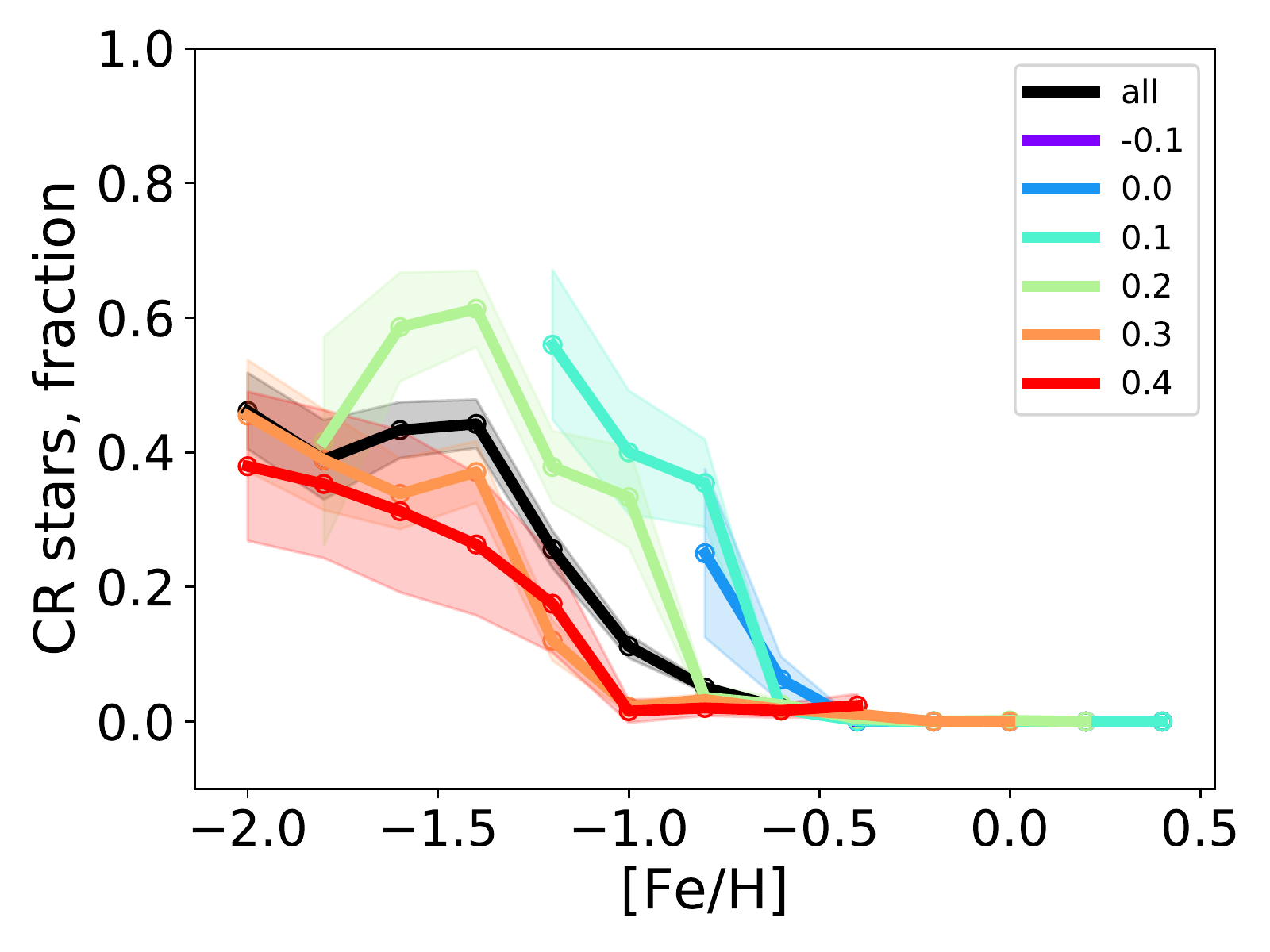}
\caption{(\emph{Top panel}): Fraction of counter-rotating stars in the [Fe/H]--[Mg/Fe] plane. The pixel size is 0.152$\times$0.044 and only pixels containing more than 10 stars are shown. (\emph{Middle panel}): Fraction of counter-rotating stars as a function of their [Mg/Fe] ratio. In each panel, colored curves indicate stars in specific bins of [Fe/H], as indicated in the legend. The black curves show the corresponding relation for the total sample. (\emph{Bottom panel}): Fraction of counter-rotating stars as a function of their [Fe/H] ratio. In each panel, colored curves indicate stars binned by their [Mg/Fe] values (see the legend in the upper left-hand corner of the panel). The uncertainties were estimated using a bootstrapping procedure as previously described. In the middle and bottom panels, only bins containing more than 10 stars are shown.}
\label{CR_map}
\end{figure}

Finally,  we analyze the fraction of stars counter-rotating with respect to the Galactic disc in the [Fe/H]--[Mg/Fe] plane, as well as their fraction as a function of [Mg/Fe] and [Fe/H] (Fig.~\ref{CR_map}).  Again, we sliced each of the two abundance planes in the same way we have done previously (Figs.~\ref{kins_vs_MgFe} and \ref{kins_vs_FeH}). The low-[Mg/Fe] sequence stands out as a distinct sequence also for its high fraction of retrograde stars, significantly higher than that of the disc and halo at similar [Fe/H], but higher [Mg/Fe]. While the fraction of counter-rotating (hereafter CR) stars is typically less than 1\% for disc stars with $\rm [Fe/H] > -0.5$, we see that by $\rm [Fe/H]=-0.5$ and $\rm [Mg/Fe]\sim0$, this fraction of CR-stars rises to more than 10\%, indicative of a significant population of accreted stars in this abundance range. As the metallicity decreases, the fraction of CR stars increases, especially for stars with low [Mg/Fe] ratios. For example, at $\rm [Fe/H]=-1$, the percentage of CR stars is few percent for the bin centred at $\rm [Mg/Fe]=0.4$, but this fraction rises to more than 50\% for  $\rm [Mg/Fe]\le0.2$. It is only for stars with $\rm [Fe/H]<-1$ that the fraction of CR stars rises for the highest [Mg/Fe] bin. This is obviously a  sign that below this metallicity, the accreted population significantly contaminates the high-$\alpha$ sequence of in-situ stars.

In all the analysis presented in this section, the statistical uncertainties in the relations have been estimated through a bootstrapping technique, over 1000 realizations. We refer the reader to Appendix~\ref{uncertainties} for a comparison of these uncertainties, with those obtained by propagating the individual uncertainties on the observables (parallaxes, proper motions, and radial velocities). 

\subsection{The fraction of in-situ and accreted stars across the [Fe/H]--[Mg/Fe] plane}\label{disentangling}

\begin{figure} 
\centering
\includegraphics[clip=true, trim = 0mm 0mm 0mm 0mm, width=\linewidth]{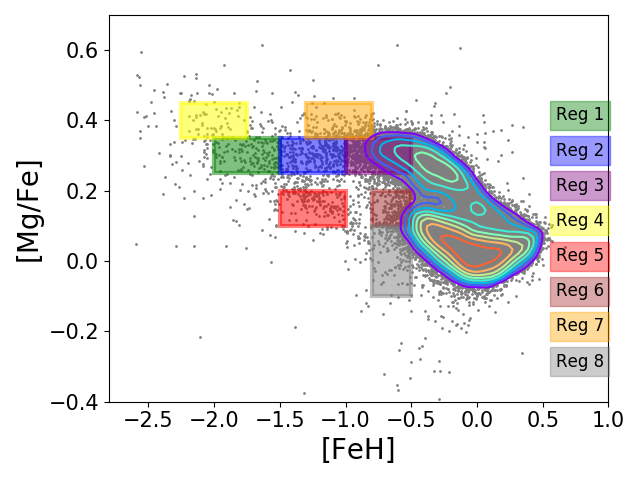}
\caption{The distribution of stars in the [Fe/H]--[Mg/Fe] plane with colored lines indicating isodensity contours (see Fig.~\ref{FeHMgFe_map}). We show eight regions which are defined in Sect.~\ref{disentangling} and are indicated with different colors as shown on the left-hand axis of the panel.}
\label{box_map}
\end{figure}


\begin{figure*}
\centering
\includegraphics[clip=true, trim = 0mm 0mm 0mm 0mm, width=0.8\linewidth]{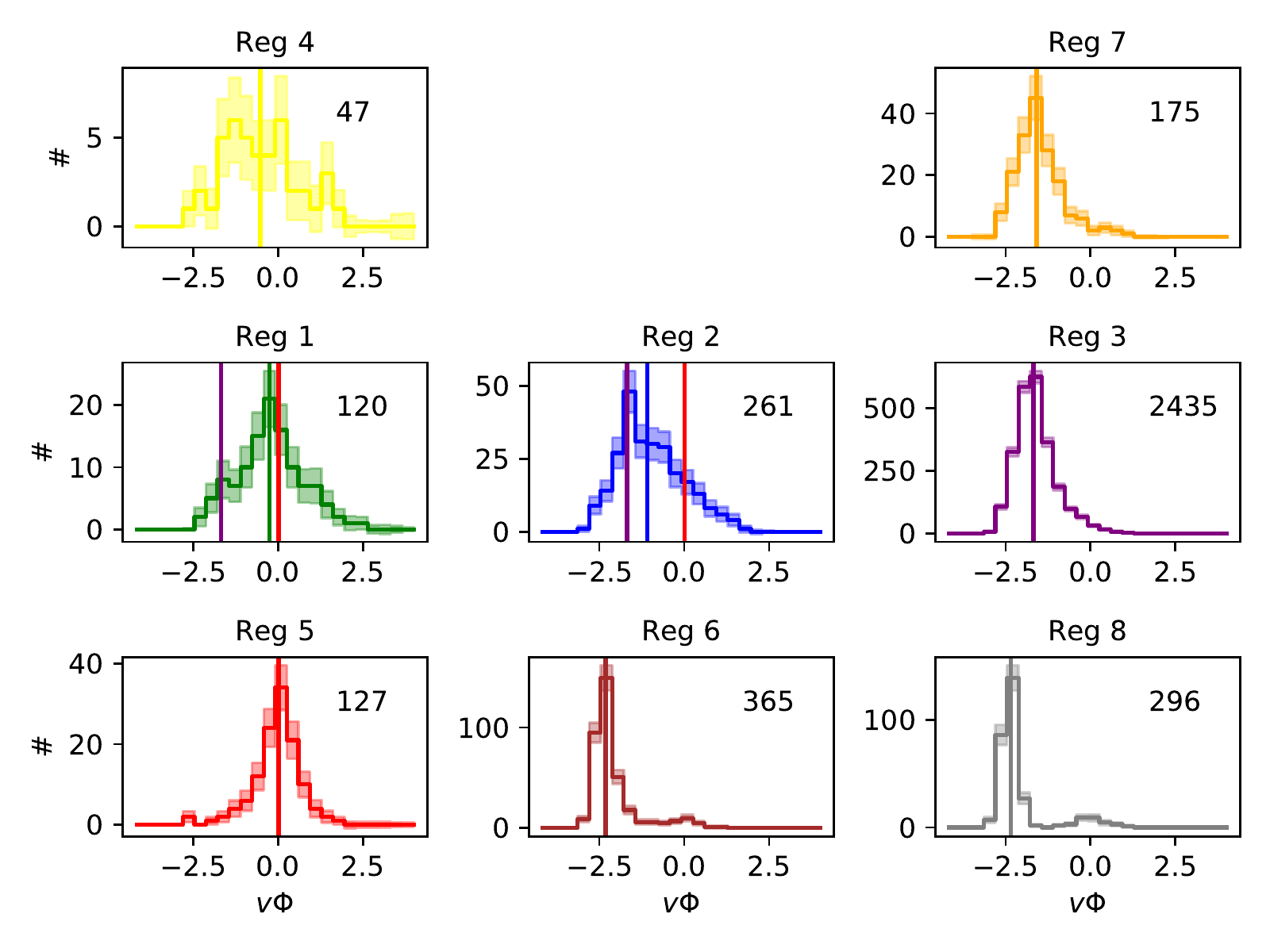}
\caption{The absolute distribution of the azimuthal velocities, $v_\Phi$, for stars in the 8 regions shown in Fig.~\ref{box_map} (see Sect.~\ref{disentangling}). In all panels: velocities are in units of 100~\kms\ and the colored solid  lines correspond to the medians of the distributions. In Regions~1 and 2 the solid purple and red lines correspond to the medians of the $v_\Phi$ distributions in Region~3 and  5, respectively. The number of stars in each region is provided in the upper left corner of each panel. The $1\sigma$ uncertainties (colored shaded regions) have been estimated by taking into account both the statistical uncertainty, through 1000 bootstrapped realizations, and the individual uncertainties propagated from the observables.  } 
\label{vPHI_histo}
\end{figure*}

\begin{figure} 
\centering
\includegraphics[clip=true, trim = 0mm 0mm  0mm 0mm, width=0.8\linewidth]{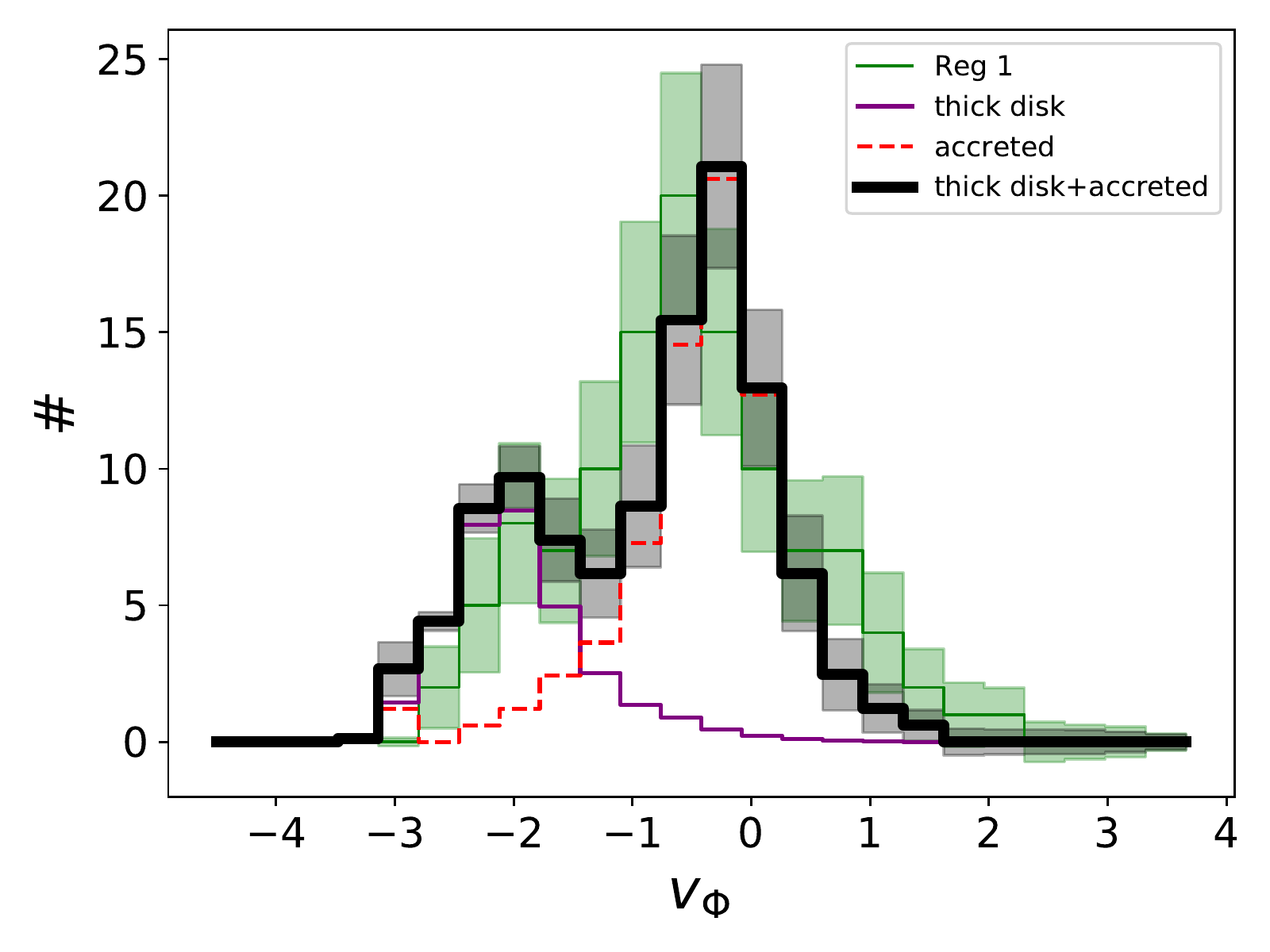}
\includegraphics[clip=true, trim = 0mm 0mm  0mm 0mm, width=0.8\linewidth]{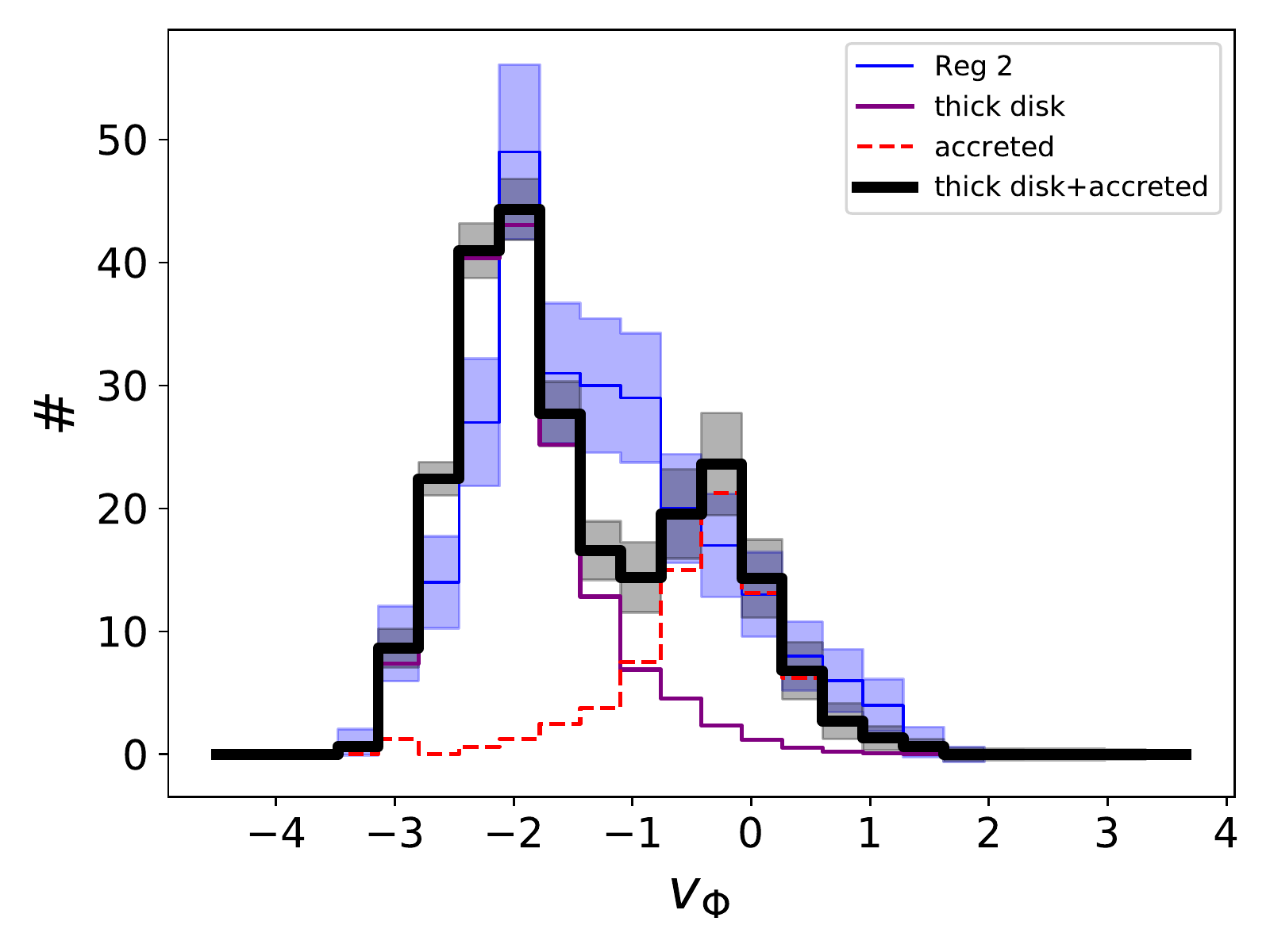}
\caption{The absolute distribution of the azimuthal velocities, $v_\Phi$, for stars in Region~1 (top panel) and 2 (bottom panel). Also shown are histograms the distributions of Regions~3 and 5 (solid purple and dashed red lines, respectively), normalized as described in the text, and their sum (thick black lines). The $1\sigma$ uncertainty in the distribution of stars in Region~1 and~2 are shown, respectively,  by green and blue shaded regions (top and bottom panels), and the uncertainty in the normalized sum of  Regions~3 and 5  is shown by a grey shaded area. All uncertainties have been estimated by taking into account both the statistical uncertainty, through 1000 bootstrapped realizations, and the individual uncertainties propagated from the observables.  In all panels: velocities are in units of 100~\kms.  } 
\label{vPHI_ISAC}
\end{figure}

\begin{figure*} 
\centering
\includegraphics[clip=true, trim = 0mm 0mm 0mm 0mm, width=0.8\linewidth]{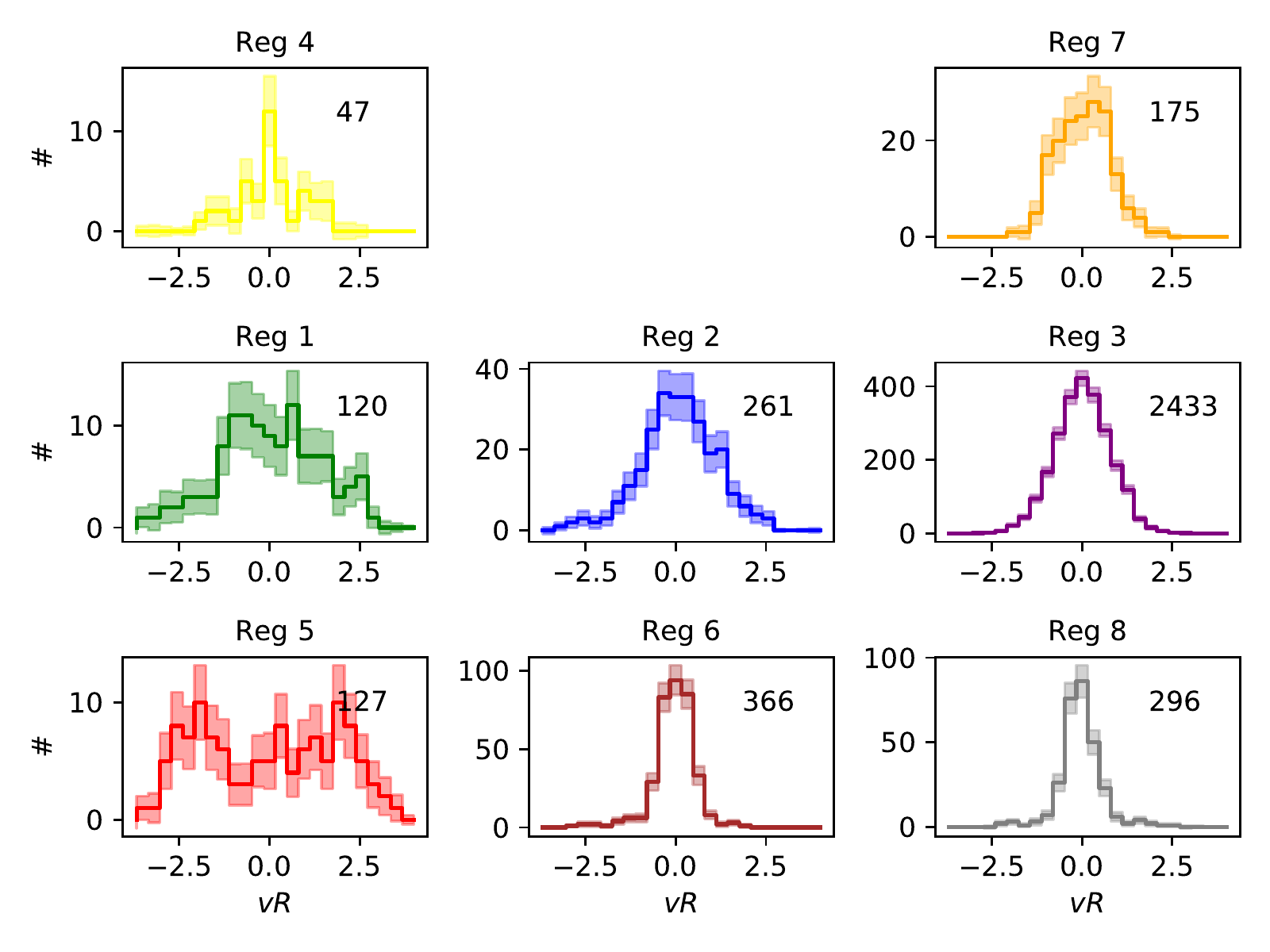}
\caption{The absolute distribution of the radial velocities, $v_R$, for stars in the 8 regions defined in Fig.~\ref{box_map} (see Sect.~\ref{disentangling}). In all panels: velocities are in units of 100~\kms. The number of stars in each region is given at the upper left of each panel. The $1\sigma$ uncertainties (colored shaded regions) have been estimated by taking into account both the statistical uncertainty, through 1000 bootstrapped realizations, and the individual uncertainties propagated from the observables.  }
\label{vRAD_histo}
\end{figure*}


\begin{figure*}
\centering
\includegraphics[clip=true, trim = 0mm 0mm 0mm 0mm, width=0.75\linewidth]{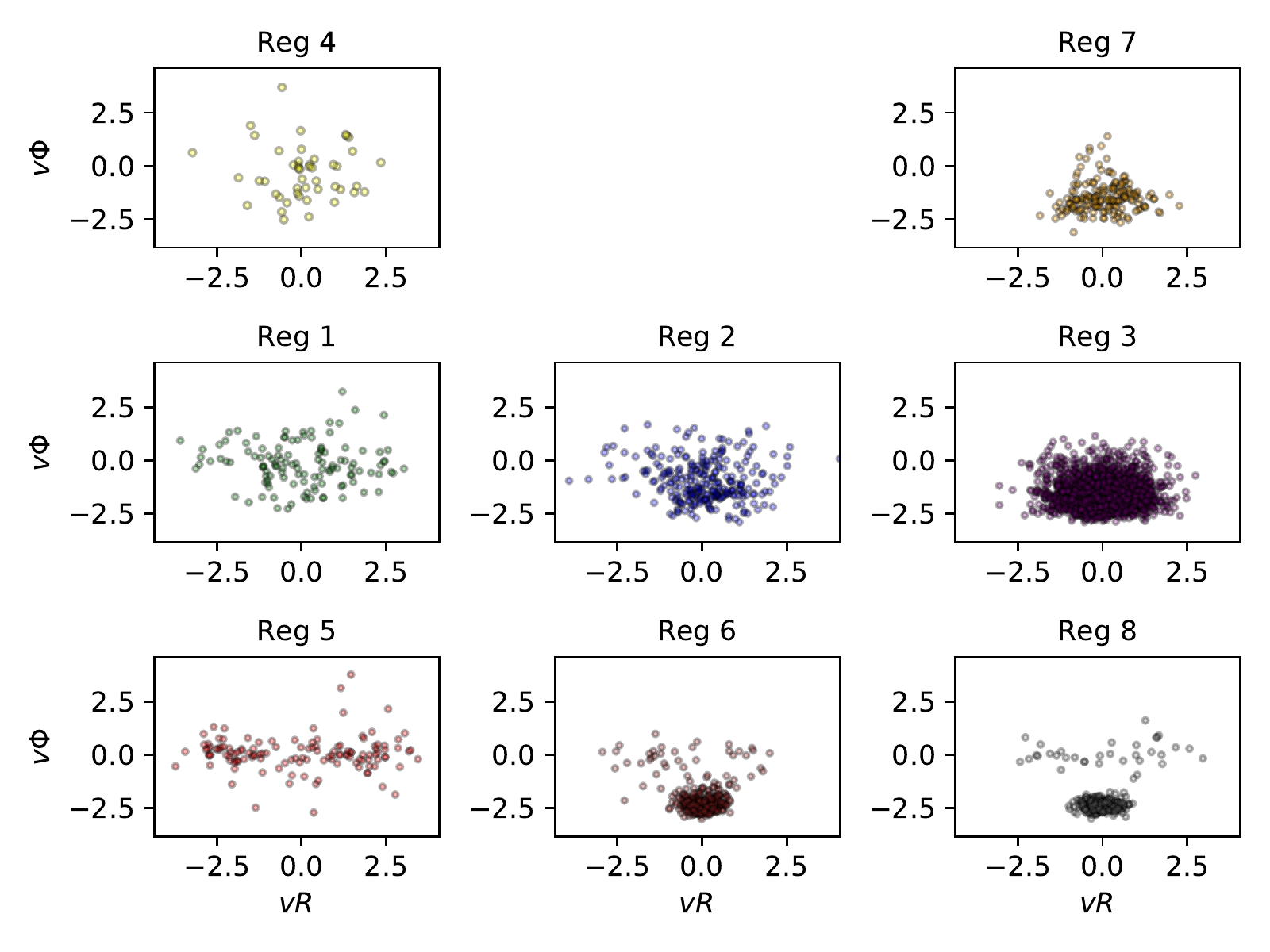}
\includegraphics[clip=true, trim = 0mm 0mm 0mm 0mm, width=0.75\linewidth]{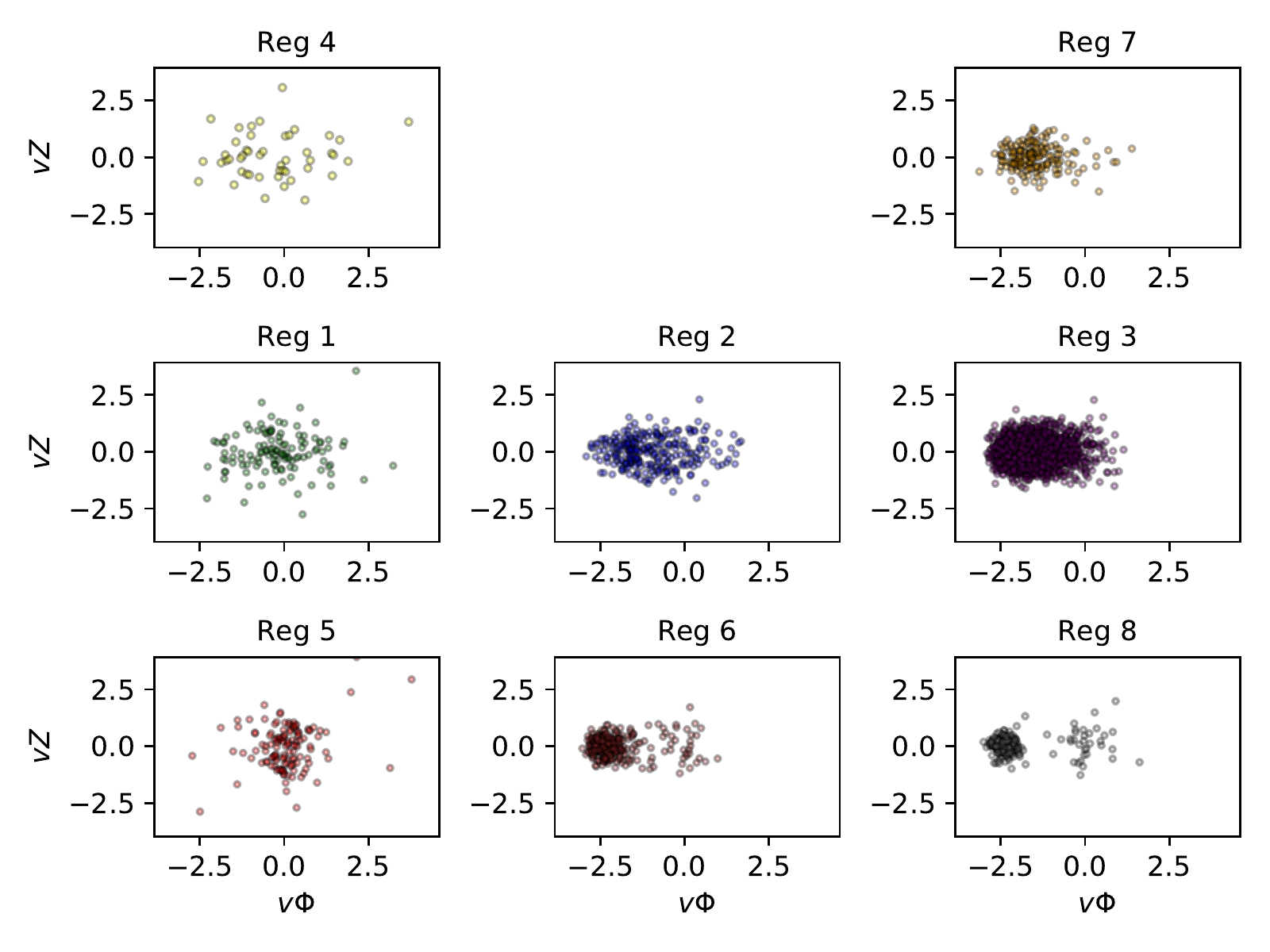}
\caption{Distribution of stars in the $v_R-v_\Phi$ plane (\emph{Top panels}) and in the $v_\Phi-v_Z$ plane (\emph{Bottom panels}) for the eight regions defined in Fig.~\ref{box_map} (see Sect.~\ref{disentangling}). In all plots, velocities are in units of 100~\kms.}
\label{vR-vZ-vPHI}
\end{figure*}

In the previous section, we have presented the chemo-kinematics relations of stars in the \gaia~DR2-APOGEE sample, addressing the role accreted stars play in shaping these relations. We have seen that already at the extreme of the metal-poor thin disc sequence, $\rm [Fe/H] = -0.5$ and $\rm [Mg/Fe] < 0.1$, the presence of accreted stars can be demonstrated due to the signature they leave on the stellar kinematics.  
The question we want to address in this Section concerns the relative fraction of accreted and in-situ stars for $\rm [Fe/H] \le-0.5$, and in particular, whether an in-situ population can be identified also at $\rm [Fe/H]\le-1.5$, in a region where the low and high-$\alpha$ sequences merge and where no differentiation between accreted and in-situ populations seems possible only on the basis of elemental abundances.

To make a step further in our understanding of accreted and in-situ populations, we now present the velocity distributions of stars in different regions of the [Fe/H]--[Mg/Fe]  plane. Our aim is to use these distributions and the way they change across the abundance plane to weight the fraction of accreted and in-situ material for different metallicity and [Mg/Fe] bins. In doing so, we will be also naturally lead to discuss the origin of the in-situ population found across the  [Fe/H]--[Mg/Fe]  plane. For this analysis, we have defined 8 regions in the [Fe/H]--[Mg/Fe] at the crossroad between the thick disc, the metal-poor thin disc, the low and high-$\alpha$ halo sequences as follows (see also Fig.~\ref{box_map}). 

\begin{itemize}

\item Three regions are chosen along the high-[Mg/Fe] sequence: Region~1 contains stars in the interval $\rm [Fe/H]-\rm [Mg/Fe] = [[-2., -1.5], [0.25, 0.35]]$, Region~2 stars in the interval  $\rm [Fe/H]-\rm [Mg/Fe] = [[-1.5, -1], [0.25, 0.35]]$  and Region~3 stars in the interval $\rm [Fe/H]-\rm [Mg/Fe] = [[-1., -0.5], [0.25, 0.35]]$.

\item Three regions are chosen along the low-[Mg/Fe] sequence of accreted stars: Region~5 contains stars in the interval $\rm [Fe/H]-\rm [Mg/Fe] = [[-1.5, -1], [0.1, 0.2]]$,  Region~6 contains stars in the interval $\rm [Fe/H]-\rm [Mg/Fe] = [[-0.8, -0.5], [0.1, 0.2]]$, thus at the upper edge of the metal-poor thin disc, and Region~8 contains stars with $\rm [Fe/H]-\rm [Mg/Fe] = [[-0.8, -0.5], [-0.1, 0.1]]$, thus at the lower edge of the metal-poor thin disc and below it.

\item Finally two regions have been chosen at the upper edge of the high-[Mg/Fe] sequence: Region~4 contains stars in the interval $\rm [Fe/H]-\rm [Mg/Fe] = [[-2.25, -1.75], [0.35, 0.45]]$ and Region~7 stars in the interval $\rm [Fe/H]-\rm [Mg/Fe] = [[-1.3, -0.8], [0.35, 0.45]]$. Despite these two regions contain fewer stars than the previous ones, they have been chosen to tentatively probe the in-situ/accreted content also at these levels of  [Mg/Fe]-enhancement.

\end{itemize}

Two regions of the 8 are particularly important for this study as they will serve as fiducial regions with which to compare and analyze all of the regions. One is Region~3, which is not contaminated by the accreted sequence of stars with low-[Mg/Fe] and which we will use as the reference for  in-situ stars belonging to the galactic thick disc. The other is Region~5 which lies on the elemental-abundance defined sequence for stars that were accreted. As we will see in the following, stars in Region~5 have peculiar kinematics, clearly different from that of Region~3. This suggests that the properties of stars in Region~5 can be used as reference for the accreted sequence. We caution that even if we refer to  Region~3 as reference for the  thick disc, it also contains a tail in its distribution of properties that includes stars that are slowing rotating and even counter-rotating relative to disc stars. Despite being not contaminated by the accreted sequence of low-[Mg/Fe] stars, the kinematics of stars in this region has been indeed significantly affected by this accretion (see Sect.~\ref{dating}).

In Fig.~\ref{vPHI_histo}, we show the distributions of the azimuthal velocities, $v_\Phi$, for stars in each of these regions. For each region, the medians of the distributions are also shown. {Uncertainties in this and in the following distributions (see Figs.~\ref{vPHI_ISAC} and \ref{vRAD_histo}) have been calculated by bootstrapping the sample, and for each bootstrap realizations we have taken into account also the individual uncertainties in the velocities, by a Monte-Carlo sampling. We have considered explicitly  the individual uncertainties to make sure that the distributions are robust also for stars with halo kinematics, whose individual uncertainties, on average, are higher than those of disc stars (see Fig.~\ref{veluncert_vs_x}).\\
The  ``pure''  accreted region (Region~5) is characterized by an average weak retrograde distribution peaking  $v_\Phi=2.5$~\kms\ (median), while the ``pure'' thick disc region (Region~3) by an average prograde rotation peaking at  $v_\Phi=-168.1$~\kms\ (median). Moving from Region~3, to Region~2 and then Region~1 and thus moving to lower metallicities, at constant [Mg/Fe] ratio, the $v_\Phi$ distribution can be qualitatively described as the weighted sum of the distributions of Regions~3 and 5. Thus it can be viewed as the combination of thick disc and accreted stars and the stellar population rotate more slowly on average. The medians of the distribution, indeed, increase to $v_\Phi= -109.4$~\kms\  in Region~2 and to  $v_\Phi=-23.8$~\kms\ (median) for Region~1. However, the comparison of the distributions in these three regions shows that the  rotation decreases not because of a decrease in the rotation of the thick disc at lower metallicities: indeed, the purple line, which indicates the median of the distribution in Region~3, always coincides with the fast rotating peak of the distributions of Region~2 and  Region~1. The decrease of the median rotation with decreasing [Fe/H], is indeed most likely due to an increase of the fraction of stars with null or retrograde motion. The evidence for this is the increasing fraction of stars in Region~2 and Region~1 with  $v_\Phi$ equal or higher than the median $v_\Phi$ value of Region~5. As we demonstrate in the following, the distribution of $v_\Phi$ in Regions~1 and 2 can be understood as a weighted sum of the $v_\Phi$ distributions of Regions~3 and 5, with the relative fraction of the accreted stars increasing when moving from Region~2 to 1. 
 It is natural to make use of this modulation of $v_\Phi$ distributions to try to derive the fraction of accreted and in-situ stars, on the high-[Mg/Fe] sequence, as a function of [Fe/H]. We caution the reader that the approach used here is simple and should only be thought of as a first attempt to estimate the relative fractions of these two populations in different regions of the abundance plane.
 
 In Fig.~\ref{vPHI_ISAC}, we show the histograms of the azimuthal velocities $v_\Phi$ for Regions~1 and 2, this time overlaying on each of them the $v_\Phi$ distributions of Region~3, which we consider as consisting purely of thick disc stars and Region~5, as purely consisting of accreted stars. To compare the distributions of Regions~3 and 5 to those of Regions~1 and 2, we have applied a $\chi^2$-minimization to find the normalization constants for the distributions of Regions~3 and 5, that minimize the quadratic difference between the  normalized sum of Regions~3 and 5, and the distribution of Regions~1 (and ~2).

 In this way, we can estimate the fractional contribution of  Region~3 and 5, to  Region~1 and 2, respectively, obtaining the following results.   For Region~2, that -- we remind the reader -- includes stars with $\rm [Fe/H]-\rm [Mg/Fe] = [[-1.5, -1], [0.25, 0.35]]$, we estimate the fraction of accreted stars to be about  25\% of the total number of stars in these abundance intervals. This fraction increases up to about 70\% in Region~1, which includes all stars in the sample with $\rm [Fe/H]-\rm [Mg/Fe] = [[-2., -1.5], [0.25, 0.35]]$. To further test the null  hypothesis that the distributions in  Region~1 and 2 can be described as the sum of the distributions of stars in the reference thick disc region, and of stars in the accreted region, we have run a Kolmogorov-Smirnov test, finding a KS-statistics and a p-value  equal to 0.08 ans 0.95 for Region~1 and 0.12 and 0.91 for Region~2. These values show that the hypothesis that Region 1 and 2 are made simply  of stars of the thick disc and of accreted material cannot be rejected with a high level of confidence. 
 
 Before discussing the $v_\Phi$ distributions in other regions of the $\rm [Fe/H]-\rm [Mg/Fe]$ plane, we would like to comment on the excess of stars at $v_\Phi\sim -100$~km/s that appears when comparing the distribution of stars in Region~2 with the normalized sum of Region~3 and 5 (see Fig.~\ref{vPHI_ISAC}, bottom panel). The current uncertainties make this excess of stars not statistically significant (at these values of $v_\Phi$, the distributions are compatible within 2-$\sigma$. If this excess will be confirmed by further studies, some hypothesis about its origin can be however proposed: 1) it could be mainly made of accreted stars belonging to the same major accretion event that makes the majority of stars found in Region~5 (in \citet{jeanbaptiste17} we have indeed shown that accreted stars which come from the same parent satellite do not necessarily all have the same kinematic properties, these latter depending on the time they were stripped from the parent satellite, at the first passage or later); 2) it could be still part of the thick disc, that is this excess could still have, for the majority,  an in-situ origin, and this would imply that  the rotation of the thick disc, at these metallicities, would be slightly lower than that of thick disc stars at higher [Fe/H]. In all cases, we emphasize that this excess of stars -- if real -- represents only  about 10\% of stars of Region~2,  that is less than 5\% of all stars with $\rm [Fe/H] \le -1$ in our sample, thus not impacting our conclusions.
 
 Because of the similarity of its $v_\Phi$ distribution with that of Region~3, and because of its location in the $\rm [Fe/H]-\rm [Mg/Fe]$ plane, we conclude that Region~7 is not contaminated by accreted stars, while some accreted stars are found in the Region~6 and Region~8: indeed while the vast majority of stars in these regions have prograde motions, a secondary, weak peak at positive $v_\Phi$ is visible in both distributions. This is further  evidence, together with the analysis presented in Sect.~\ref{mean},  that accreted stars are present up to the edge of the metal-poor thin disc, $\rm [Fe/H] \sim -0.5$, at the level of few percent. We do not draw any conclusions about Region~4 because of very poor statistics.
 
 By summing the contribution of in-situ stars in the Regions~1, 2 and 5 and by comparing it to the total number of stars in those regions, we can derive a rough estimate of their fraction in our sample for $\rm [Fe/H]\le-1$.  We estimate that the contribution of in-situ stars is $\sim$40-45\%. While we must be cautious about the actual values, our estimate indicates that thick disc stars should constitute a non-negligible fraction of the stars found at metallicities $-2 \le \rm [Fe/H] \le -1$. These stars constitute the metal-poor tail of the thick disc, the so-called metal-weak thick disc (see Sect~\ref{discussion}).
 
Finally, the distributions of radial velocities, $v_R$, shown in Fig.~\ref{vRAD_histo} reinforces our previous conclusions. Region~5, which we take as the reference for a region dominated by accreted stars, has most of its stars in two high velocity peaks ($|v_R| >$200~\kms). 
The regions that, according to our analysis, are contaminated or dominated by accreted stars, Regions 1, 2,  6,  and 8, all show a tail in their distribution of stars with high radial velocities, $|v_R| >$200~\kms. These high radial velocity stars are clearly visible in Regions~1 and 2 and in less significant numbers in Regions~6 and 8.

In the $v_R-v_\Phi$ plane (Fig.~\ref{vR-vZ-vPHI}, top panel), the different kinematics of accreted and in-situ populations is even more evident. While in Region~5, stars are distributed along a horizontal line of $v_\Phi \approx$0~\kms, with absolute values of $v_R$ as high as 400~\kms, in Regions~6 and 8 most of the stars are grouped in a clump of fast prograde rotating stars, with $v_\Phi \le -$200\kms. Only few tens of stars are found in the region of accreted stars (i.e., having kinematics like stars in Region~5). In Regions~1 and 2, both populations co-exist, while the distribution of stars of Region~3 in the  $v_R-v_\Phi$ does not show any clear contamination from accreted stars. Also in the $v_\Phi-v_Z$ plane (see Fig.~\ref{vR-vZ-vPHI}, bottom panel), the main kinematic populations can be easily distinguished (e.g., Regions 6, 8).
 
\begin{figure}
\centering
\includegraphics[clip=true, trim = 0mm 0mm 0mm 0mm, width=0.9\linewidth]{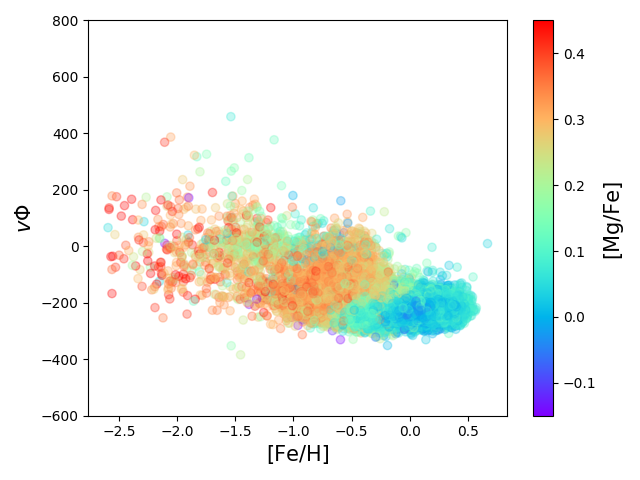}
\caption{The distribution of our sample of stars in the $[Fe/H]- v_\Phi$ plane. The colors of each point represent their [Mg/Fe] ratio as indicated by the color bar at the right. }
\label{dating_ALL}
\end{figure}

\begin{figure*}
\centering
\includegraphics[clip=true, trim = 0mm 0mm 0mm 0mm, width=0.8\linewidth]{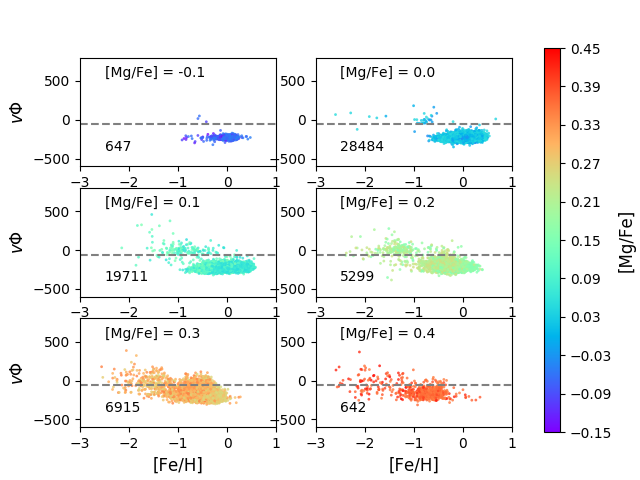}
\caption{Distribution of the sample stars in the $[Fe/H]- v_\Phi$ plane, for different intervals of [Mg/Fe] abundances, as indicated in the legends at the top of each panel. [Mg/Fe] bins have been defined as in Fig.~\ref{kins_vs_FeH}. The number of stars in each metallicity interval are also given in the lower left corner of each panel. Colors of each point represent the [Mg/Fe] ratio as indicated in the color bar on the right-side of the figure.}
\label{dating_main}
\end{figure*}

\begin{figure}
\centering
\includegraphics[clip=true, trim = 0mm 0mm 0mm 0mm,, width=0.98\linewidth]{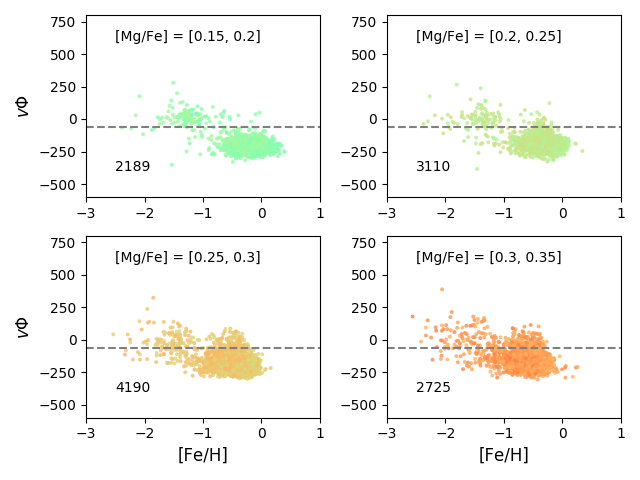}
\caption{Same as in Fig.~\ref{dating_main}, but for narrower intervals of fixed [Mg/Fe] as indicated at the top of each panel and now extending from [Mg/Fe]=0.15 to 0.35. The colors of each point represent their [Mg/Fe] ratios just as in Fig.~\ref{dating_main}. The number of stars in each [Mg/Fe] bin is indicated at the lower left in each panel. }
\label{dating_zoom}
\end{figure}

\begin{figure*}
\centering
\includegraphics[clip=true, trim = 0mm 0mm 0mm 0mm, width=0.8\linewidth]{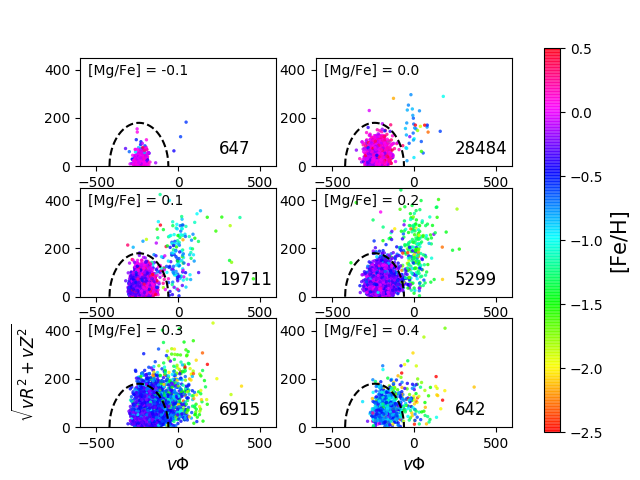}
\caption{Distribution of the sample stars in the Toomre diagram, for different intervals of [Mg/Fe] abundances, as indicated at the top left of each panel. [Mg/Fe] bins have been defined as in Fig.~\ref{kins_vs_FeH}. The number of stars in each metallicity interval is given in the lower right corner of each panel. The color of each point indicates its [Fe/H] ratio as given in the color bar shown on the right of the figure.}
\label{Toomre_main}
\end{figure*}

\begin{figure}
\centering
\includegraphics[clip=true, trim = 0mm 0mm 0mm 0mm,, width=0.98\linewidth]{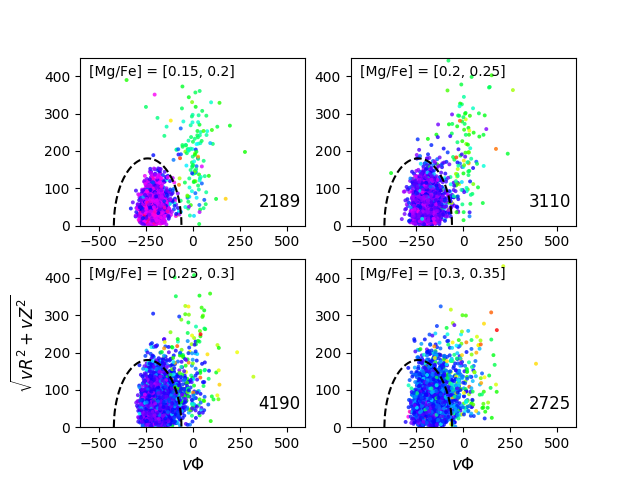}
\caption{Same as in Fig.~\ref{Toomre_main}, but for narrower [Mg/Fe] intervals, extending from [Mg/Fe]=0.15 to 0.35. Colors of each point represent their [Fe/H] ratios, as in Fig.~\ref{Toomre_main}. The number of stars in each [Mg/Fe] bin is indicated at the lower right of each panel. }
\label{Toomre_zoom}
\end{figure}

\begin{figure}
\centering
\includegraphics[clip=true, trim = 0mm 0mm 0mm 0mm,, width=0.98\linewidth]{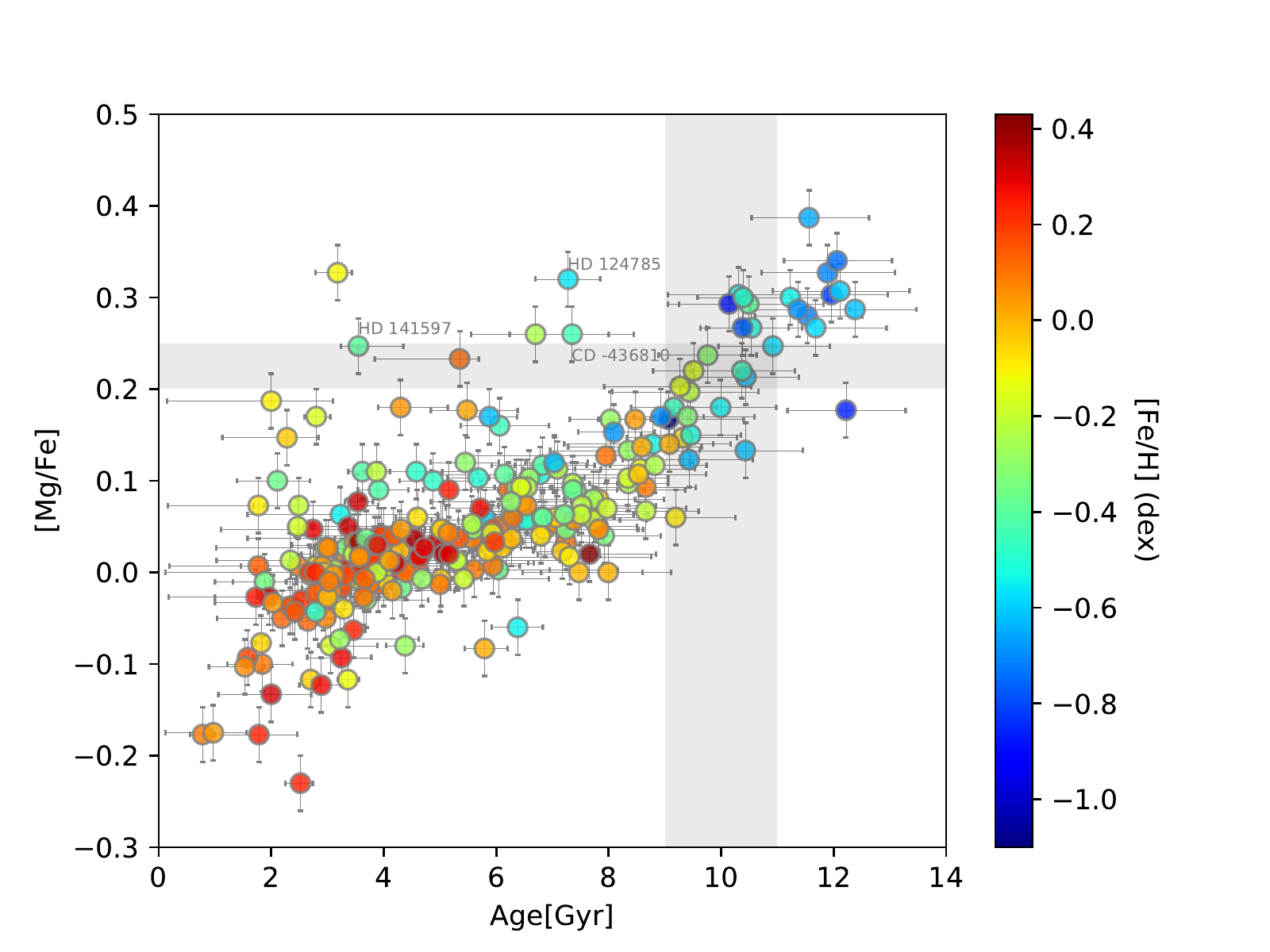}
\caption{Age--[Mg/Fe] distribution for stars in the Adibekyan sample for which we could determine ages and that have Mv<4.5. 
The horizontal grey band corresponds to the Mg abundance where kinematically heated stars appear. It corresponds 
to an age range from 9 to 11 Gyr. Three stars are confirmed to be `young', alpha-rich objects, their name is indicated in the figure (see text for details).}
\label{age}
\end{figure}

\begin{figure}
\centering
\includegraphics[clip=true, trim = 0mm  0mm 0mm 0mm, width=0.6\linewidth]{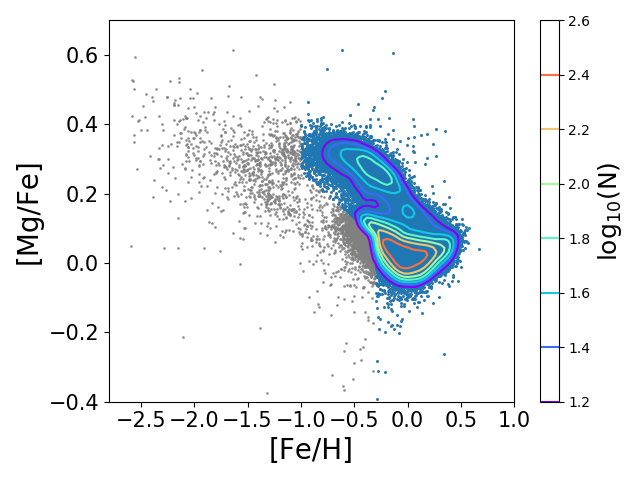}\\
\includegraphics[clip=true, trim = 0mm  0mm 0mm 0mm, width=0.6\linewidth]{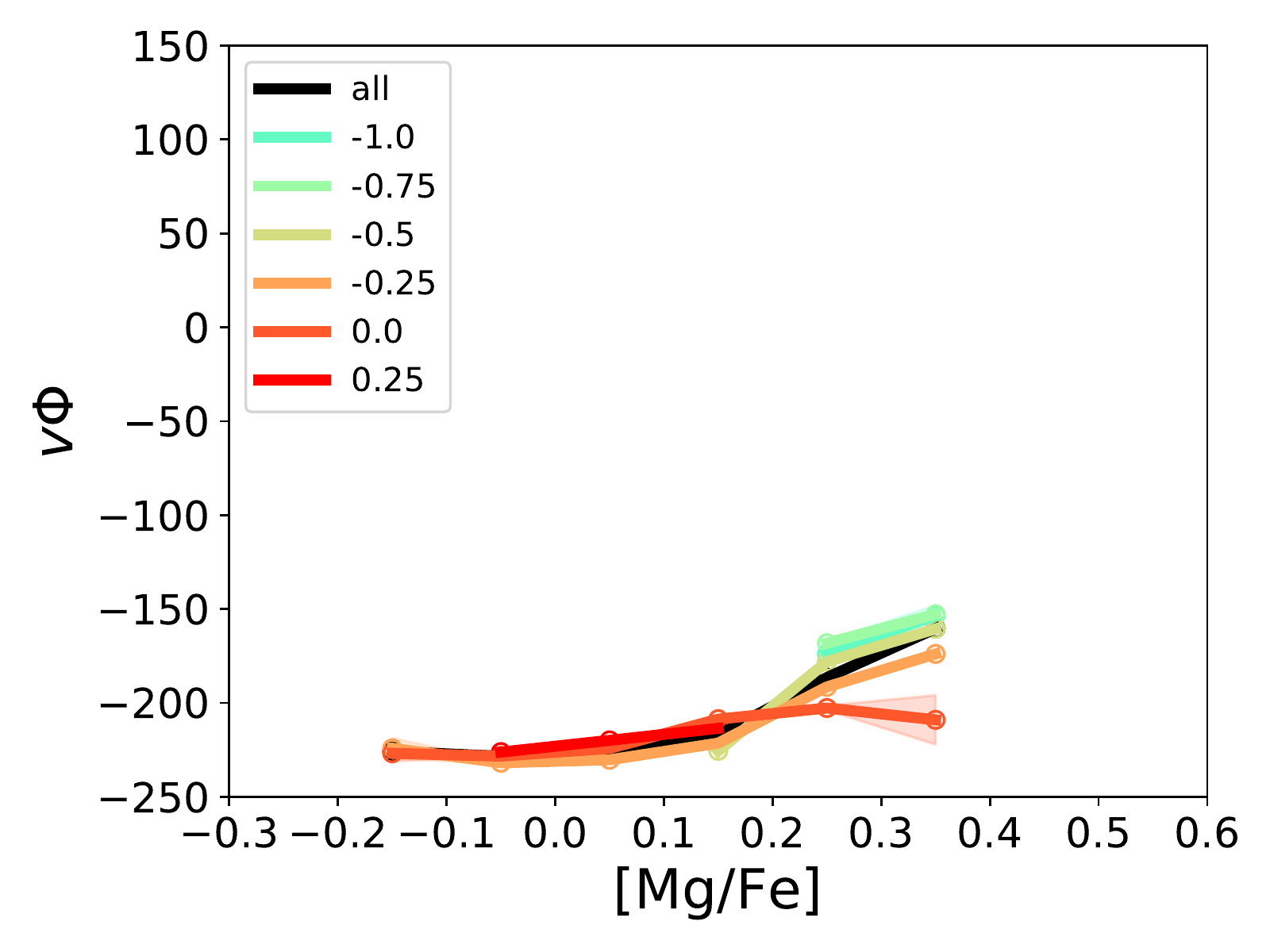}
\includegraphics[clip=true, trim = 0mm  0mm 0mm 0mm, width=0.6\linewidth]{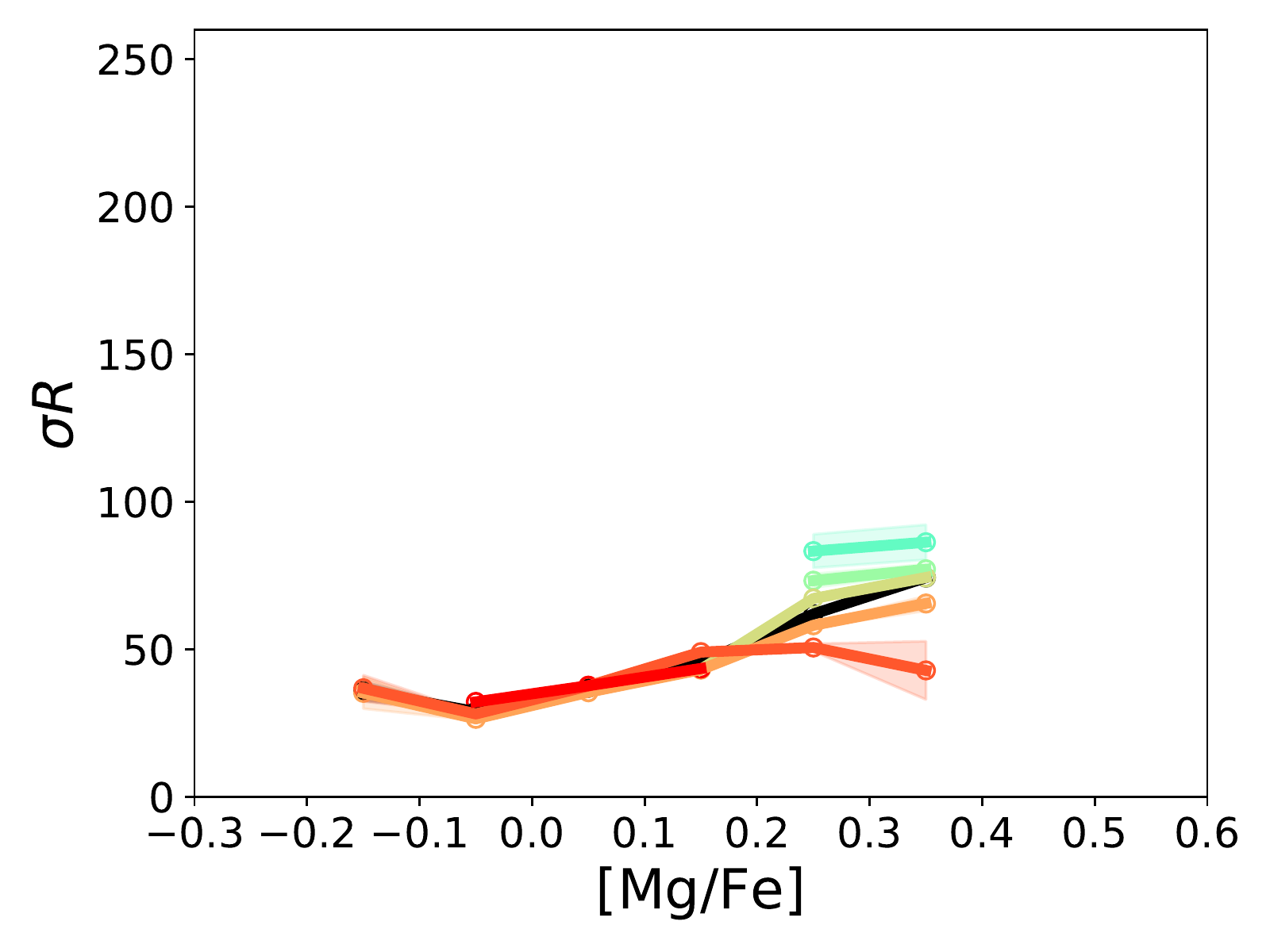}
\includegraphics[clip=true, trim = 0mm  0mm 0mm 0mm, width=0.6\linewidth]{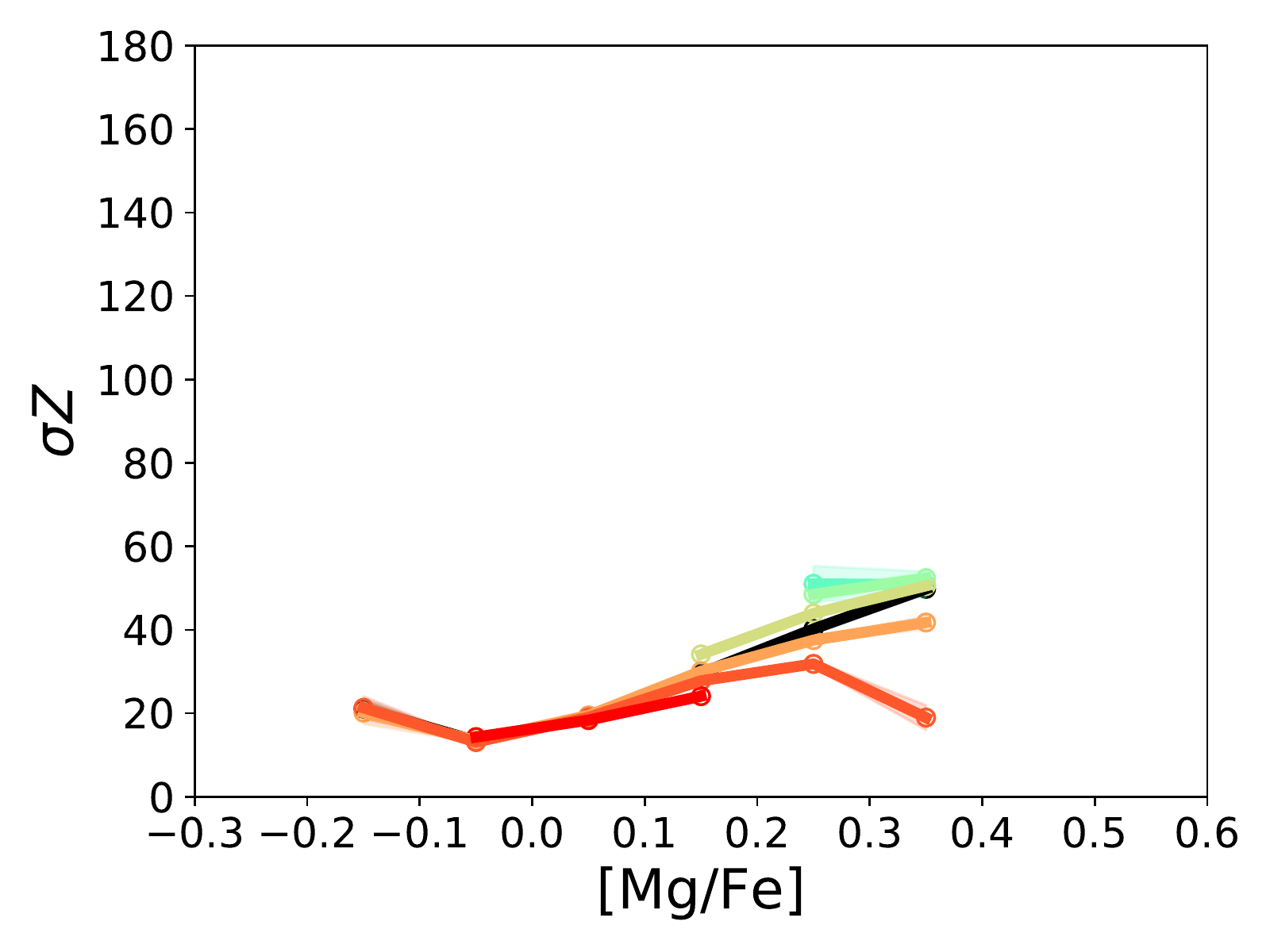}
\includegraphics[clip=true, trim = 0mm  0mm 0mm 0mm, width=0.6\linewidth]{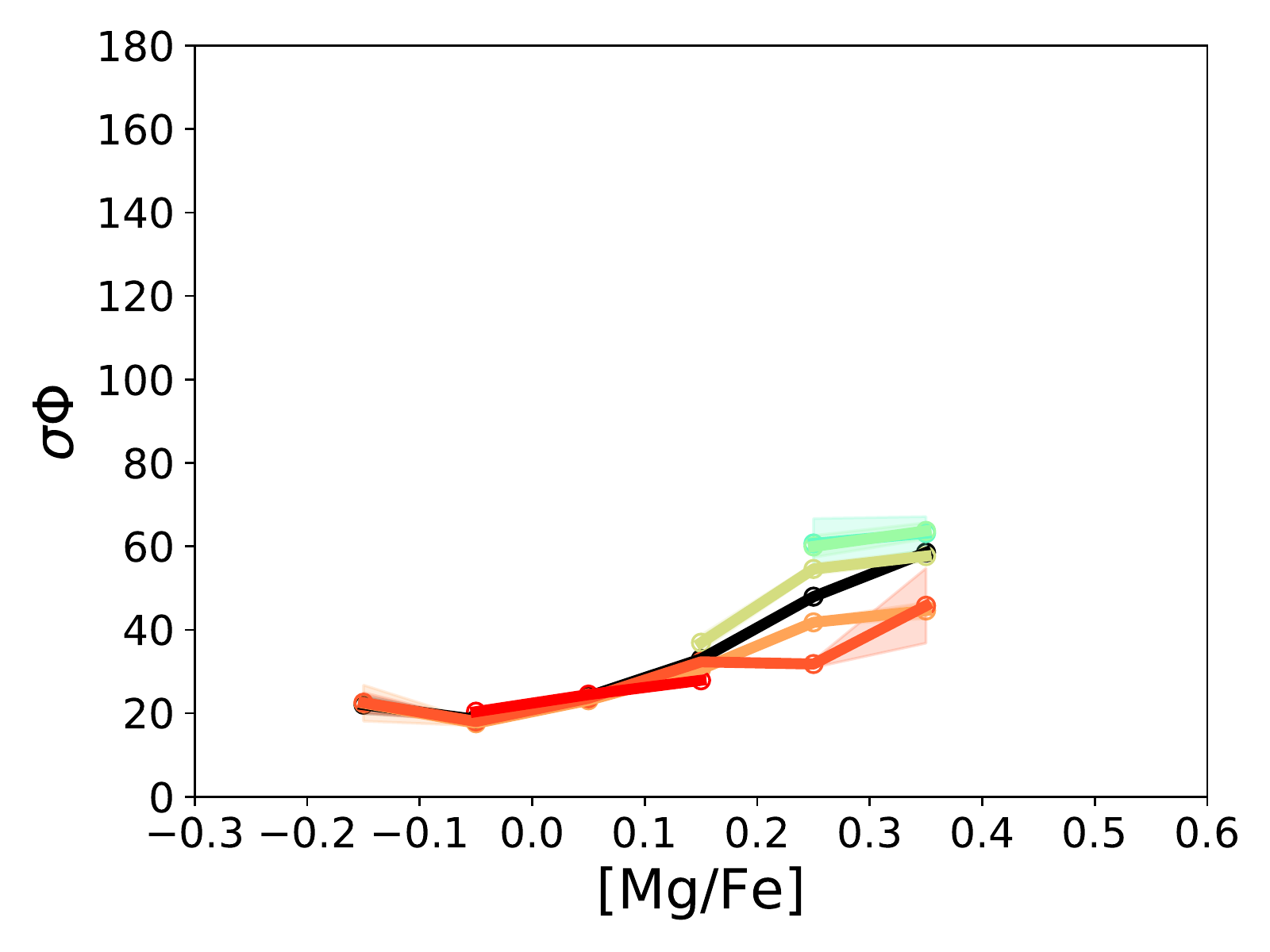}
\caption{(\emph{Top panel}): Distribution of stars of the in-situ sample (blue points) in the  [Fe/H]--[Mg/Fe] plane. Colored lines correspond to isodensity contours, in logarithmic scale, as indicated in the color bar on the right-hand side of the panel. For comparison, all stars of the main sample studied in this paper, and shown in Fig.~\ref{FeHMgFe_map}, are shown by gray points.  (\emph{From the second to the bottom panel}) Mean azimuthal velocity, radial, vertical and azimuthal velocity dispersion of stars of the in-situ sample, as a function of their [Mg/Fe] ratio. In each panel, the relations are given for bins in [Fe/H], as indicated in the legend in the top-left panel. The black curves show the corresponding relation, for the total sample not binned in [Fe/H]. The $1\sigma$ uncertainty in each relation (colored shaded regions) is estimated through 1000 bootstrapped realizations. In all panels, only bins containing more than 10 stars are shown. In all plots, we have adopted the same  $y$-axis limits, as adopted in Fig.~\ref{kins_vs_MgFe}, to facilitate the comparison. }
\label{kins_vs_MgFe_noACC}
\end{figure}

\begin{figure*}
\centering
\includegraphics[clip=true, trim = 2mm 0mm 0mm 0mm,, width=0.47\linewidth]{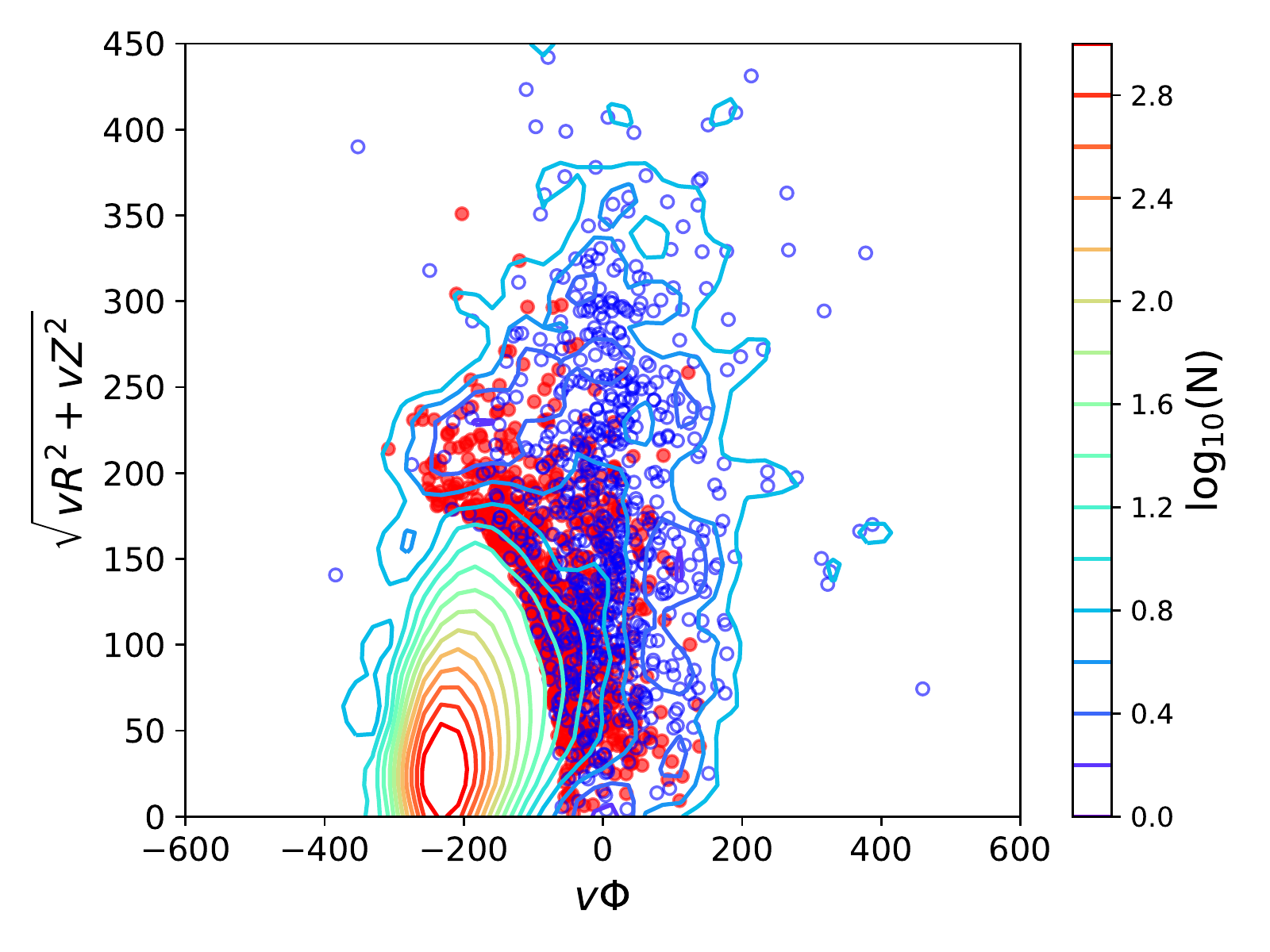}
\includegraphics[clip=true, trim = 2mm 0mm 0mm 0mm,, width=0.47\linewidth]{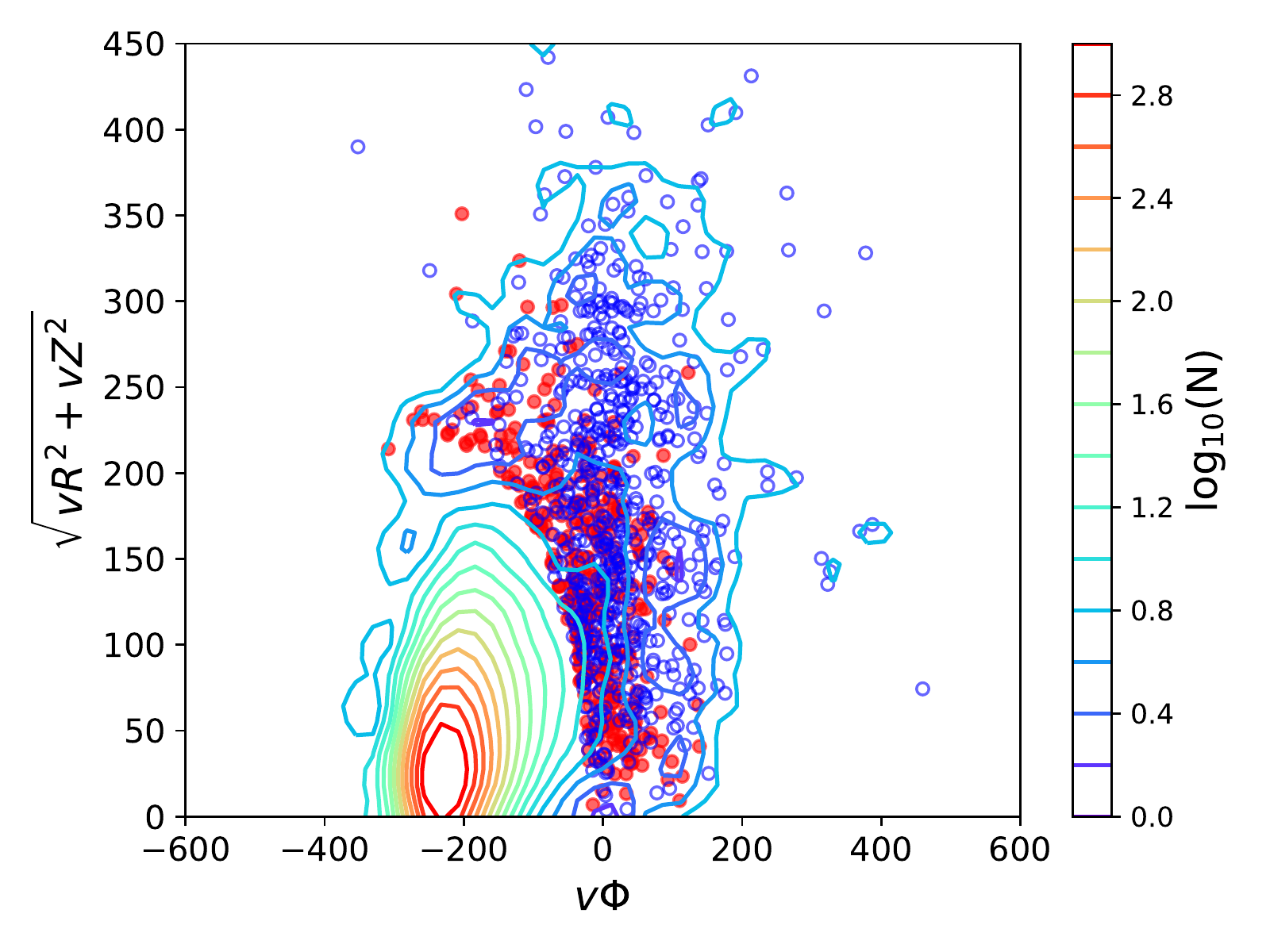}

\includegraphics[clip=true, trim = 2mm 0mm 0mm 0mm,, width=0.47\linewidth]{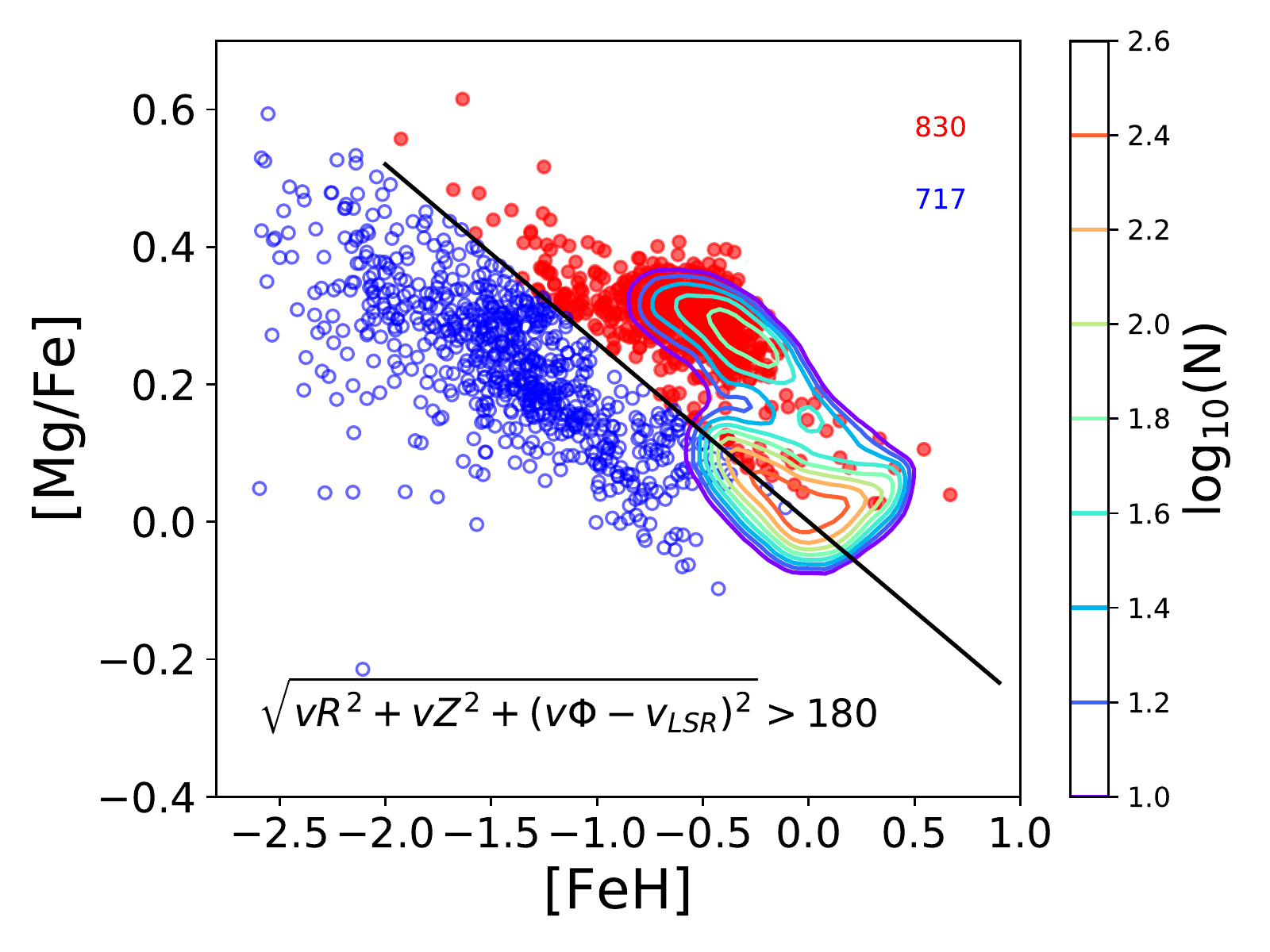}
\includegraphics[clip=true, trim = 2mm 0mm 0mm 0mm,, width=0.47\linewidth]{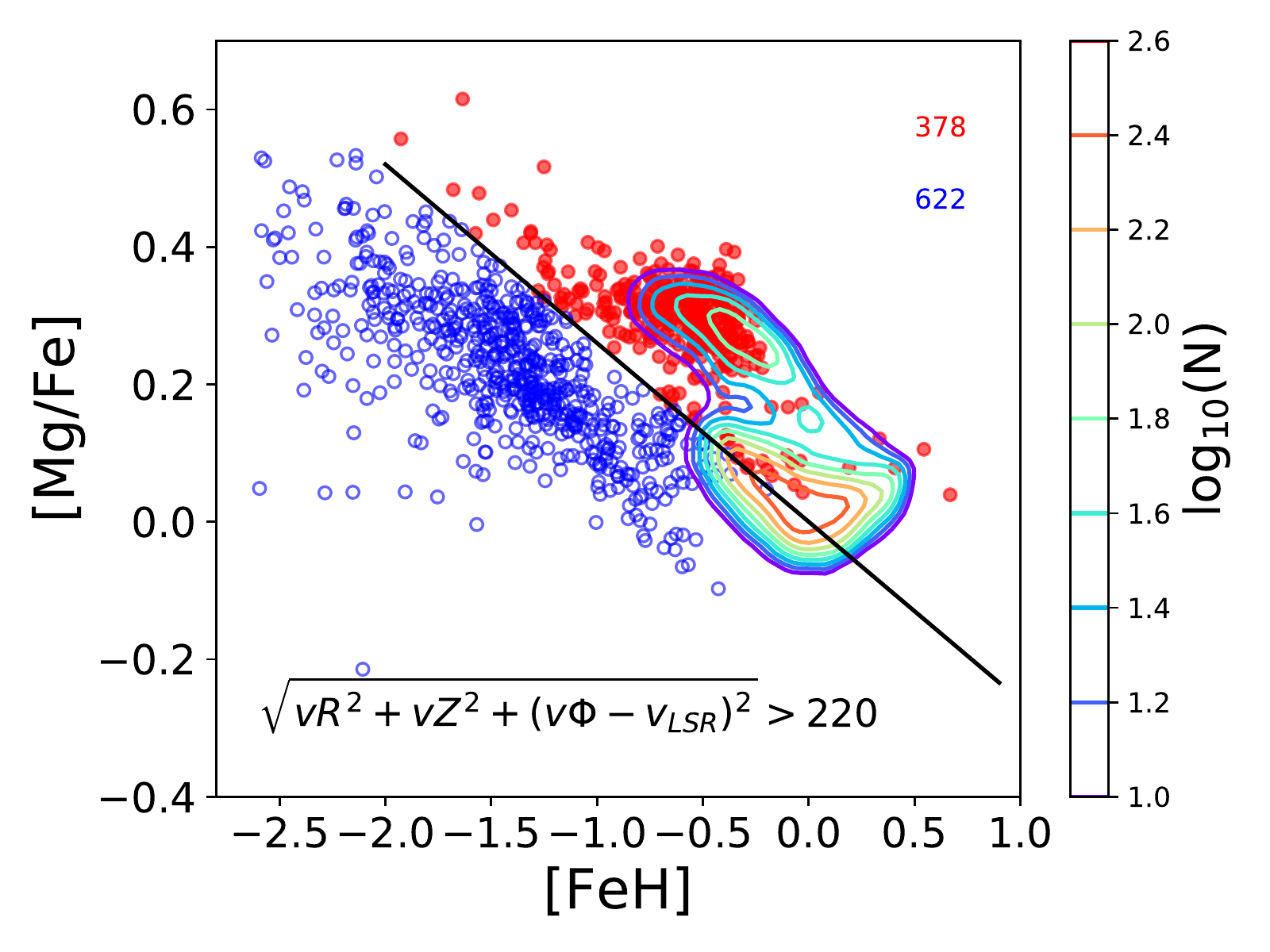}

\includegraphics[clip=true, trim = 2mm 0mm 0mm 0mm,, width=0.47\linewidth]{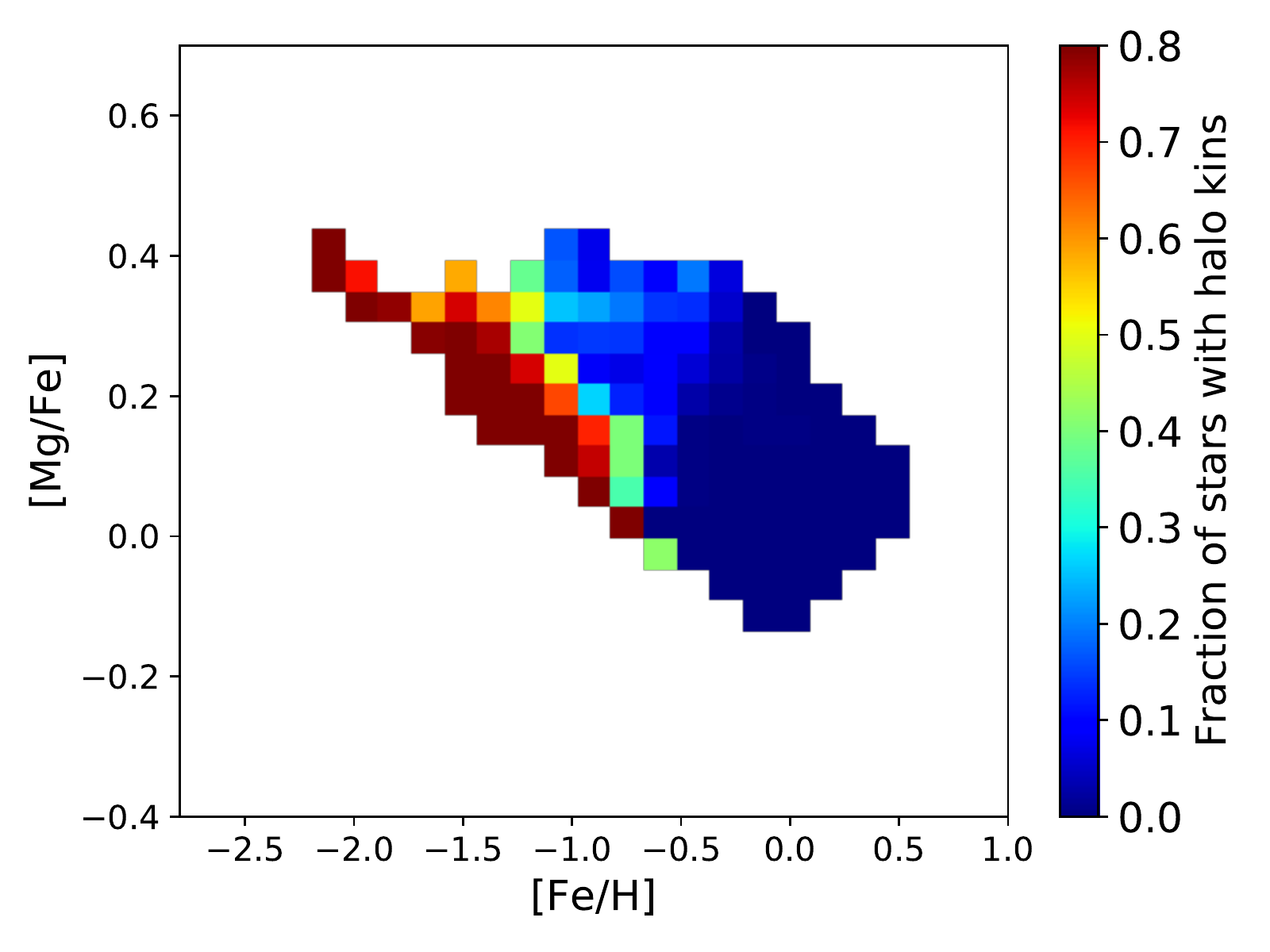}
\includegraphics[clip=true, trim = 2mm 0mm 0mm 0mm,, width=0.47\linewidth]{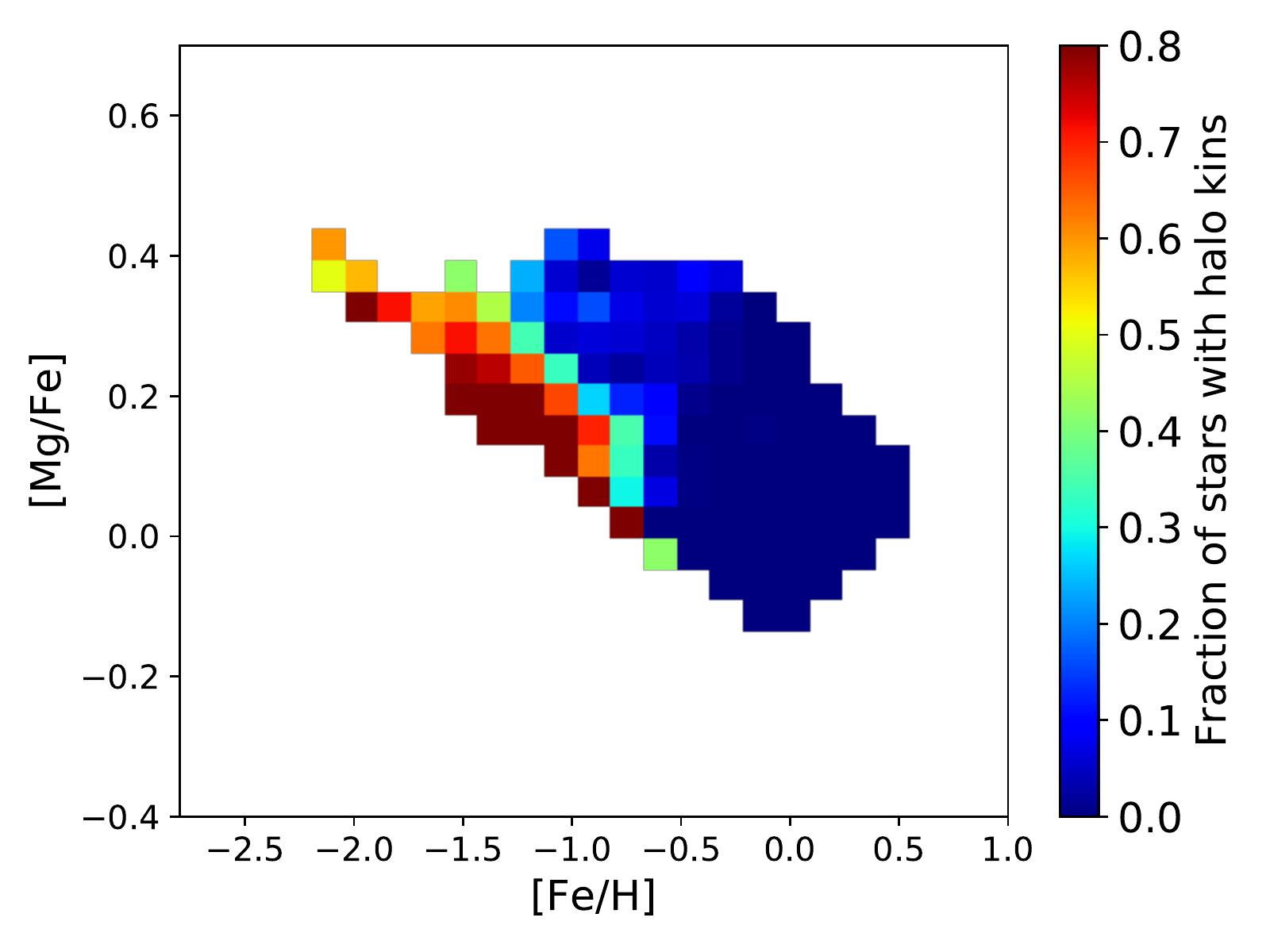},
\caption{\emph{First row: }Toomre diagram of all stars in our sample with halo-like kinematics. The distributions in the right and left panels differ for the threshold adopted in defining halo stars, as reported in the middle row. Red points indicate metal-rich thick disc stars, blue points indicate accreted and metal-poor thick disc stars (see middle row and text for their definition). Colored lines correspond to isodensity contours for the whole sample analyzed in this paper. 
\emph{Middle row:} Distribution, in the  [Fe/H]--[Mg/Fe] plane, of stars of the sample with halo-like kinematics. The diagonal line separates the metal-rich thick disc sequence, in red on the right, from the accreted and metal-poor thick disc, in blue on the left. The distributions in the right and left panels differ for the threshold adopted in defining halo stars, as reported in the plots. The number of stars on the left and right of the diagonal lines are given in both panels, in blue and red respectively. Colored lines correspond to isodensity contours, as defined in Fig.~\ref{FeHMgFe_map}. \emph{Bottom row:} Fraction of stars with halo-like kinematics, in the  [Fe/H]--[Mg/Fe] plane, for the two different definitions of the halo, adopted for the left panels. In both plots, the fraction is normalized to the total number of stars in a given pixel.  }
\label{insituhalo}
\end{figure*}

\begin{figure}
\centering
\includegraphics[clip=true, trim = 2mm 0mm 0mm 0mm,, width=0.8\linewidth]{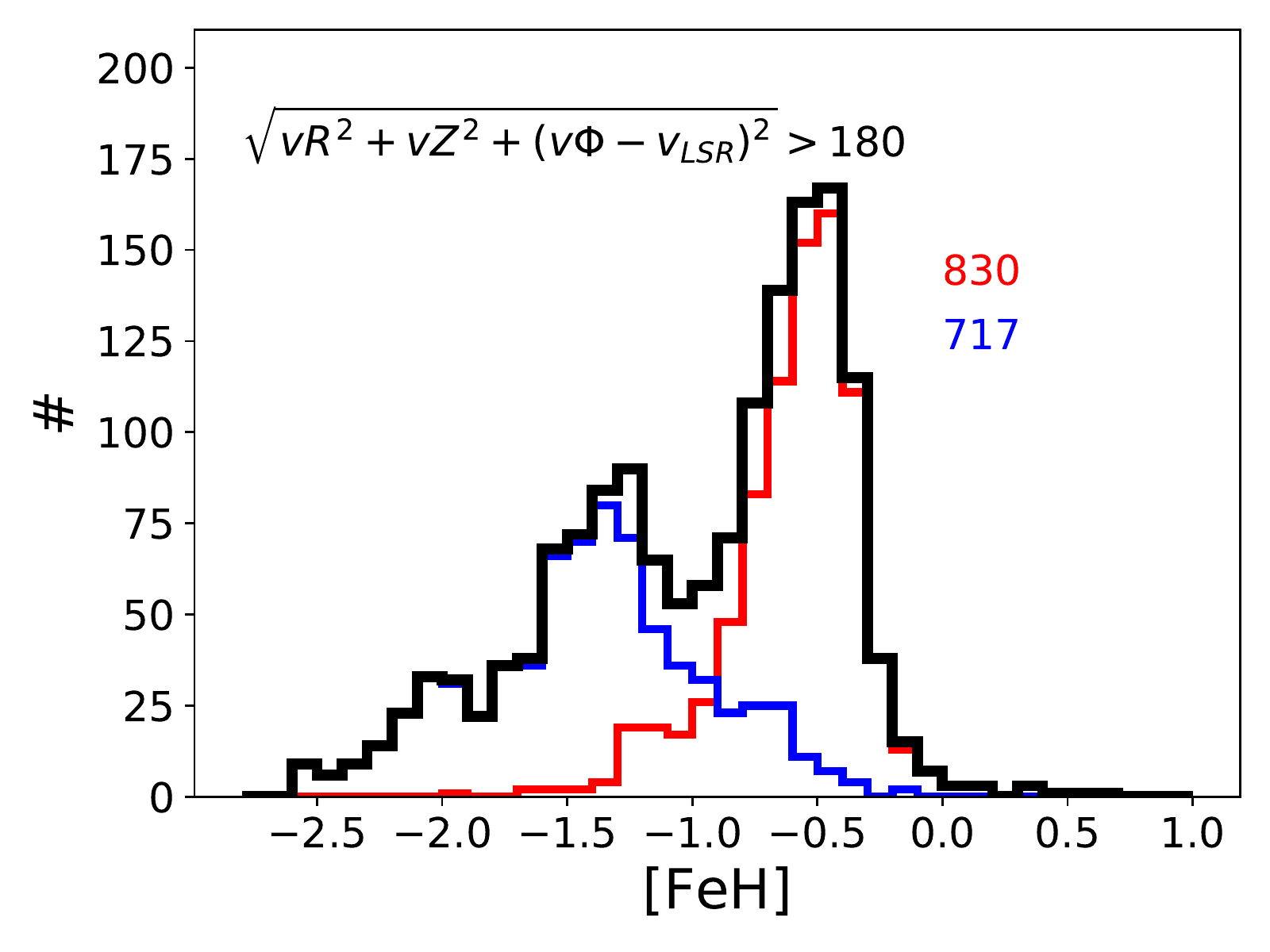}
\includegraphics[clip=true, trim = 2mm 0mm 0mm 0mm,, width=0.8\linewidth]{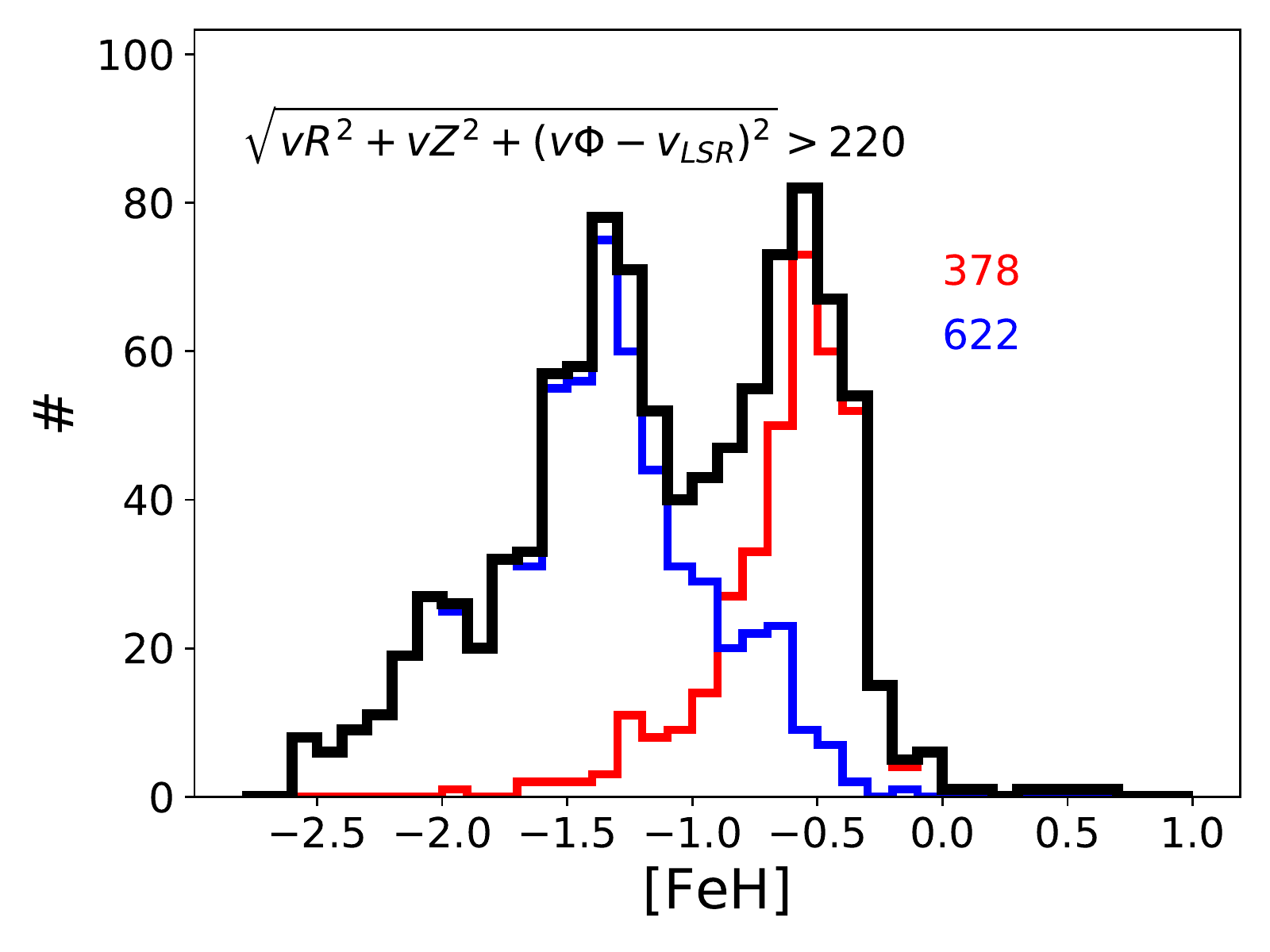}
\caption{Metallicity distribution of stars in the sample with halo-like kinematics (black curve). The contribution of metal-rich thick disc stars, and of  accreted and metal-poor thick disc stars to the total distribution is shown, respectively, by the red and blue curves. Two different limiting velocities for discriminating halo  stars is adopted in the top and bottom panels, as reported in the legends.  }
\label{mdf}
\end{figure}

\subsection{Age-dating the accretion event}\label{dating}

When did this accretion event occur? If the fusion of the Milky Way with another galaxy happened and some of the stars that were heated belong to already formed components of the Milky Way, then it should be possible to estimate when the merger occurred. 
The accretion of satellites onto Milky Way-type galaxies has been extensively studied through simulations over the last decades. Two main results of simulations are interesting within this context: \textit{(1)} the disc of the Milky Way should have heated during the accretion -- that is its radial, vertical and azimuthal velocity dispersions should have increased of an amount that depends on the inclination of the satellites, its mass, and on the gas content of the main galaxy at the time of the accretion \citep{quinn93, walker96, velazquez99, font01, benson04, moster10, villalobos08, kazantzidis08, villalobos09, qu10, qu11a, qu11b, house11} -- and, importantly, the disc stars that were impacted by the merger should slow down \citep{qu10};  \textit{(2)} there should be a significant overlap in the kinematics of the accreted stars and the in-situ stars present in the early Milky Way disc at the time of the merger. For example, a merger with a mass ratio of about 1:10, possibly similar to the mass ratio of the accreted satellite studied here and which may define the nature of the low-$\alpha$ sequence at the time of its accretion \citep{haywood18, helmi18}, is expected to slow the rotation of some of this stars in the disc significantly \citep{jeanbaptiste17}. 

In Figs.~\ref{dating_ALL}, \ref{dating_main} and \ref{dating_zoom}, we now make use of these arguments to look for signatures of the kinematic heating of the Milky Way disc that must have been generated at the time when stars were accreted onto the Galaxy.
We start by showing the distribution in the $v_\Phi-\rm [Fe/H]$ of all sample stars (Fig.~\ref{dating_ALL}). This is the same plane already shown in Fig.~\ref{kins_vs_FeH}, but here we can see the gain in looking at the whole distribution of stars, instead of looking at the simple mean relations.  Stars redistribute in this plane in a sort of ``shoe''-like shape, with the tip of the shoe at high metallicities, $\rm [Fe/H] \ge -0.2$, and the instep of the shoe at $\rm [Fe/H] \le -0.3$. Moving from the tip to the instep, [Mg/Fe] increases, as expected for populations whose [Fe/H] decreases. Also, the accreted sequence stands out clearly, as a sequence of $<v_\Phi>\sim0$~\kms\ and lower [Mg/Fe] values compared to the surrounding (red/brown in the plot) stars. The striking feature in this plot, however, is the rapid decline of the rotation (i.e steep rise of $v_\Phi$) with decreasing [Fe/H], for $\rm [Fe/H] \le -0.3$, something not clearly visible in mean relation shown in Fig.~\ref{kins_vs_FeH}, where the decline in the mean rotation starts at lower [Fe/H]. To better understand this trend, we show, in Fig.~\ref{dating_main}, the same $v_\Phi-\rm [Fe/H]$ plane but this time for stars binned by their [Mg/Fe] ratios. The amplitude of the bins and their mean values are the same as in Fig.~\ref{kins_vs_FeH}. For $\rm  [Mg/Fe]=-0.1, 0. $ and $0.1$, we can clearly distinguish two separated groups in their kinematics, one with $<v_\Phi> \sim 0$~\kms and the second one with prograde rotation (i.e., negative $v_\Phi$). At $\rm  [Mg/Fe]=0.2$, however, the separation between the two groups is no longer evident and stars of the in-situ sequence with null or retrograde rotation appear at $\rm  [Fe/H]\sim-0.3$. The same thing is evident at  $\rm  [Mg/Fe]=0.3$, where, again, one sees clearly that the in-situ population extends up to positive $v_\Phi$ (retrograde motions). At $\rm  [Mg/Fe]=0.4$,  the distinction between the two groups is no longer evident in this plane but the limited number of stars in this bin does not allow us to draw any robust conclusions. However, the remarkable result we glean from this figure is that, while at $\rm  [Mg/Fe] < 0.2$, the group with prograde rotation has distinct $v_\Phi$ from that of accreted stars, at $\rm [Mg/Fe]\ga0.2$, a tail of zero or positive $v_\Phi$ -- null or retrograde motions -- appears among the metal-rich stars, $\rm [Fe/H] > -1$. 

To investigate this change further in the $v_\Phi$ properties of stars with [Mg/Fe], in Fig.~\ref{dating_zoom}, we present the same analysis as in Fig.~\ref{dating_main}, but this time only for the interval $\rm 0.15 \le [Mg/Fe] \le 0.35$, that has been divided in four bins. We can clearly see a plume of stars extending from $v_\Phi$ typical of the thick disc, to zero and positive $v_\Phi$, reaching values of the azimuthal velocity typical of the accreted component.  This plume is not evident at $\rm [Mg/Fe]<0.2$, appears at $\rm 0.2 \le [Mg/Fe] \le  0.25$, and  $\rm [Fe/H] \sim -0.3$, and it is still present at higher [Mg/Fe] ratios, and slightly lower metallicities. We interpret this excess of stars with thick disc abundances but accreted-like $v_\Phi$ as the signature of the heating of the early Milky Way disc, by the satellite. And to show that this kinematic heating does not affect the azimuthal velocities only, but also the radial and vertical motions, we show in Fig.~\ref{Toomre_main} the distribution in the Toomre diagram -- that is in the $v_\Phi-\sqrt{{v_R}^2+{v_Z}^2}$ -- of stars binned according to their [Mg/Fe] abundance, as in Fig.~\ref{dating_main}. In each panel of the plot, the dashed curve represents the locus of stars with $\sqrt{{(v_\Phi-v_{LSR})}^2+{v_R}^2+{v_Z}^2}=180$~\kms, commonly used as separator between thick disc and halo kinematics \citep[see, for example][]{nissen10}, and the colors code the metallicity, [Fe/H].  For $\rm [Mg/Fe] \le  0.1$, two distinct groups appear:  a sequence of metal-poor stars (with typical $\rm [Fe/H] \le -0.5$), which redistributes vertically in this plane, along the null $v_\Phi$ line, and which is thus characterized by halo-like kinematics, and a sequence (with $\rm [Fe/H] \ge -1$) -- the last majority of stars - with $\sqrt{{(v_\Phi-v_{LSR})}^2+{v_R}^2+{v_Z}^2} < 180$~\kms, and thus with thin and thick disc-like kinematics. The first of the two sequences is made of accreted stars, as has been discussed in the previous sections,  and shown also by \citet{haywood18} and \citet{helmi18}. At $\rm [Mg/Fe] = 0.2$, however, the distinction between the two sequences becomes less evident, and at $\rm [Mg/Fe] > 0.2$ the kinematic borders of the two sequences can no longer be distinguished. In particular, clearly, a sequence of stars with thick disc metallicities but halo kinematics is found at $\rm [Mg/Fe] = 0.2$ (see also Fig.~\ref{Toomre_zoom}). These stars do not only overlap with the accreted sequence in $v_\Phi$, as previously shown, but also in their coupled radial and vertical motions. This overlap of kinematic properties of accreted and in-situ stars, predicted by simulations \citep{jeanbaptiste17}, is thus clearly present at the high-[Mg/Fe] end of the sample and constitutes the smoking gun of the accretion event experienced by the Galaxy in its early evolution and of the resulting kinematic heating.

Having defined the lower [Mg/Fe] limit of the heating event, we can now convert the abundance ratio into an age, to age-date the merger. For this, we need to quantify  the evolution of the [Mg/Fe] ratio with age. Because our cross-matched \gaia~DR2-APOGEE sample is made of giants  (we remind the reader that only stars with $1 < log(g) < 3.5$ have been retained for this analysis), we cannot date these stars directly by isochrone fitting techniques, as done, for example, in \citet{haywood13}. However, as done in \citet{haywood13}, we derive ages for a subsample of  273  dwarf stars from the \citet{adibekyan12} sample, retaining only stars with $Mv<4.5$, the only difference with  \citet{haywood13} work being that here we make use of parallaxes from \gaia~DR2. We refer the reader to \citet{haywood13} for all the details concerning the adopted methodology for the age determination (bayesian method, set of isochrones, estimates on the age errors), and to a more extensive work in preparation (Haywood et al, in prep) for the analysis and discussion of ages with \gaia~DR2. In Fig.~\ref{age}, we report the derived age-[Mg/Fe]  relation for this subsample\footnote{Note that in this plot the error bars on the magnesium abundances are fixed to 0.03~dex for all stars, this error coming from the estimate given in Adibekyan et al. (2012) for solar type stars. Cooler and hotter stars have mean uncertainties on magnesium abundance of 0.07 and 0.05 dex, see their table 3.  Three stars (HD 124785, HD 141597 and CD -436810) are identified has `young', alpha-rich stars, having magnesium, silicon and titanium 
abundance higher than 0.15~dex and ages less than 8~Gyr. }. We emphasize that, even if the sample used to estimate the ages is local, that is, it is confined to stars at few hundred parsecs from the Sun, these stars come from different parts of the disc. We have indeed shown in previous works that the distribution of their pericentres  \citep[][Fig.~4]{haywood15} and apocentres \citep[][Fig.~3]{snaith15}  is very broad, in particular for stars at the [Mg/Fe] levels relevant for the dating of the accretion. The large extent of their orbits implies that these stars (currently observed at the solar vicinity) are indeed representative of a large portion of the inner disc. 

For a [Mg/Fe] interval equal to [0.2, 0.25], which corresponds to the end of the disc heating phase, the corresponding age interval is  9 --11~Gyr\footnote{The biggest uncertainty on this estimate comes from possible systematics between the magnesium abundance scales of APOGEE and \cite{adibekyan12}. There are 8 stars in common between these two surveys, and which have magnesium abundances between 0.13 and 0.36. The mean of the differences between the [Mg/Fe] abundances of the two samples for these 8 stars is 0.002~dex (and a dispersion around this value of 0.059~dex), suggesting (although the strength of the evidence is limited because of to the small number of stars), that there is no serious bias between the two abundance scales.}.
We emphasize that this estimate coincides with the end of the satellite merging process -- or better with the time when the kinematic heating caused by the decaying satellite to the early Milky Way disc became negligible. According to the N-body models by \citet{qu11b}, indeed, the heating of the early Galactic disc may have started already at the first close passage of the satellite to the Milky Way, and indeed we can expect the heating to be significant in the first phases of the accretion - see Fig.~3 in their paper -- and continued up to the final phases of the accretion.


\section{Discussion}\label{discussion}

\subsection{The overlapping borders of Galactic stellar populations: revising  \citet{minchev14} results }

In the previous section, we have seen that discs and halo populations can overlap both in chemical and kinematic spaces:  1) stars with thick disc kinematics can be found at metallicities as low as $\rm [Fe/H]\sim -2$, thus at metallicities typical of the stellar halo; 2) stars with halo-like kinematics  can be found also at metallicities typical of the thick disc and of the metal-poor thin disc as well (see, for example, Figs.~\ref{vPHI_histo} -- \ref{vR-vZ-vPHI}). 

One of the consequence of this absence of well-defined borders is on the Galactic chemo-kinematic relations at  $\rm [Fe/H] \ge -1$, that is at metallicities traditionally associated to disc populations. \citet{minchev14}, for example, used RAVE and SEGUE data to derive these relations in an extended solar vicinity volume (about 1~kpc in radius from the Sun position). 

Their derived velocity dispersions--[Mg/Fe] relations, for different metallicity bins, show two main characteristics: 
1)  the velocity dispersions increase with [Mg/Fe] for all metallicities   $\rm [Fe/H] > -0.4$, while the opposite trend is found for   $\rm [Fe/H] < -0.4$, where at the highest magnesium abundances the dispersions reach their minima 
2)  the maxima reached in the velocity dispersion--[Mg/Fe] relations move to lower [Mg/Fe] with decreasing metallicities.

In \cite{minchev14}, these properties were interpreted as the signature of a significant merger in the early disc evolution, which would have increased the disc velocity dispersions to high values, and generated at the same time an inside-out migration of stars with cold kinematics, both  processes responsible of  causing the reversal in the velocity dispersions--[Mg/Fe]  trends observed at $\rm [Fe/H] < -0.4$. 
More in detail, in their interpretation, based on a detailed match to a chemodynamical model lacking, however, accreted populations, the metal-poor stars with high velocity dispersions and low [Mg/Fe] ratios were considered in-situ, early disc stars heated by the interaction, while metal-poor stars with velocity dispersions as low as those of the metal-rich populations, and high [Mg/Fe] were associated to in-situ stars migrated from the innermost regions of the early disc, as a result of the gravitational perturbation induced by the accreted satellite. 
Moreover, the displacement of the maximum of the velocity dispersion at lower  [Mg/Fe], the higher the metallicity, was suggested to be related to mergers of decreasing mass ratios occurred during the Galaxy lifetime. 

The present analysis allow us to revisit  \citet{minchev14} findings -- see also \citet{guiglion15} for  similar results -- thanks to the exquisite details in kinematics and abundances of our sample (also $\sim$ 13 times larger), and in particular permits us to address the role of accreted stars in shaping these relations,  a role underestimated in previous works. 

First, as explained in Sect.~\ref{mean}, \emph{it is the accreted population which drives the reversal of the velocity dispersions-- [Mg/Fe] relations} at $\rm [Fe/H] < -0.4$, where dispersions decrease with magnesium abundance.
Because accreted stars describe a sequence of decreasing [Mg/Fe] with increasing [Fe/H], and are also characterized by the highest velocity dispersions,  this naturally explains the displacement of the maximum of the velocity dispersions to higher [Fe/H], as the [Mg/Fe] decreases. 
Hence, this displacement is not related to a series of accretions, but is most probably due to the presence of stars from one accreted satellite only.
Moreover, the reversal of the  velocity dispersion--[Mg/Fe] relation, found at $\rm [Fe/H] < -0.4$, is due to the contamination by these accreted stars, which have very high velocities, 
and is not due to an intrinsic decrease of the velocity dispersion of the in-situ population.
In bins that are not contaminated by accreted stars (at $\rm [Fe/H]>-0.5$), the dispersions show no sign of significant 
decrease in the highest abundance bins. At metallicities of -0.75 and -1, 
low abundances ($\rm [Mg/Fe]<0.2$) are contaminated by accreted stars, and show high dispersions, which decrease to lower values as the [Mg/Fe] ratios increase, 
but simply reaching levels of non-contaminated standard thick disc stars, not the very low values found in \cite{minchev14}. 
At still lower metallicities, even high-[Mg/Fe] stars are contaminated by accreted stars, 
and show higher velocity dispersions, even if the decreasing trend remains visible in many cases. 
These trends are even clearer in Fig. \ref{FeHMgFe_map}, which show that the velocity dispersion is uniform or even 
rising at the upper limit of the [Fe/H]-[Mg/Fe] distribution, except in the region of the accreted sequence, in the lower
right side of the distribution.
To further probe the importance of the accreted component in shaping the Galactic chemo-kinematic relations, and show the impact they have on the latter, in Fig.~\ref{kins_vs_MgFe_noACC} we show the same chemo-kinematic relations already presented in Fig.~\ref{kins_vs_MgFe}, where this time we  select only stars with $\rm [Fe/H] > -1$ and with $\rm [Mg/Fe] > -0.26 \times [Fe/H]$, and stars with $\rm [Mg/Fe] \le  -0.26 \times [Fe/H]$, but $\rm [Fe/H] > -0.3$ (this is the same diagonal line adopted in Fig.~\ref{insituhalo}, middle row).  The selection appears  severe, but it allows to have a clean in-situ sample, not contaminated by the accreted sequence (see Fig.~\ref{kins_vs_MgFe_noACC}, top panel, blue points). The mean chemo-kinematic relations of this sample are also shown in  Fig.~\ref{kins_vs_MgFe_noACC} and demonstrate that when a clean in-situ sample is selected, no reversal is anymore found in any of the velocity dispersions-[Mg/Fe] relations, thus further probing that this reversal is entirely generated by accreted stars.

Therefore, we cannot confirm the conclusions reached by \cite{minchev14}.
In particular, the shift of maxima  in the velocity dispersion--[Mg/Fe] relations to lower [Mg/Fe] with decreasing metallicities  interpreted  by \cite{minchev14} as possibly being the 
consequence of successive accretions, is indeed most probably due to the accretion of a single satellite. Likewise, the decrease in velocity dispersions found by these authors at high-[Mg/Fe] ratios, and interpreted as the signature of radial
migration of kinematically colder stars from the inner to the outer disc, is also absent from our sample.

We conclude by emphasizing that Galactic chemo-kinematic relations, also in a metallicity regime often associated to the Galactic disc ($\rm [Fe/H] > -1$) are affected and reshaped by the accreted component(s), and that it is thus necessary to make selections as careful as possible to be able to separate the contribution of these accreted stars from that of in-situ populations. The astrometric and spectroscopic quality of the data now available, and not achievable only few years ago, can now make these selections possible.

\subsection{The in-situ population below $\rm [Fe/H] = -1$ is the Galactic thick disc}

The halo population is, at first sight, an heterogeneous collection of stars. If the halo is defined as all stars with   $\rm [Fe/H] < -1$, our analysis shows that our sample, in this metallicity interval, is made for about 55-60\% of accreted material, with weak prograde, null or retrograde rotation, and for the remaining 40-45\% of in-situ stars, with kinematics similar to that of thick disc stars in the metallicity range  $\rm [Fe/H]=[-1., -0.5]$. If the halo is defined on the basis of its kinematics, halo stars are found not only at $\rm [Fe/H] < -1$, but also among stars with $\rm [Fe/H] > -1$: at [Mg/Fe] typical of the thick disc, we find the in-situ heated disc that we will discuss further in the next section,  and at [Mg/Fe] typical of the metal-poor thin disc, we find the most metal-rich stars of the accreted population.
 
This complexity 
-- and also ambiguity in the definition -- of the halo population is not new.
The existence of metal-poor, $\rm [Fe/H] < -1$,  stars with thick disc kinematics, usually referred to as the metal-weak thick disc, has been known for decades \citep[][but see subsequent criticisms \citealt{twarog94, twarog96}]{norris85, morrison90}. More recently, there has been tremendous effort expended to understand this
thick disc population \citep{chiba00, beers02, reddy08, brown08, kordopatis13, hawkins15a, li17, hayes18}. 
\citet{beers02} estimated the fraction of stars in the metal-weak thick disc to be between 30\% and 40\%  at $1.6<\rm [Fe/H]<-1$ and suggested that this population could extend to metallicities as low as $\rm [Fe/H]\la-2$. These estimates are in very good agreement with ours, since we find the fraction of stars with $\rm [Fe/H]<-1$ and thick disc kinematics to be 40-45\% of the total fraction of stars at these metallicites and with a  contribution that decreases from 70\% at [Fe/H]--[Mg/Fe]=[[-1.5, -1], [0.25, 0.35]] to 30\% at [Fe/H]--[Mg/Fe]=[[-2., -1.5], [0.25, 0.35]] (see Sect.~\ref{disentangling}).  Thus, our results confirm previous findings by Beers and collaborators that the fraction of metal-poor stars with thick disc kinematics is significant, and it extends up to the lower limit of our sample, at $\rm [Fe/H]\sim-2$. The recent finding by \citet{sestito18} of several very low 
metallicity stars with disc kinematics is a first step towards extending these estimates to the very metal-poor stars.

Interestingly, the fraction of metal-poor stars with thick disc kinematics is also in very good agreement with early estimates by \citet{sommer90}, who found a metal-poor halo consisting of two components. One component is highly flattened, confined in the inner parts of the Galaxy, and contributes $\approx$40\% of the density of metal-poor stars near the Sun. 
For all these reasons, we conclude that \emph{stars with thick disc kinematics below and above $\rm [Fe/H]=-1$ constitute the same population, commonly referred to as the thick disc}, thus strengthening previous findings of a continuity between the thick disc and its metal-poor extension \citep{hawkins15a, hayes18}. 
Among the thick disc population,  stars with $\rm [Fe/H]\le-0.3$ have been significantly heated by the accretion of the satellite whose debris constitute the dominant halo population, at $\rm [Fe/H]<-1$ and at few kpc from the Sun. 
 
The second component that we find at $\rm [Fe/H]<-1$ belongs to the accreted population discovered by \citet{nissen10},  and -- in the metallicity range explored in this study -- this component appears to be the dominant one, constituting about 55-60\% of the total metal-poor populations at few kpc from the Sun. Currently, we find no evidence in our study  for a third, non-rotating, in-situ component : \emph{the non-rotating inner halo by \citet{carollo07, carollo10} is made, in our interpretation, only of stars belonging to the accreted population and of the low velocity tail of the thick disc at these metallicities.} As already discussed by \citet{haywood18}, the stars in the halo at metallicities below $\rm [Fe/H]=-1$ are dominated by stars accreted from the merging galaxy. The only other significant component that we find at these metallicities is the metal-poor thick disc, partly heated by the accretion to kinematics typical of halo stars, and thus at least partially overlapping with the population of accreted stars. The existence of  in-situ original non-rotating halo stars, i.e. the collapsed halo -- once accreted stars and metal-poor thick disc stars heated by the interaction are removed --  remains to be demonstrated.

\subsection{The kinematically defined halo is predominantly made of in situ, metal-rich, $\rm [Fe/H]>-1$, disc stars, heated by mergers}

In \citet{jeanbaptiste17}, we analysed N-body simulations of the accretion of one or several satellites onto a Milky Way-type galaxy to investigate the possibility to discriminate the origin of stars -- in-situ or accreted -- on the basis of their kinematics alone, as suggested by other studies \citep[see, for example][]{helmi00, gomez10}. One of the main conclusions of \citeauthor{jeanbaptiste17} was that distinguishing accreted and in-situ populations on
the basis of kinematics alone would be virtually impossible without detailed chemical abundances. They reached this conclusions because the populations have overlapping kinematics. In their simulations, indeed, in solar volumes of few kpc in size, the dominant halo component -- defined as stars with hot kinematics -- was found  to be made of in-situ material \citep[see also][]{font11, mccarthy12}. Figs.~14--17 already indicate that an in-situ halo, made of stars with metallicities typical of the Galactic thick disc with $\rm [Fe/H]>-1$ are present in the data. 
That stars with halo kinematics and  $\rm [Fe/H]>-1$ exist in the Galaxy is not a new finding, as it has been discussed in a number of previous studies \citep[see, for example][]{nissen10, nissen11, schuster12, jacksonjones14, bonaca17, fernandezalvar18, gaiababusiaux18}. Here we want to make a step further, by quantifying how significant this population is to the total halo population at few kpc from the Sun.

Fig.~\ref{insituhalo} (top-left panel) shows the Toomre diagram of all stars in our sample with halo kinematics, that is of all stars in the sample with $\sqrt{{(v_\Phi-v_{LSR})}^2+{v_R}^2+{v_Z}^2} >180$~\kms. To isolate the contribution of thick disc stars with  $\rm [Fe/H] > -1$, we have drawn an oblique line in the [Fe/H]--[Mg/Fe] plane that intercepts the thick disc at $\rm [Fe/H] = -1$ and $\rm [Mg/Fe] = 0.3$, and which separate the low-$\alpha$ portion of the accreted sequence from the thin and thick discs relatively well (middle-left panel).
Among stars with thick disc abundances and $\rm [Fe/H] > -1$, 830 stars have halo kinematics. For comparison, the number of stars of the accreted and in-situ sequence on the left of the diagonal line, with halo kinematics, is 717. Our sample thus indicates that at few kpc from the Sun, the majority  of the kinematically defined halo  is made of stars with $\rm [Fe/H] > -1$ and  [Mg/Fe]-abundances of the early thick disc, heated by the interaction. This is a very nice confirmation of the predictions  by \citet{jeanbaptiste17}, based on the analysis of N-body models, which suggested that in volumes of few  kpc around the Sun, the kinematically-defined halo should be dominated by in-situ disc stars heated by one or several merger(s).\\
 These heated disc stars do not only constitute the majority of the kinematically-defined halo in the region under study,  but also a not negligible fraction of $\alpha$-abundant stars at  $\rm [Fe/H] > -1$.  Fig.~\ref{insituhalo} (bottom-left panel) indeed shows the fraction of stars with halo kinematics, in various regions of the [Fe/H]--[Mg/Fe] plane. At high   [Mg/Fe] and $-1 < \rm [Fe/H] < -0.3$, thus in a range typical of  the thick disc, the fraction of halo stars can be as high as  20\%. This implies that any selection of thick disc stars made only on the basis of chemical abundances could lead to a significant contamination by stars with significantly hotter kinematics, on halo-like orbits. Of course the above fractions change if we change the limiting velocities discriminating disc stars from halo stars. If, for example, we adopt a kinematic definition for the halo as all stars with $\sqrt{{(v_\Phi-v_{LSR})}^2+{v_R}^2+{v_Z}^2} >220$~\kms,  the fractional contribution of the heated thick disc diminishes, and conversely the contribution of stars on the left of the diagonal line increases (top, middle and bottom-right panels).
With this selection, the fraction of metal-rich thick disc stars with halo kinematics decreases indeed to 38\% of the total sample of kinematically selected halo stars.  Still, such a selection is not able to completely eliminate the contamination of these stars to the canonical thick disc population, and indeed, as shown on the bottom-right panel of this Figure, the fraction of halo stars in the region of the [Fe/H]--[Mg/Fe] plane where  thick disc stars with  $-1 < \rm [Fe/H] < -0.3$ are  located can be as high as 10\%.\\
The significant contribution of the metal-rich thick disc population to the kinematically-defined halo is also shown in Fig.~\ref{mdf}, where the metallicity distribution of kinematically-defined halo stars is shown, for the two limiting velocities discriminating disc stars from halo stars, as described above. Even adopting the restrictive threshold of $\sqrt{{(v_\Phi-v_{LSR})}^2+{v_R}^2+{v_Z}^2} >220$~\kms, the peak associated to metal-rich thick disc population stands out clearly. Interestingly, this metallicity distribution is strongly reminiscent of the distribution of metallicity versus distance from the plane found by \citet{ibata17} and of that found by \citet{gallart19}. \\
These metal-rich, heated disc stars have a velocity ellipsoid $\rm (\sigma_R, \sigma_\Phi, \sigma_Z)=  (124\pm 3., 72.\pm 2., 72.\pm 2.)$~\kms, when a   threshold of $180$~\kms\ is adopted in defining halo stars, and $\rm (\sigma_R, \sigma_\Phi, \sigma_Z)=  (134 \pm 5., 70.  \pm 3., 81.\pm 3.)$~\kms, for a   threshold of $220$~\kms.

As we have seen in Sect.~\ref{sample}, our sample is still relatively local and biased against the inner populations of the Galaxy, because it lacks stars in the fourth quadrant, and because most of its stars are beyond the solar circle. However, we know from previous studies that the $\alpha$-abundant thick disc is massive \citep{snaith14, snaith15} and mostly confined in the inner Galactic regions  \citep{bensby11, bovy12b, cheng12, bovy16}, i.e. inside the solar circle. Under the hypothesis that the fraction of stars with halo-like kinematics, but thick disc abundances, as found here,  can be representative of the whole inner Galaxy, this would imply that the mass of thick disc stars at $\rm [Fe/H] > -1$ with halo-like kinematics would be significant, being between 10\%--20\% of the total thick disc mass, that is between $2-5\times10^9M_\odot$\footnote{For a total stellar mass of the Galaxy of  $5\times10^{10}M_\odot$}. This would make of this component, by far, the dominant halo component in the inner Galaxy, at $R\le 8-10$~kpc. These estimates must, of course, be taken with caution, because they are derived by extrapolating local fractions, but however suggest that the  in-situ heated disc may be the dominant halo component not only in the region under study, but in the inner Galaxy as well. 

\subsection{Heating versus cooling scenario}
So far, we have interpreted our findings as the signature of a major heating event in the Galaxy. Here we want to briefly discuss why we favour this interpretation to the one where the in-situ population at retrograde or null rotation (i.e. the plume found in Figs.~\ref{dating_main} and \ref{dating_zoom}) is made of stars formed from a cooling gaseous disc.
While the distinction between a heated disc and a cooling one could be difficult to establish, we believe that some conclusions can be derived on the basis of the metallicities of stars which make up the plume. The plume is visible up to [Fe/H] as high as -0.3, where the very large majority of stars  has already  disc kinematics. If the cooling of the gaseous disc left signatures in the stellar populations, we expect them to be found at [Fe/H] which pre-date the disc formation, not at metallicities where the disc is already fully formed. Conversely, in the case of a heating scenario, we expect exactly the signature we observe : a fraction of disc stars - with same chemical abundances - to have halo- (prograde and retrograde) kinematics. As discussed in the previous sections, indeed, in this scenario, an initial (before the merger(s)) same in-situ disc  population would end up (after the merger(s)) having mostly disc-kinematics, while a fraction of it would be heated up to the halo. The prediction is thus to have same abundances, but different kinematics, which is exactly what we see in the data. Moreover, the prediction of a heating scenario is that part of the heated halo stars should have kinematic properties  overlapping with those of the accreted retrograde material \citep{jeanbaptiste17}, which is what we find: stars with in-situ chemistry and retrograde motions. The presence of retrograde motions in the plume, in particular, seems to favor the heating scenario, because in a cooling disc, while an increase of the angular momentum with time is expected,  it is difficult to explain how to maintain part of a cooling gaseous disc in counter-rotation, because of its dissipative nature.

\subsection{The epoch of the last significant accretion and the question of an in-situ stellar halo}

The epoch of accretion that we measure is in agreement with what has been inferred in various studies \citep{belokurov18,mackereth18,helmi18, gallart19}. It is also consistent with the epoch of last significant merging experienced by the Galaxy, derived by  \citet{kruijssen19}  based on the analysis of the globular cluster population. The difference, however, 
is that we find the fossil record left  by the accretion in the kinematic properties of the disc at $\rm [Fe/H]\le -0.3$
and $\rm [Mg/Fe] \ge 0.2$, corresponding to 9-11 Gyr ago. This is in particular  in agreement with the inference made by \cite{belokurov18} that the accretion must have occurred at a time when the disc 
was already substantially massive, in order to produce the velocity anisotropy that is observed.
This is also in agreement with the results found by \citet{mackereth18}, who show, analysing the 
Eagle simulations, that material on high eccentricities cannot have been accreted too early.
Hence, an accretion occurring 9 to 11 Gyr ago falls at the right time, that is when the formation of thick disc was  ongoing or approaching 
near completion, reaching a stellar mass between 1 and $\rm 2.10^{10} M_{\odot}$ \citep[see][]{snaith15}. It also corresponds to a time when the star formation in the Galaxy has been particularly intense \citep[see][]{snaith14, lehnert14}.\\
As anticipated in \cite{haywood18}, the accretion event has been responsible for heating the disc and even pushing 
a fraction of the thick disc stars on counter-rotating orbits and orbits on higher eccentricities, populating the 
kinematically defined halo. 
With the results presented here, the question of the existence of an in-situ halo, other than the heated early disc, as raised already in \cite{haywood18}, is even more acute.

Even on larger scales, the evidence of an in-situ halo other than the heated early Milky Way disc is becoming weaker by the day. For instance, the recent work by \cite{iorio18}
shows that the distribution of RR Lyrae over the entire Galaxy -- after being cleaned of known structures -- has kinematic properties suggesting a strong 
anisotropy compatible with accreted material  -- the so-called ``Gaia Sausage'', in their study.  
On the large scale, the Gaia Sausage stars would be distributed in a triaxial structure whose major axis would be 
aligned at $\rm 70^{\circ}$ with the Sun-Galactic centre direction. 
According to \cite{simion19}, the Hercules-Aquila cloud and the Virgo Over-Density, possibly aligned with major axis of the Gaia Sausage,  
could be part of the same accretion event. 
\cite{iorio18} conclude from their analysis that the bulk of the triaxial structure made by RR Lyrae
stars must have been left by this accretion event.

An entirely new and unexpected result also support the view that a spherical isotropic in-situ halo
is  a very small component of the Milky Way (much less massive than the $\sim$ 1\% of stars that was 
up to now viewed as the contribution of the Galactic halo).
The discovery of stars with disc orbits at very low metallicities ($\rm [Fe/H]< -4$, see \citet{sestito18}) suggests that the dissipative 
collapse that led to the formation of the thick disc probably was extremely rapid (within 10$^8$ years?), 
leaving very few time to build a spherical, in-situ halo.
This argument, which is reminiscent of the rapid collapse advocated by \cite{eggen62}, suggests that the Galaxy may have
started to form stars in a disc configuration very soon after the Big-Bang, leaving only a very short amount of time 
to form stars in a more spheroidal configuration.

\section{Conclusions}\label{conclusions}

By coupling astrometric data from \gaia~DR2, with elemental abundances from APOGEE~DR14, we continue our study on the nature of the Galactic halo at few kpc from the Sun, started in \citet{haywood18}. Here, in particular, we characterize the kinematics and chemistry of in-situ and accreted populations in the Galaxy up to [Fe/H]$\sim -2$. Our results can be summarized as follows.
\begin{itemize}
\item In the [Fe/H]--[Mg/Fe] plane, the chemical sequence of accreted stars  -- firstly identified by \citet{nissen10}, and later shown by \citet{haywood18} to be the dominant component among stars with $\rm [Fe/H] < -1$ -- remarkably appears as a distinct sequence in terms of its kinematics, being characterized by a mean null or retrograde motion and significantly higher velocity dispersions than those of disc stars and halo stars with higher [Mg/Fe] ratios, but same metallicities. 
\item Accreted stars appear to significantly impact the Galactic chemo-kinematic relations, not only at [Fe/H]$\le -1$, but also at metallicities typical of the thick and  metal-poor thin discs.  
In this context, we can revisit the finding of \citet{minchev14} \citep[see also][]{guiglion15} and conclude that: \textit{(1)} the inversion they found in velocity dispersions--[Mg/Fe] relations at $\rm [Fe/H] < -0.4$ -- where stars with  the lowest [Mg/Fe] ratios have also the highest velocity dispersions -- is exclusively driven by the accreted population, and not by in-situ disc stars heated by the accretion. In-situ disc stars heated by the interaction are indeed characterized by high [Mg/Fe] ratios, and not  low  [Mg/Fe], as expected in their scenario;  \textit{(2)} there is no sign of a drop in the velocity dispersion of stars at the high  [Mg/Fe]  end, indicative, in  \citet{minchev14} interpretation, of a significant radial migration from the inner to the outer disc.
\item We show that stars with thick disc kinematics are present up to the low  [Fe/H] end of our sample ($\rm [Fe/H] \sim -2$) , and that they constitute about 40\% of all stars with $\rm [Fe/H] < -1$. This estimate is in excellent agreement with the findings of \citet{beers02}.  The remaining 60\% is made of accreted stars, which thus -- confirming the finding in \citet{haywood18} -- represent the majority of stars at these metallicities.
\item While stars at $\rm [Fe/H] < -1$ are a mix of accreted and thick disc stars, when the halo is defined on the basis of the kinematics of its stars, the majority is made, at few kpc from the Sun, of metal-rich  (i.e. $\rm [Fe/H] > -1$) stars heated by the interaction. These stars have the same chemical composition of the thick disc, but hotter kinematics. Their existence in large proportions in the inner Galaxy has been predicted by a number of N-body models  \citep{zolotov10, purcell10, font11, qu11b, mccarthy12, jeanbaptiste17} and here we show that they constitute indeed a significant kinematically `hot' component in the volume under study. 
\item Based on these findings, we tentatively estimate this early disc component heated by the interaction  to have a total stellar mass of  about $2-5 \times10^9 M_\odot$ in the inner Galaxy, i.e. inside $R=8-10$~kpc. If these estimates will find confirmations in future studies, this would make, of this component, by far the dominant halo component of the inner Galaxy. 
\item By constraining the metallicity and [Mg/Fe] ratio of these disc stars heated to halo kinematics, we can date the major accretion event now reported in a number of studies \citep{nissen10,  nissen11, schuster12, ramirez12, hawkins15a, hayes18, belokurov18, haywood18, iorio18, helmi18, mackereth18, gallart19} to have occurred between 9 and 11~Gyr ago. To our knowledge, this is the first time this accretion  is dated on the basis of the kinematic imprints it left on disc stars present in the Galaxy at the time it occurred.
\end{itemize}

The picture that emerges from our study, and that confirms the earliest conclusions in \citet{haywood18}, is that an in-situ halo, other than the heated thick disc,   possibly does not exist in our Galaxy, or it represents much less than  some few percent of stars, as usually reported in the literature. \emph{The only in-situ population that we find in great proportions both among the kinematically defined, and the chemically defined, halo stars is the thick disc, that is the early disc of the Galaxy heated to hot kinematics}. This population constitutes indeed a significant or dominant contributor to the inner halo, and cannot be neglected in all discussions about its origins. 
As for a distinctive in-situ halo population, if it exists, is further beyond our reach.

\section*{Acknowledgments}
The authors wish to thank P.~Bonifacio, E.~Caffau, D.~Kruijssen and I.~Minchev for valuable discussions and remarks on this work. The authors are grateful to the referee, for their very constructive report which much improved the manuscript.\\
This work has been supported by the ANR (Agence Nationale de la Recherche) through the MOD4Gaia project (ANR-15-CE31-0007, P.I.: P. Di Matteo). \\
SK was supported by the Russian Science Foundation, project no. 19-72-20089. \\
This work has made use of data
from the European Space Agency (ESA) mission Gaia
(https://www.cosmos.esa.int/gaia), processed by the
Gaia Data Processing and Analysis Consortium (DPAC,
https://www.cosmos.esa.int/web/gaia/dpac/consortium).
Funding for the DPAC has been provided by national
institutions, in particular the institutions participating
in the Gaia Multilateral Agreement. \\This research has
made use of the SIMBAD database, operated at CDS,
Strasbourg, France.\\ Funding for the Sloan Digital Sky
Survey IV has been provided by the Alfred P. Sloan
Foundation, the U.S. Department of Energy Office of
Science, and the Participating Institutions. SDSS-IV
acknowledges support and resources from the Center
for High-Performance Computing at the University of
Utah. The SDSS web site is www.sdss.org. SDSS-IV
is managed by the Astrophysical Research Consortium
for the Participating Institutions of the SDSS Collaboration
including the Brazilian Participation Group,
the Carnegie Institution for Science, Carnegie Mellon
University, the Chilean Participation Group, the French
Participation Group, Harvard-Smithsonian Center for
Astrophysics, Instituto de Astrof$\rm \acute{i}$sica de Canarias,
The Johns Hopkins University, Kavli Institute for the
Physics and Mathematics of the Universe (IPMU)
/ University of Tokyo, Lawrence Berkeley National
Laboratory, Leibniz Institut f$\rm\ddot u$r Astrophysik Potsdam
(AIP), Max-Planck-Institut f$\rm\ddot u$r Astronomie (MPIA
Heidelberg), Max-Planck-Institut f$\rm\ddot u$r Astrophysik (MPA
Garching), Max-Planck-Institut f$\rm\ddot u$r Extraterrestrische
Physik (MPE), National Astronomical Observatories of
China, New Mexico State University, New York University,
University of Notre Dame, Observat$\rm\acute{a}$rio Nacional /
MCTI, The Ohio State University, Pennsylvania State
University, Shanghai Astronomical Observatory, United
Kingdom Participation Group, Universidad Nacional
Aut$\rm\acute{o}$noma de M$\rm\acute{e}$xico, University of Arizona, University
of Colorado Boulder, University of Oxford, University of
Portsmouth, University of Utah, University of Virginia,
University of Washington, University of Wisconsin,
Vanderbilt University, and Yale University.

\bibliographystyle{aa}
\bibliography{biblio}

\begin{thebibliography}{105}
\expandafter\ifx\csname natexlab\endcsname\relax\def\natexlab#1{#1}\fi

\bibitem[{{Adibekyan} {et~al.}(2012){Adibekyan}, {Sousa}, {Santos}, {Delgado
  Mena}, {Gonz{\'a}lez Hern{\'a}ndez}, {Israelian}, {Mayor}, \&
  {Khachatryan}}]{adibekyan12}
{Adibekyan}, V.~Z., {Sousa}, S.~G., {Santos}, N.~C., {et~al.} 2012, \aap, 545,
  A32

\bibitem[{{Arenou} {et~al.}(2018){Arenou}, {Luri}, {Babusiaux}, {Fabricius},
  {Helmi}, {Muraveva}, {Robin}, {Spoto}, {Vallenari}, {Antoja},
  {Cantat-Gaudin}, {Jordi}, {Leclerc}, {Reyl{\'e}}, {Romero-G{\'o}mez}, {Shih},
  {Soria}, {Barache}, {Bossini}, {Bragaglia}, {Breddels}, {Fabrizio},
  {Lambert}, {Marrese}, {Massari}, {Moitinho}, {Robichon}, {Ruiz-Dern},
  {Sordo}, {Veljanoski}, {Eyer}, {Jasniewicz}, {Pancino}, {Soubiran}, {Spagna},
  {Tanga}, {Turon}, \& {Zurbach}}]{arenou18}
{Arenou}, F., {Luri}, X., {Babusiaux}, C., {et~al.} 2018, \aap, 616, A17

\bibitem[{{Beers} {et~al.}(2002){Beers}, {Drilling}, {Rossi}, {Chiba}, {Rhee},
  {F{\"u}hrmeister}, {Norris}, \& {von Hippel}}]{beers02}
{Beers}, T.~C., {Drilling}, J.~S., {Rossi}, S., {et~al.} 2002, \aj, 124, 931

\bibitem[{{Bell} {et~al.}(2008){Bell}, {Zucker}, {Belokurov}, {Sharma},
  {Johnston}, {Bullock}, {Hogg}, {Jahnke}, {de Jong}, {Beers}, {Evans},
  {Grebel}, {Ivezi{\'c}}, {Koposov}, {Rix}, {Schneider}, {Steinmetz}, \&
  {Zolotov}}]{bell08}
{Bell}, E.~F., {Zucker}, D.~B., {Belokurov}, V., {et~al.} 2008, \apj, 680, 295

\bibitem[{{Belokurov} {et~al.}(2018){Belokurov}, {Erkal}, {Evans}, {Koposov},
  \& {Deason}}]{belokurov18}
{Belokurov}, V., {Erkal}, D., {Evans}, N.~W., {Koposov}, S.~E., \& {Deason},
  A.~J. 2018, \mnras, 478, 611

\bibitem[{{Bensby} {et~al.}(2011){Bensby}, {Alves-Brito}, {Oey}, {Yong}, \&
  {Mel{\'e}ndez}}]{bensby11}
{Bensby}, T., {Alves-Brito}, A., {Oey}, M.~S., {Yong}, D., \& {Mel{\'e}ndez},
  J. 2011, \apjl, 735, L46

\bibitem[{{Benson} {et~al.}(2004){Benson}, {Lacey}, {Frenk}, {Baugh}, \&
  {Cole}}]{benson04}
{Benson}, A.~J., {Lacey}, C.~G., {Frenk}, C.~S., {Baugh}, C.~M., \& {Cole}, S.
  2004, \mnras, 351, 1215

\bibitem[{{Bonaca} {et~al.}(2017){Bonaca}, {Conroy}, {Wetzel}, {Hopkins}, \&
  {Kere{\v s}}}]{bonaca17}
{Bonaca}, A., {Conroy}, C., {Wetzel}, A., {Hopkins}, P.~F., \& {Kere{\v s}}, D.
  2017, \apj, 845, 101

\bibitem[{{Bovy} {et~al.}(2012){Bovy}, {Rix}, {Liu}, {Hogg}, {Beers}, \&
  {Lee}}]{bovy12b}
{Bovy}, J., {Rix}, H.-W., {Liu}, C., {et~al.} 2012, \apj, 753, 148

\bibitem[{{Bovy} {et~al.}(2016){Bovy}, {Rix}, {Schlafly}, {Nidever},
  {Holtzman}, {Shetrone}, \& {Beers}}]{bovy16}
{Bovy}, J., {Rix}, H.-W., {Schlafly}, E.~F., {et~al.} 2016, \apj, 823, 30

\bibitem[{{Brown} {et~al.}(2008){Brown}, {Beers}, {Wilhelm}, {Allende Prieto},
  {Geller}, {Kenyon}, \& {Kurtz}}]{brown08}
{Brown}, W.~R., {Beers}, T.~C., {Wilhelm}, R., {et~al.} 2008, \aj, 135, 564

\bibitem[{{Bullock} \& {Johnston}(2004)}]{bullock04}
{Bullock}, J.~S. \& {Johnston}, K.~V. 2004, in Astronomical Society of the
  Pacific Conference Series, Vol. 327, Satellites and Tidal Streams, ed.
  F.~{Prada}, D.~{Martinez Delgado}, \& T.~J. {Mahoney}, 80

\bibitem[{{Carollo} {et~al.}(2010){Carollo}, {Beers}, {Chiba}, {Norris},
  {Freeman}, {Lee}, {Ivezi{\'c}}, {Rockosi}, \& {Yanny}}]{carollo10}
{Carollo}, D., {Beers}, T.~C., {Chiba}, M., {et~al.} 2010, \apj, 712, 692

\bibitem[{{Carollo} {et~al.}(2007){Carollo}, {Beers}, {Lee}, {Chiba}, {Norris},
  {Wilhelm}, {Sivarani}, {Marsteller}, {Munn}, {Bailer-Jones}, {Fiorentin}, \&
  {York}}]{carollo07}
{Carollo}, D., {Beers}, T.~C., {Lee}, Y.~S., {et~al.} 2007, \nat, 450, 1020

\bibitem[{{Chen} {et~al.}(2001){Chen}, {Stoughton}, {Smith}, {Uomoto}, {Pier},
  {Yanny}, {Ivezi{\'c}}, {York}, {Anderson}, {Annis}, {Brinkmann}, {Csabai},
  {Fukugita}, {Hindsley}, {Lupton}, {Munn}, \& {SDSS Collaboration}}]{chen01}
{Chen}, B., {Stoughton}, C., {Smith}, J.~A., {et~al.} 2001, \apj, 553, 184

\bibitem[{{Cheng} {et~al.}(2012){Cheng}, {Rockosi}, {Morrison}, {Lee}, {Beers},
  {Bizyaev}, {Harding}, {Malanushenko}, {Malanushenko}, {Oravetz}, {Pan},
  {Schlesinger}, {Schneider}, {Simmons}, \& {Weaver}}]{cheng12}
{Cheng}, J.~Y., {Rockosi}, C.~M., {Morrison}, H.~L., {et~al.} 2012, \apj, 752,
  51

\bibitem[{{Chiba} \& {Beers}(2000)}]{chiba00}
{Chiba}, M. \& {Beers}, T.~C. 2000, \aj, 119, 2843

\bibitem[{{Cole}(1991)}]{cole91}
{Cole}, S. 1991, \apj, 367, 45

\bibitem[{{De Lucia} \& {Helmi}(2008)}]{delucia08}
{De Lucia}, G. \& {Helmi}, A. 2008, \mnras, 391, 14

\bibitem[{{Eggen} {et~al.}(1962){Eggen}, {Lynden-Bell}, \& {Sandage}}]{eggen62}
{Eggen}, O.~J., {Lynden-Bell}, D., \& {Sandage}, A.~R. 1962, \apj, 136, 748

\bibitem[{{Fattahi} {et~al.}(2018){Fattahi}, {Belokurov}, {Deason}, {Frenk},
  {Gomez}, {Grand}, {Marinacci}, {Pakmor}, \& {Springel}}]{fattahi18}
{Fattahi}, A., {Belokurov}, V., {Deason}, A.~J., {et~al.} 2018, ArXiv e-prints
  [\eprint[arXiv]{1810.07779}]

\bibitem[{{Fern{\'a}ndez-Alvar}
  {et~al.}(2018{\natexlab{a}}){Fern{\'a}ndez-Alvar}, {Fern{\'a}ndez-Trincado},
  {Moreno}, {Schuster}, {Carigi}, {Recio-Blanco}, {Beers}, {Chiappini},
  {Anders}, {Santiago}, {Queiroz}, {P{\'e}rez-Villegas}, {Zamora}, \&
  {Garc{\'{\i}}a-Hern{\'a}ndez}}]{fernandezalvar18}
{Fern{\'a}ndez-Alvar}, E., {Fern{\'a}ndez-Trincado}, J.~G., {Moreno}, E.,
  {et~al.} 2018{\natexlab{a}}, ArXiv e-prints [\eprint[arXiv]{1807.07269}]

\bibitem[{{Fern{\'a}ndez-Alvar}
  {et~al.}(2018{\natexlab{b}}){Fern{\'a}ndez-Alvar}, {Tissera}, {Carigi},
  {Schuster}, {Beers}, \& {Belokurov}}]{fernandez18}
{Fern{\'a}ndez-Alvar}, E., {Tissera}, P.~B., {Carigi}, L., {et~al.}
  2018{\natexlab{b}}, ArXiv e-prints [\eprint[arXiv]{1809.02368}]

\bibitem[{{Font} {et~al.}(2006){Font}, {Johnston}, {Bullock}, \&
  {Robertson}}]{font06}
{Font}, A.~S., {Johnston}, K.~V., {Bullock}, J.~S., \& {Robertson}, B.~E. 2006,
  \apj, 646, 886

\bibitem[{{Font} {et~al.}(2011){Font}, {McCarthy}, {Crain}, {Theuns}, {Schaye},
  {Wiersma}, \& {Dalla Vecchia}}]{font11}
{Font}, A.~S., {McCarthy}, I.~G., {Crain}, R.~A., {et~al.} 2011, \mnras, 416,
  2802

\bibitem[{{Font} {et~al.}(2001){Font}, {Navarro}, {Stadel}, \&
  {Quinn}}]{font01}
{Font}, A.~S., {Navarro}, J.~F., {Stadel}, J., \& {Quinn}, T. 2001, \apjl, 563,
  L1

\bibitem[{{Forbes} \& {Bridges}(2010)}]{forbes10}
{Forbes}, D.~A. \& {Bridges}, T. 2010, \mnras, 404, 1203

\bibitem[{{Gaia Collaboration} {et~al.}(2018{\natexlab{a}}){Gaia
  Collaboration}, {Babusiaux}, {van Leeuwen}, {Barstow}, {Jordi}, {Vallenari},
  {Bossini}, {Bressan}, {Cantat-Gaudin}, {van Leeuwen}, \&
  et~al.}]{gaiababusiaux18}
{Gaia Collaboration}, {Babusiaux}, C., {van Leeuwen}, F., {et~al.}
  2018{\natexlab{a}}, \aap, 616, A10

\bibitem[{{Gaia Collaboration} {et~al.}(2018{\natexlab{b}}){Gaia
  Collaboration}, {Brown}, {Vallenari}, {Prusti}, {de Bruijne}, {Babusiaux},
  {Bailer-Jones}, {Biermann}, {Evans}, {Eyer}, \& et~al.}]{gaia18}
{Gaia Collaboration}, {Brown}, A.~G.~A., {Vallenari}, A., {et~al.}
  2018{\natexlab{b}}, \aap, 616, A1

\bibitem[{{Gaia Collaboration} {et~al.}(2017){Gaia Collaboration},
  {Clementini}, {Eyer}, {Ripepi}, {Marconi}, {Muraveva}, {Garofalo}, {Sarro},
  {Palmer}, {Luri}, \& et~al.}]{gaia17}
{Gaia Collaboration}, {Clementini}, G., {Eyer}, L., {et~al.} 2017, \aap, 605,
  A79

\bibitem[{{Gaia Collaboration} {et~al.}(2016){Gaia Collaboration}, {Prusti},
  {de Bruijne}, {Brown}, {Vallenari}, {Babusiaux}, {Bailer-Jones}, {Bastian},
  {Biermann}, {Evans}, \& et~al.}]{gaia16}
{Gaia Collaboration}, {Prusti}, T., {de Bruijne}, J.~H.~J., {et~al.} 2016,
  \aap, 595, A1

\bibitem[{{Gallart} {et~al.}(2019){Gallart}, {Bernard}, {Brook}, {Ruiz-Lara},
  {Cassisi}, {Hill}, \& {Monelli}}]{gallart19}
{Gallart}, C., {Bernard}, E.~J., {Brook}, C.~B., {et~al.} 2019, Nature
  Astronomy, 407

\bibitem[{{G{\'o}mez} {et~al.}(2010){G{\'o}mez}, {Helmi}, {Brown}, \&
  {Li}}]{gomez10}
{G{\'o}mez}, F.~A., {Helmi}, A., {Brown}, A.~G.~A., \& {Li}, Y.-S. 2010,
  \mnras, 408, 935

\bibitem[{{Guiglion} {et~al.}(2015){Guiglion}, {Recio-Blanco}, {de Laverny},
  {Kordopatis}, {Hill}, {Mikolaitis}, {Minchev}, {Chiappini}, {Wyse},
  {Gilmore}, {Randich}, {Feltzing}, {Bensby}, {Flaccomio}, {Koposov},
  {Pancino}, {Bayo}, {Costado}, {Franciosini}, {Hourihane}, {Jofr{\'e}},
  {Lardo}, {Lewis}, {Lind}, {Magrini}, {Morbidelli}, {Sacco}, {Ruchti},
  {Worley}, \& {Zaggia}}]{guiglion15}
{Guiglion}, G., {Recio-Blanco}, A., {de Laverny}, P., {et~al.} 2015, \aap, 583,
  A91

\bibitem[{{Hawkins} {et~al.}(2015){Hawkins}, {Jofr{\'e}}, {Masseron}, \&
  {Gilmore}}]{hawkins15a}
{Hawkins}, K., {Jofr{\'e}}, P., {Masseron}, T., \& {Gilmore}, G. 2015, \mnras,
  453, 758

\bibitem[{{Hayes} {et~al.}(2018){Hayes}, {Majewski}, {Shetrone},
  {Fern{\'a}ndez-Alvar}, {Allende Prieto}, {Schuster}, {Carigi}, {Cunha},
  {Smith}, {Sobeck}, {Almeida}, {Beers}, {Carrera}, {Fern{\'a}ndez-Trincado},
  {Garc{\'{\i}}a-Hern{\'a}ndez}, {Geisler}, {Lane}, {Lucatello}, {Matthews},
  {Minniti}, {Nitschelm}, {Tang}, {Tissera}, \& {Zamora}}]{hayes18}
{Hayes}, C.~R., {Majewski}, S.~R., {Shetrone}, M., {et~al.} 2018, \apj, 852, 49

\bibitem[{{Haywood}(2008)}]{haywood08}
{Haywood}, M. 2008, \mnras, 388, 1175

\bibitem[{{Haywood} {et~al.}(2013){Haywood}, {Di Matteo}, {Lehnert}, {Katz}, \&
  {G{\'o}mez}}]{haywood13}
{Haywood}, M., {Di Matteo}, P., {Lehnert}, M.~D., {Katz}, D., \& {G{\'o}mez},
  A. 2013, \aap, 560, A109

\bibitem[{{Haywood} {et~al.}(2018){Haywood}, {Di Matteo}, {Lehnert}, {Snaith},
  {Khoperskov}, \& {G{\'o}mez}}]{haywood18}
{Haywood}, M., {Di Matteo}, P., {Lehnert}, M.~D., {et~al.} 2018, \apj, 863, 113

\bibitem[{{Haywood} {et~al.}(2015){Haywood}, {Di Matteo}, {Snaith}, \&
  {Lehnert}}]{haywood15}
{Haywood}, M., {Di Matteo}, P., {Snaith}, O., \& {Lehnert}, M.~D. 2015, \aap,
  579, A5

\bibitem[{{Helmi} {et~al.}(2018){Helmi}, {Babusiaux}, {Koppelman}, {Massari},
  {Veljanoski}, \& {Brown}}]{helmi18}
{Helmi}, A., {Babusiaux}, C., {Koppelman}, H.~H., {et~al.} 2018, ArXiv e-prints
  [\eprint[arXiv]{1806.06038}]

\bibitem[{{Helmi} \& {de Zeeuw}(2000)}]{helmi00}
{Helmi}, A. \& {de Zeeuw}, P.~T. 2000, \mnras, 319, 657

\bibitem[{{Helmi} {et~al.}(1999){Helmi}, {White}, {de Zeeuw}, \&
  {Zhao}}]{helmi99}
{Helmi}, A., {White}, S.~D.~M., {de Zeeuw}, P.~T., \& {Zhao}, H. 1999, \nat,
  402, 53

\bibitem[{{House} {et~al.}(2011){House}, {Brook}, {Gibson},
  {S{\'a}nchez-Bl{\'a}zquez}, {Courty}, {Few}, {Governato}, {Kawata}, {Ro{\v
  s}kar}, {Steinmetz}, {Stinson}, \& {Teyssier}}]{house11}
{House}, E.~L., {Brook}, C.~B., {Gibson}, B.~K., {et~al.} 2011, \mnras, 415,
  2652

\bibitem[{{Ibata} {et~al.}(2018){Ibata}, {Malhan}, {Martin}, \&
  {Starkenburg}}]{ibata18}
{Ibata}, R.~A., {Malhan}, K., {Martin}, N.~F., \& {Starkenburg}, E. 2018, ArXiv
  e-prints [\eprint[arXiv]{1806.01195}]

\bibitem[{{Ibata} {et~al.}(2017){Ibata}, {McConnachie}, {Cuillandre}, {Fantin},
  {Haywood}, {Martin}, {Bergeron}, {Beckmann}, {Bernard}, {Bonifacio},
  {Caffau}, {Carlberg}, {C{\^o}t{\'e}}, {Cabanac}, {Chapman}, {Duc}, {Durret},
  {Famaey}, {Fabbro}, {Gwyn}, {Hammer}, {Hill}, {Hudson}, {Lan{\c c}on},
  {Lewis}, {Malhan}, {di Matteo}, {McCracken}, {Mei}, {Mellier}, {Navarro},
  {Pires}, {Pritchet}, {Reyl{\'e}}, {Richer}, {Robin}, {S{\'a}nchez-Janssen},
  {Sawicki}, {Scott}, {Scottez}, {Spekkens}, {Starkenburg}, {Thomas}, \&
  {Venn}}]{ibata17}
{Ibata}, R.~A., {McConnachie}, A., {Cuillandre}, J.-C., {et~al.} 2017, \apj,
  848, 129

\bibitem[{{Iorio} \& {Belokurov}(2019)}]{iorio18}
{Iorio}, G. \& {Belokurov}, V. 2019, \mnras, 482, 3868

\bibitem[{{Jackson-Jones} {et~al.}(2014){Jackson-Jones}, {Jofr{\'e}},
  {Hawkins}, {Hourihane}, {Gilmore}, {Kordopatis}, {Worley}, {Randich},
  {Vallenari}, {Bensby}, {Bragaglia}, {Flaccomio}, {Korn}, {Recio-Blanco},
  {Smiljanic}, {Costado}, {Heiter}, {Hill}, {Lardo}, {de Laverny}, {Guiglion},
  {Mikolaitis}, {Zaggia}, \& {Tautvai{\v s}ien{\.e}}}]{jacksonjones14}
{Jackson-Jones}, R., {Jofr{\'e}}, P., {Hawkins}, K., {et~al.} 2014, \aap, 571,
  L5

\bibitem[{{Jean-Baptiste} {et~al.}(2017){Jean-Baptiste}, {Di Matteo},
  {Haywood}, {G{\'o}mez}, {Montuori}, {Combes}, \& {Semelin}}]{jeanbaptiste17}
{Jean-Baptiste}, I., {Di Matteo}, P., {Haywood}, M., {et~al.} 2017, \aap, 604,
  A106

\bibitem[{{Johnson} \& {Soderblom}(1987)}]{johnson87}
{Johnson}, D. R.~H. \& {Soderblom}, D.~R. 1987, The Astronomical Journal, 93,
  864

\bibitem[{{Johnston} {et~al.}(2008){Johnston}, {Bullock}, {Sharma}, {Font},
  {Robertson}, \& {Leitner}}]{johnston08}
{Johnston}, K.~V., {Bullock}, J.~S., {Sharma}, S., {et~al.} 2008, \apj, 689,
  936

\bibitem[{{Kazantzidis} {et~al.}(2008){Kazantzidis}, {Bullock}, {Zentner},
  {Kravtsov}, \& {Moustakas}}]{kazantzidis08}
{Kazantzidis}, S., {Bullock}, J.~S., {Zentner}, A.~R., {Kravtsov}, A.~V., \&
  {Moustakas}, L.~A. 2008, \apj, 688, 254

\bibitem[{{Kepley} {et~al.}(2007){Kepley}, {Morrison}, {Helmi}, {Kinman}, {Van
  Duyne}, {Martin}, {Harding}, {Norris}, \& {Freeman}}]{kepley07}
{Kepley}, A.~A., {Morrison}, H.~L., {Helmi}, A., {et~al.} 2007, The
  Astronomical Journal, 134, 1579

\bibitem[{{Koppelman} {et~al.}(2018){Koppelman}, {Helmi}, \&
  {Veljanoski}}]{koppelman18}
{Koppelman}, H., {Helmi}, A., \& {Veljanoski}, J. 2018, \apjl, 860, L11

\bibitem[{{Kordopatis} {et~al.}(2013){Kordopatis}, {Gilmore}, {Wyse},
  {Steinmetz}, {Siebert}, {Bienaym{\'e}}, {McMillan}, {Minchev}, {Zwitter},
  {Gibson}, {Seabroke}, {Grebel}, {Bland-Hawthorn}, {Boeche}, {Freeman},
  {Munari}, {Navarro}, {Parker}, {Reid}, \& {Siviero}}]{kordopatis13}
{Kordopatis}, G., {Gilmore}, G., {Wyse}, R.~F.~G., {et~al.} 2013, \mnras, 436,
  3231

\bibitem[{{Kruijssen} {et~al.}(2019){Kruijssen}, {Pfeffer}, {Reina-Campos},
  {Crain}, \& {Bastian}}]{kruijssen19}
{Kruijssen}, J.~M.~D., {Pfeffer}, J.~L., {Reina-Campos}, M., {Crain}, R.~A., \&
  {Bastian}, N. 2019, \mnras, 486, 3180

\bibitem[{{Leaman} {et~al.}(2013){Leaman}, {VandenBerg}, \&
  {Mendel}}]{leaman13}
{Leaman}, R., {VandenBerg}, D.~A., \& {Mendel}, J.~T. 2013, \mnras, 436, 122

\bibitem[{{Lee} {et~al.}(2011){Lee}, {Beers}, {An}, {Ivezi{\'c}}, {Just},
  {Rockosi}, {Morrison}, {Johnson}, {Sch{\"o}nrich}, {Bird}, {Yanny},
  {Harding}, \& {Rocha-Pinto}}]{lee11}
{Lee}, Y.~S., {Beers}, T.~C., {An}, D., {et~al.} 2011, \apj, 738, 187

\bibitem[{{Lehnert} {et~al.}(2014){Lehnert}, {Di Matteo}, {Haywood}, \&
  {Snaith}}]{lehnert14}
{Lehnert}, M.~D., {Di Matteo}, P., {Haywood}, M., \& {Snaith}, O.~N. 2014,
  \apjl, 789, L30

\bibitem[{{Li} \& {Zhao}(2017)}]{li17}
{Li}, C. \& {Zhao}, G. 2017, \apj, 850, 25

\bibitem[{{Lindegren} {et~al.}(2018){Lindegren}, {Hern{\'a}ndez}, {Bombrun},
  {Klioner}, {Bastian}, {Ramos-Lerate}, {de Torres}, {Steidelm{\"u}ller},
  {Stephenson}, {Hobbs}, {Lammers}, {Biermann}, {Geyer}, {Hilger}, {Michalik},
  {Stampa}, {McMillan}, {Casta{\~n}eda}, {Clotet}, {Comoretto}, {Davidson},
  {Fabricius}, {Gracia}, {Hambly}, {Hutton}, {Mora}, {Portell}, {van Leeuwen},
  {Abbas}, {Abreu}, {Altmann}, {Andrei}, {Anglada}, {Balaguer-N{\'u}{\~n}ez},
  {Barache}, {Becciani}, {Bertone}, {Bianchi}, {Bouquillon}, {Bourda},
  {Br{\"u}semeister}, {Bucciarelli}, {Busonero}, {Buzzi}, {Cancelliere},
  {Carlucci}, {Charlot}, {Cheek}, {Crosta}, {Crowley}, {de Bruijne}, {de
  Felice}, {Drimmel}, {Esquej}, {Fienga}, {Fraile}, {Gai}, {Garralda},
  {Gonz{\'a}lez-Vidal}, {Guerra}, {Hauser}, {Hofmann}, {Holl}, {Jordan},
  {Lattanzi}, {Lenhardt}, {Liao}, {Licata}, {Lister}, {L{\"o}ffler},
  {Marchant}, {Martin-Fleitas}, {Messineo}, {Mignard}, {Morbidelli}, {Poggio},
  {Riva}, {Rowell}, {Salguero}, {Sarasso}, {Sciacca}, {Siddiqui}, {Smart},
  {Spagna}, {Steele}, {Taris}, {Torra}, {van Elteren}, {van Reeven}, \&
  {Vecchiato}}]{lindegren18}
{Lindegren}, L., {Hern{\'a}ndez}, J., {Bombrun}, A., {et~al.} 2018, \aap, 616,
  A2

\bibitem[{{Mackereth} {et~al.}(2018){Mackereth}, {Schiavon}, {Pfeffer},
  {Hayes}, {Bovy}, {Anguiano}, {Allende Prieto}, {Hasselquist}, {Holtzman},
  {Johnson}, {Majewski}, {O'Connell}, {Shetrone}, {Tissera}, \&
  {Fern{\'a}ndez-Trincado}}]{mackereth18}
{Mackereth}, J.~T., {Schiavon}, R.~P., {Pfeffer}, J., {et~al.} 2018, ArXiv
  e-prints [\eprint[arXiv]{1808.00968}]

\bibitem[{{Majewski} {et~al.}(1996){Majewski}, {Munn}, \&
  {Hawley}}]{majewski96}
{Majewski}, S.~R., {Munn}, J.~A., \& {Hawley}, S.~L. 1996, \apjl, 459, L73

\bibitem[{{Majewski} {et~al.}(2017){Majewski}, {Schiavon}, {Frinchaboy},
  {Allende Prieto}, {Barkhouser}, {Bizyaev}, {Blank}, {Brunner}, {Burton},
  {Carrera}, {Chojnowski}, {Cunha}, {Epstein}, {Fitzgerald}, {Garc{\'{\i}}a
  P{\'e}rez}, {Hearty}, {Henderson}, {Holtzman}, {Johnson}, {Lam}, {Lawler},
  {Maseman}, {M{\'e}sz{\'a}ros}, {Nelson}, {Nguyen}, {Nidever}, {Pinsonneault},
  {Shetrone}, {Smee}, {Smith}, {Stolberg}, {Skrutskie}, {Walker}, {Wilson},
  {Zasowski}, {Anders}, {Basu}, {Beland}, {Blanton}, {Bovy}, {Brownstein},
  {Carlberg}, {Chaplin}, {Chiappini}, {Eisenstein}, {Elsworth}, {Feuillet},
  {Fleming}, {Galbraith-Frew}, {Garc{\'{\i}}a}, {Garc{\'{\i}}a-Hern{\'a}ndez},
  {Gillespie}, {Girardi}, {Gunn}, {Hasselquist}, {Hayden}, {Hekker}, {Ivans},
  {Kinemuchi}, {Klaene}, {Mahadevan}, {Mathur}, {Mosser}, {Muna}, {Munn},
  {Nichol}, {O'Connell}, {Parejko}, {Robin}, {Rocha-Pinto}, {Schultheis},
  {Serenelli}, {Shane}, {Silva Aguirre}, {Sobeck}, {Thompson}, {Troup},
  {Weinberg}, \& {Zamora}}]{majewski17}
{Majewski}, S.~R., {Schiavon}, R.~P., {Frinchaboy}, P.~M., {et~al.} 2017, \aj,
  154, 94

\bibitem[{{Malhan} {et~al.}(2018){Malhan}, {Ibata}, \& {Martin}}]{malhan18}
{Malhan}, K., {Ibata}, R.~A., \& {Martin}, N.~F. 2018, ArXiv e-prints
  [\eprint[arXiv]{1804.11339}]

\bibitem[{{McCarthy} {et~al.}(2012){McCarthy}, {Font}, {Crain}, {Deason},
  {Schaye}, \& {Theuns}}]{mccarthy12}
{McCarthy}, I.~G., {Font}, A.~S., {Crain}, R.~A., {et~al.} 2012, Monthly
  Notices of the Royal Astronomical Society, 420, 2245

\bibitem[{{Minchev} {et~al.}(2014){Minchev}, {Chiappini}, {Martig},
  {Steinmetz}, {de Jong}, {Boeche}, {Scannapieco}, {Zwitter}, {Wyse}, {Binney},
  {Bland-Hawthorn}, {Bienaym{\'e}}, {Famaey}, {Freeman}, {Gibson}, {Grebel},
  {Gilmore}, {Helmi}, {Kordopatis}, {Lee}, {Munari}, {Navarro}, {Parker},
  {Quillen}, {Reid}, {Siebert}, {Siviero}, {Seabroke}, {Watson}, \&
  {Williams}}]{minchev14}
{Minchev}, I., {Chiappini}, C., {Martig}, M., {et~al.} 2014, \apjl, 781, L20

\bibitem[{{Morrison} {et~al.}(1990){Morrison}, {Flynn}, \&
  {Freeman}}]{morrison90}
{Morrison}, H.~L., {Flynn}, C., \& {Freeman}, K.~C. 1990, \aj, 100, 1191

\bibitem[{{Moster} {et~al.}(2010){Moster}, {Macci{\`o}}, {Somerville},
  {Johansson}, \& {Naab}}]{moster10}
{Moster}, B.~P., {Macci{\`o}}, A.~V., {Somerville}, R.~S., {Johansson}, P.~H.,
  \& {Naab}, T. 2010, \mnras, 403, 1009

\bibitem[{{Myeong} {et~al.}(2018){Myeong}, {Evans}, {Belokurov}, {Sanders}, \&
  {Koposov}}]{myeong18}
{Myeong}, G.~C., {Evans}, N.~W., {Belokurov}, V., {Sanders}, J.~L., \&
  {Koposov}, S.~E. 2018, \apjl, 863, L28

\bibitem[{{Navarro} {et~al.}(2011){Navarro}, {Abadi}, {Venn}, {Freeman}, \&
  {Anguiano}}]{navarro11}
{Navarro}, J.~F., {Abadi}, M.~G., {Venn}, K.~A., {Freeman}, K.~C., \&
  {Anguiano}, B. 2011, \mnras, 412, 1203

\bibitem[{{Nissen} \& {Schuster}(2010)}]{nissen10}
{Nissen}, P.~E. \& {Schuster}, W.~J. 2010, \aap, 511, L10

\bibitem[{{Nissen} \& {Schuster}(2011)}]{nissen11}
{Nissen}, P.~E. \& {Schuster}, W.~J. 2011, \aap, 530, A15

\bibitem[{{Norris} {et~al.}(1985){Norris}, {Bessell}, \& {Pickles}}]{norris85}
{Norris}, J., {Bessell}, M.~S., \& {Pickles}, A.~J. 1985, \apjs, 58, 463

\bibitem[{{Pillepich} {et~al.}(2015){Pillepich}, {Madau}, \&
  {Mayer}}]{pillepich15}
{Pillepich}, A., {Madau}, P., \& {Mayer}, L. 2015, \apj, 799, 184

\bibitem[{{Price-Whelan} \& {Bonaca}(2018)}]{price-whelan18}
{Price-Whelan}, A.~M. \& {Bonaca}, A. 2018, \apjl, 863, L20

\bibitem[{{Purcell} {et~al.}(2010){Purcell}, {Bullock}, \&
  {Kazantzidis}}]{purcell10}
{Purcell}, C.~W., {Bullock}, J.~S., \& {Kazantzidis}, S. 2010, \mnras, 404,
  1711

\bibitem[{{Qu} {et~al.}(2010){Qu}, {Di Matteo}, {Lehnert}, {van Driel}, \&
  {Jog}}]{qu10}
{Qu}, Y., {Di Matteo}, P., {Lehnert}, M., {van Driel}, W., \& {Jog}, C.~J.
  2010, \aap, 515, A11

\bibitem[{{Qu} {et~al.}(2011{\natexlab{a}}){Qu}, {Di Matteo}, {Lehnert}, \&
  {van Driel}}]{qu11b}
{Qu}, Y., {Di Matteo}, P., {Lehnert}, M.~D., \& {van Driel}, W.
  2011{\natexlab{a}}, \aap, 530, A10

\bibitem[{{Qu} {et~al.}(2011{\natexlab{b}}){Qu}, {Di Matteo}, {Lehnert}, {van
  Driel}, \& {Jog}}]{qu11a}
{Qu}, Y., {Di Matteo}, P., {Lehnert}, M.~D., {van Driel}, W., \& {Jog}, C.~J.
  2011{\natexlab{b}}, \aap, 535, A5

\bibitem[{{Quinn} {et~al.}(1993){Quinn}, {Hernquist}, \& {Fullagar}}]{quinn93}
{Quinn}, P.~J., {Hernquist}, L., \& {Fullagar}, D.~P. 1993, \apj, 403, 74

\bibitem[{{Ram{\'{\i}}rez} {et~al.}(2012){Ram{\'{\i}}rez}, {Mel{\'e}ndez}, \&
  {Chanam{\'e}}}]{ramirez12}
{Ram{\'{\i}}rez}, I., {Mel{\'e}ndez}, J., \& {Chanam{\'e}}, J. 2012, \apj, 757,
  164

\bibitem[{{Reddy} \& {Lambert}(2008)}]{reddy08}
{Reddy}, B.~E. \& {Lambert}, D.~L. 2008, \mnras, 391, 95

\bibitem[{{Reid} {et~al.}(2014){Reid}, {Menten}, {Brunthaler}, {Zheng}, {Dame},
  {Xu}, {Wu}, {Zhang}, {Sanna}, {Sato}, {Hachisuka}, {Choi}, {Immer},
  {Moscadelli}, {Rygl}, \& {Bartkiewicz}}]{reid14}
{Reid}, M.~J., {Menten}, K.~M., {Brunthaler}, A., {et~al.} 2014, \apj, 783, 130

\bibitem[{{Sahlholdt} {et~al.}(2019){Sahlholdt}, {Casagrande}, \&
  {Feltzing}}]{Sahlholdt19}
{Sahlholdt}, C.~L., {Casagrande}, L., \& {Feltzing}, S. 2019, \apjl, 881, L10

\bibitem[{{Sch{\"o}nrich} {et~al.}(2010){Sch{\"o}nrich}, {Binney}, \&
  {Dehnen}}]{schonrich10}
{Sch{\"o}nrich}, R., {Binney}, J., \& {Dehnen}, W. 2010, \mnras, 403, 1829

\bibitem[{{Schuster} {et~al.}(2012){Schuster}, {Moreno}, {Nissen}, \&
  {Pichardo}}]{schuster12}
{Schuster}, W.~J., {Moreno}, E., {Nissen}, P.~E., \& {Pichardo}, B. 2012, \aap,
  538, A21

\bibitem[{{Searle} \& {Zinn}(1978)}]{searle78}
{Searle}, L. \& {Zinn}, R. 1978, \apj, 225, 357

\bibitem[{{Sestito} {et~al.}(2018){Sestito}, {Longeard}, {Martin},
  {Starkenburg}, {Fouesneau}, {Gonzalez Hernandez}, {Arentsen}, {Ibata},
  {Aguado}, {Carlberg}, {Jablonka}, {Navarro}, \& {Tolstoy}}]{sestito18}
{Sestito}, F., {Longeard}, N., {Martin}, N.~F., {et~al.} 2018, ArXiv e-prints
  [\eprint[arXiv]{1811.03099}]

\bibitem[{{Simion} {et~al.}(2019){Simion}, {Belokurov}, \&
  {Koposov}}]{simion19}
{Simion}, I.~T., {Belokurov}, V., \& {Koposov}, S.~E. 2019, \mnras, 482, 921

\bibitem[{{Snaith} {et~al.}(2015){Snaith}, {Haywood}, {Di Matteo}, {Lehnert},
  {Combes}, {Katz}, \& {G{\'o}mez}}]{snaith15}
{Snaith}, O., {Haywood}, M., {Di Matteo}, P., {et~al.} 2015, \aap, 578, A87

\bibitem[{{Snaith} {et~al.}(2014){Snaith}, {Haywood}, {Di Matteo}, {Lehnert},
  {Combes}, {Katz}, \& {G{\'o}mez}}]{snaith14}
{Snaith}, O.~N., {Haywood}, M., {Di Matteo}, P., {et~al.} 2014, \apjl, 781, L31

\bibitem[{{Sommer-Larsen} \& {Zhen}(1990)}]{sommer90}
{Sommer-Larsen}, J. \& {Zhen}, C. 1990, \mnras, 242, 10

\bibitem[{{Twarog} \& {Anthony-Twarog}(1994)}]{twarog94}
{Twarog}, B.~A. \& {Anthony-Twarog}, B.~J. 1994, \aj, 107, 1371

\bibitem[{{Twarog} \& {Anthony-Twarog}(1996)}]{twarog96}
{Twarog}, B.~A. \& {Anthony-Twarog}, B.~J. 1996, \aj, 111, 220

\bibitem[{{Velazquez} \& {White}(1999)}]{velazquez99}
{Velazquez}, H. \& {White}, S.~D.~M. 1999, \mnras, 304, 254

\bibitem[{{Venn} {et~al.}(2004){Venn}, {Irwin}, {Shetrone}, {Tout}, {Hill}, \&
  {Tolstoy}}]{venn04}
{Venn}, K.~A., {Irwin}, M., {Shetrone}, M.~D., {et~al.} 2004, \aj, 128, 1177

\bibitem[{{Villalobos} \& {Helmi}(2008)}]{villalobos08}
{Villalobos}, {\'A}. \& {Helmi}, A. 2008, \mnras, 391, 1806

\bibitem[{{Villalobos} \& {Helmi}(2009)}]{villalobos09}
{Villalobos}, {\'A}. \& {Helmi}, A. 2009, \mnras, 399, 166

\bibitem[{{Walker} {et~al.}(1996){Walker}, {Mihos}, \& {Hernquist}}]{walker96}
{Walker}, I.~R., {Mihos}, J.~C., \& {Hernquist}, L. 1996, \apj, 460, 121

\bibitem[{{White} \& {Frenk}(1991)}]{white91}
{White}, S.~D.~M. \& {Frenk}, C.~S. 1991, \apj, 379, 52

\bibitem[{{Xue} {et~al.}(2011){Xue}, {Rix}, {Yanny}, {Beers}, {Bell}, {Zhao},
  {Bullock}, {Johnston}, {Morrison}, {Rockosi}, {Koposov}, {Kang}, {Liu},
  {Luo}, {Lee}, \& {Weaver}}]{xue11}
{Xue}, X.-X., {Rix}, H.-W., {Yanny}, B., {et~al.} 2011, \apj, 738, 79

\bibitem[{{Zinn}(1993)}]{zinn93}
{Zinn}, R. 1993, in Astronomical Society of the Pacific Conference Series,
  Vol.~48, The Globular Cluster-Galaxy Connection, ed. G.~H. {Smith} \& J.~P.
  {Brodie}, 38

\bibitem[{{Zinn}(1996)}]{zinn96}
{Zinn}, R. 1996, in Astronomical Society of the Pacific Conference Series,
  Vol.~92, Formation of the Galactic Halo...Inside and Out, ed. H.~L.
  {Morrison} \& A.~{Sarajedini}, 211

\bibitem[{{Zolotov} {et~al.}(2010){Zolotov}, {Willman}, {Brooks}, {Governato},
  {Hogg}, {Shen}, \& {Wadsley}}]{zolotov10}
{Zolotov}, A., {Willman}, B., {Brooks}, A.~M., {et~al.} 2010, \apj, 721, 738

\end{thebibliography}

\begin{appendix}

\section{Velocity uncertainties}\label{uncertainties}

\begin{figure}
\centering
\includegraphics[clip=true, trim = 0mm 0mm 0mm 0mm, width=0.7\linewidth]{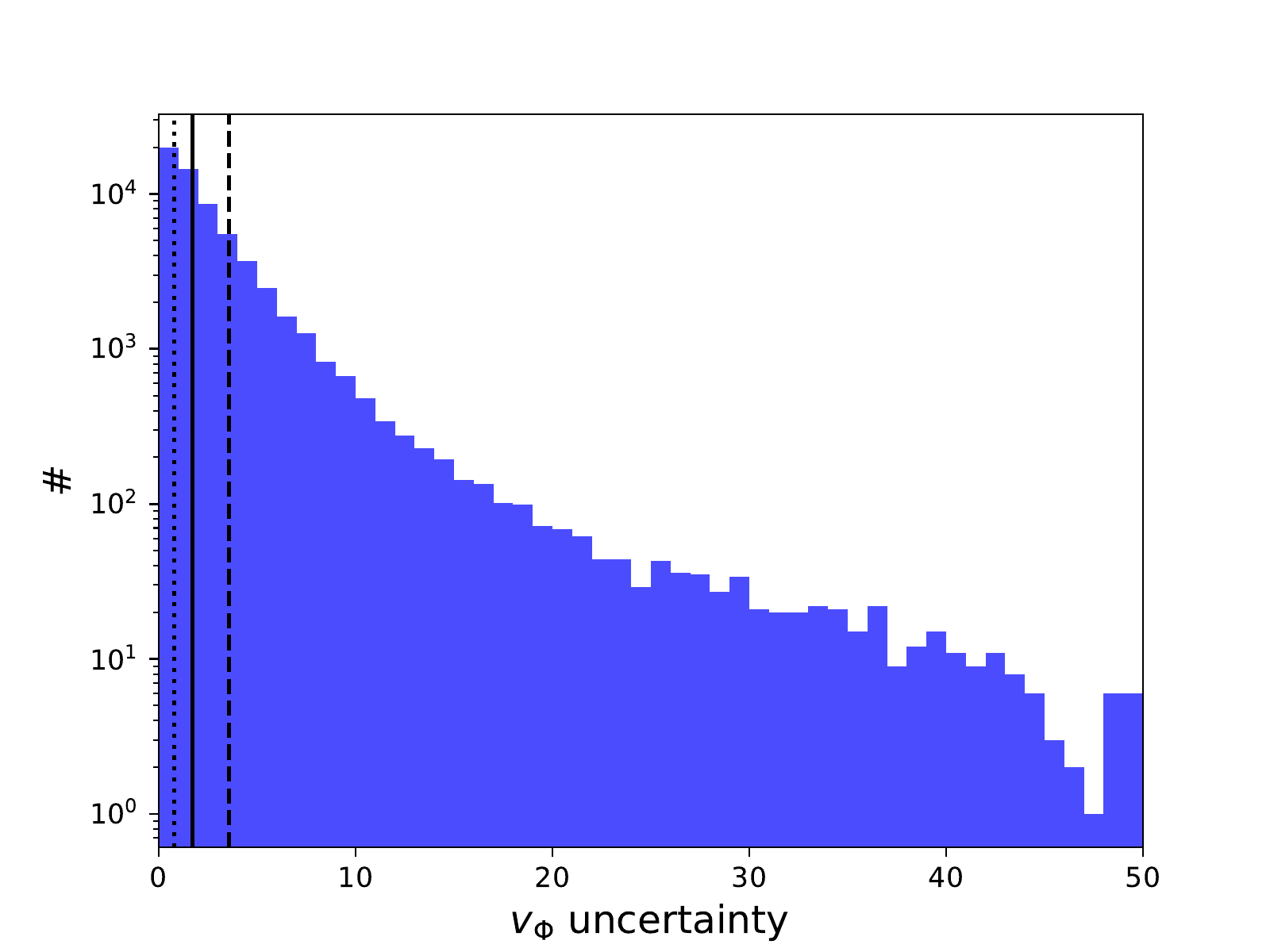}
\includegraphics[clip=true, trim = 0mm 0mm 0mm 0mm, width=0.7\linewidth]{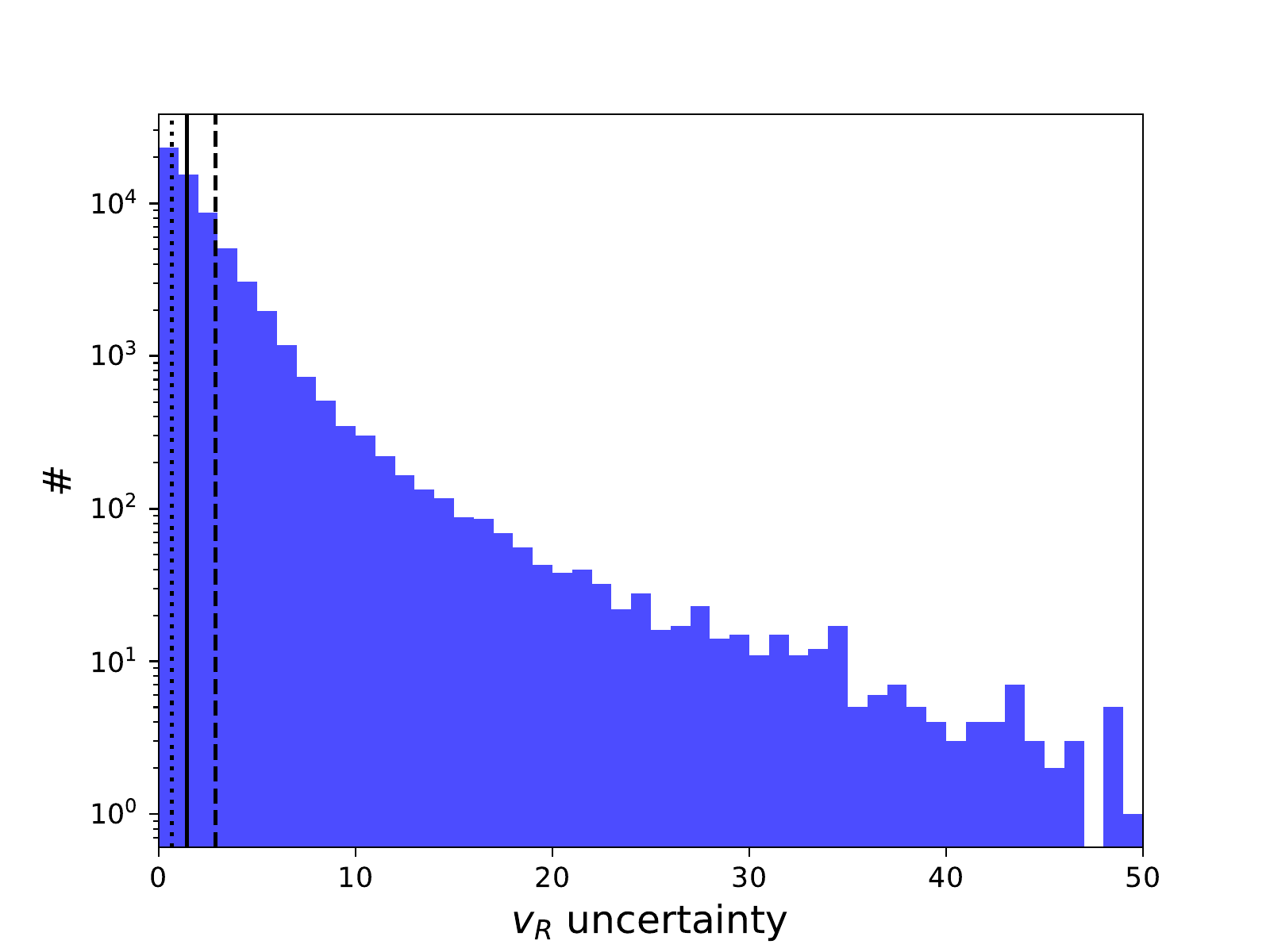}
\includegraphics[clip=true, trim = 0mm 0mm 0mm 0mm, width=0.7\linewidth]{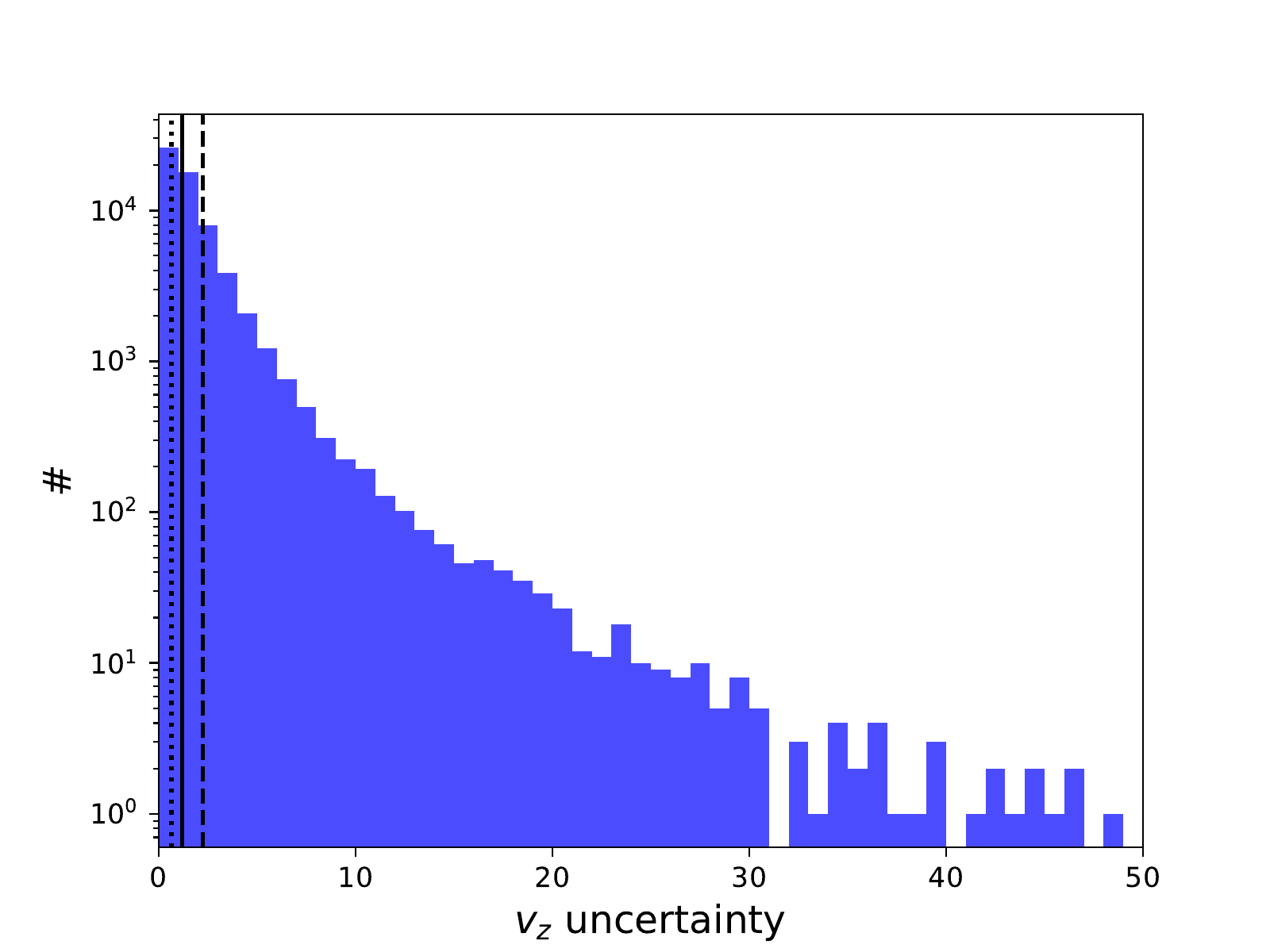}
\caption{Distribution of the velocity uncertainties on  $v_\Phi$ (\emph{top panel}),  $v_r$ (\emph{middle panel}) and $v_z$ (\emph{bottom panel}), due to the propagation of the individual errors on the observables. In each panel, the dottted, solid and dashed lines indicate, respectively, the 25th, 50th (median) and 75th percentile of the distribution. All uncertainties are given in units of km/s.}
\label{veluncert_histo}
\end{figure}

\begin{figure*}
\centering
\includegraphics[clip=true, trim = 5mm 0mm 10mm 2mm, width=0.22\linewidth]{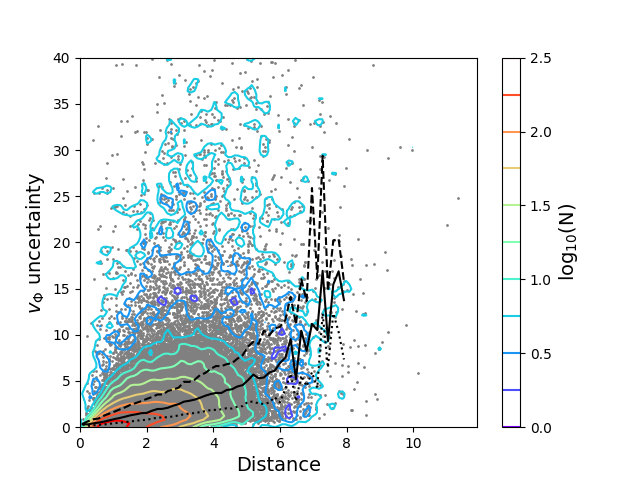}
\includegraphics[clip=true, trim = 5mm 0mm 10mm 2mm, width=0.22\linewidth]{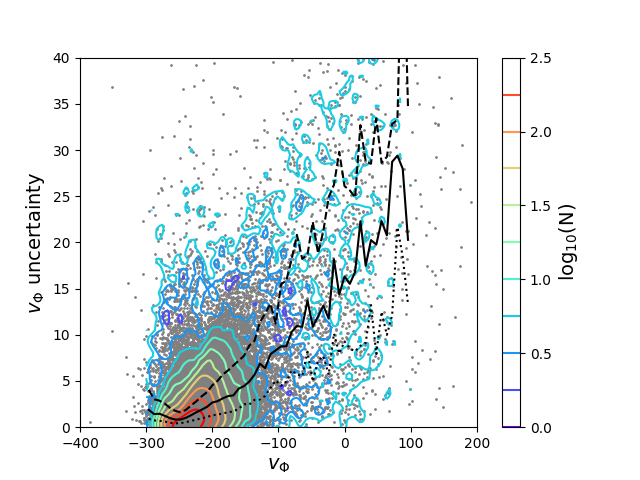}
\includegraphics[clip=true, trim = 5mm 0mm 10mm 2mm, width=0.22\linewidth]{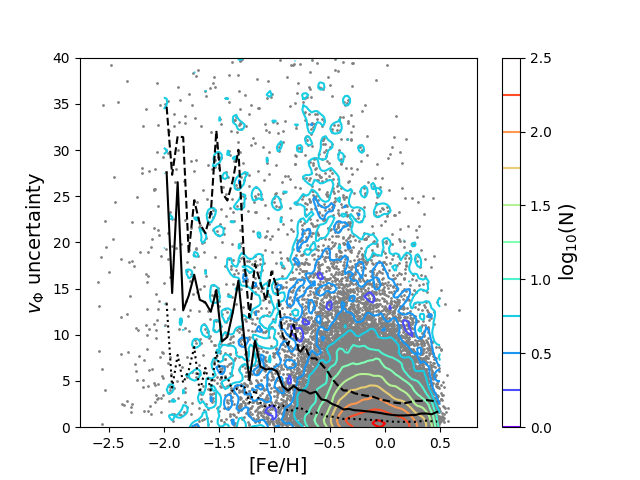}
\includegraphics[clip=true, trim = 5mm 0mm 10mm 2mm, width=0.22\linewidth]{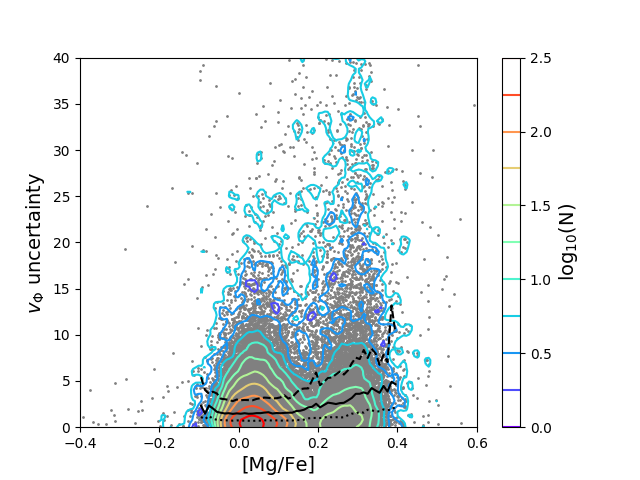}

\includegraphics[clip=true, trim = 5mm 0mm 10mm 2mm, width=0.22\linewidth]{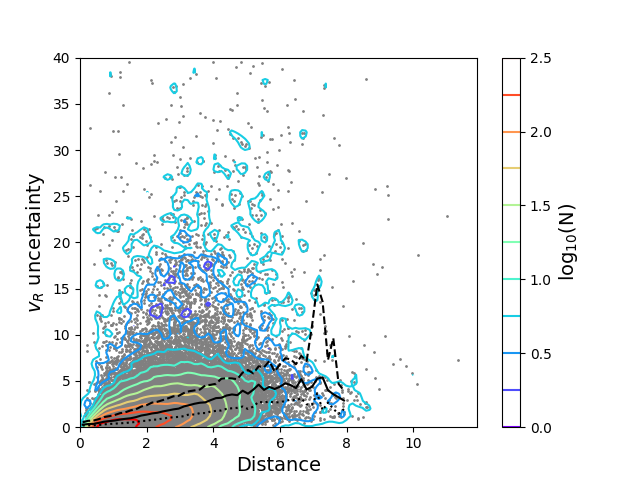}
\includegraphics[clip=true, trim = 5mm 0mm 10mm 2mm, width=0.22\linewidth]{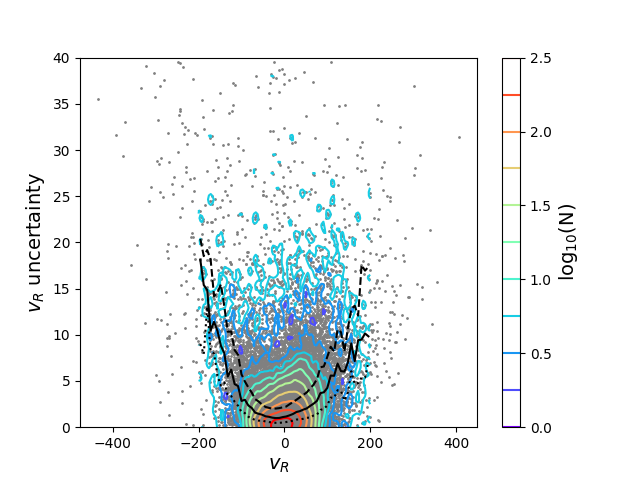}
\includegraphics[clip=true, trim = 5mm 0mm 10mm 2mm, width=0.22\linewidth]{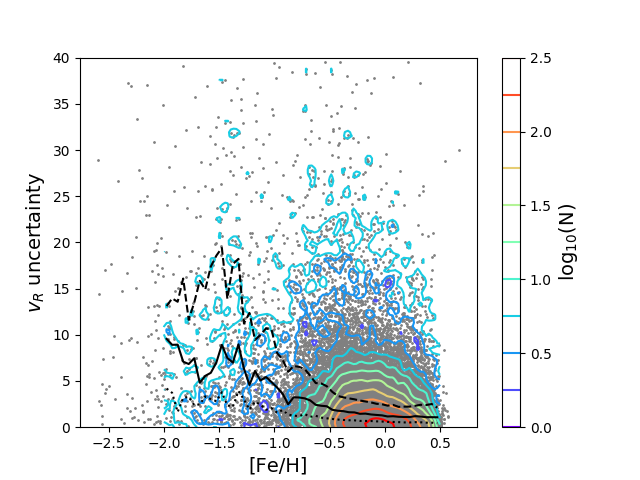}
\includegraphics[clip=true, trim = 5mm 0mm 10mm 2mm, width=0.22\linewidth]{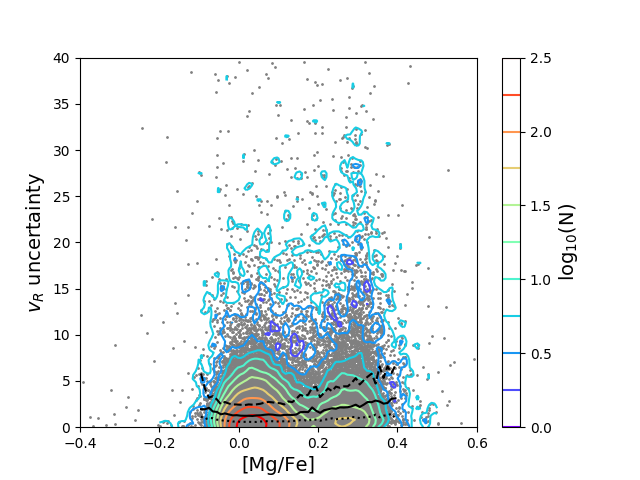}

\includegraphics[clip=true, trim = 5mm 0mm 10mm 2mm, width=0.22\linewidth]{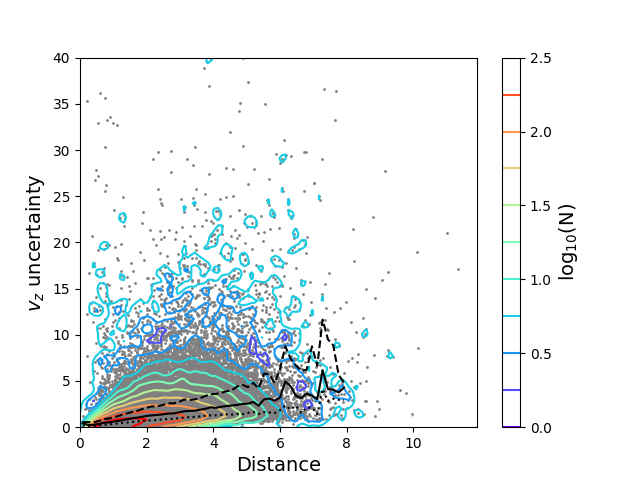}
\includegraphics[clip=true, trim = 5mm 0mm 10mm 2mm, width=0.22\linewidth]{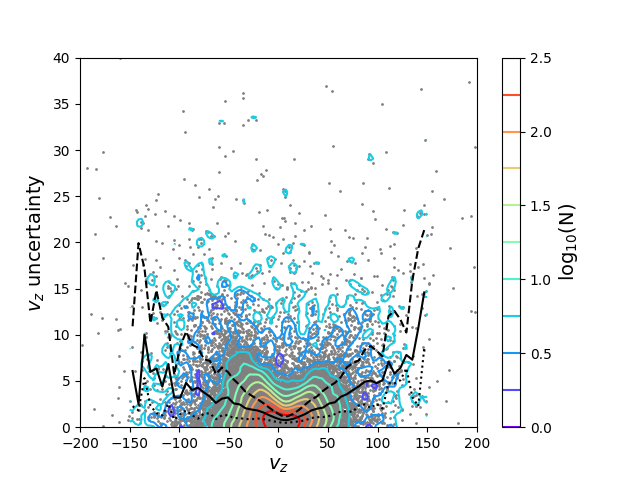}
\includegraphics[clip=true, trim = 5mm 0mm 10mm 2mm, width=0.22\linewidth]{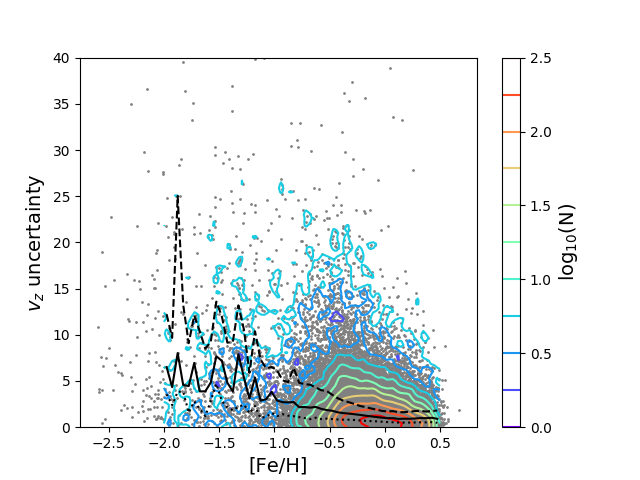}
\includegraphics[clip=true, trim = 5mm 0mm 10mm 2mm, width=0.22\linewidth]{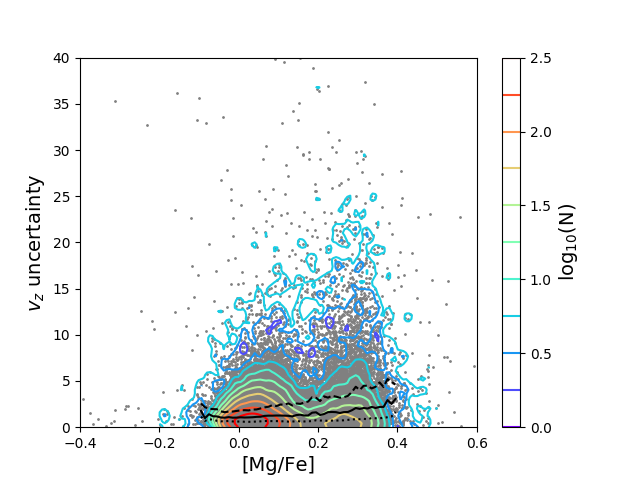}

\caption{\emph{First column:} Uncertainties in the $v_\Phi$ (\emph{first row}), $v_R$ (\emph{second row}) and  $v_z$ (\emph{third row}) of stars as a function of their distance $D$ from the Sun. Distances are given in kpc. \emph{Second column:} Uncertainties in the $v_\Phi$ (\emph{first row}), $v_R$ (\emph{second row}) and  $v_z$ (\emph{third row}) of stars as a function of the corresponding velocity (given in km/s). \emph{Third column:} Uncertainties in the $v_\Phi$ (\emph{first row}), $v_R$ (\emph{second row}) and  $v_z$ (\emph{third row}) of stars as a function of [Fe/H].   \emph{Fourth column:} Uncertainties in the $v_\Phi$ (\emph{first row}), $v_R$ (\emph{second row}) and  $v_z$ (\emph{third row}) of stars as a function of [Mg/Fe].  
In all panels: the individual uncertainties are shown by gray points; their density distributions, in logarithmic scale, are indicated by colored contours, whose corresponding values are reported in the error bars; the dotted, solid and dashed black lines show, respectively, the 25th, 50th (median) and 75th percentile of the distribution of uncertainties, as a function of the value reported on the x-axis. All velocity uncertainties are given in units of km/s. }
\label{veluncert_vs_x}
\end{figure*}

\begin{figure}
\centering
\includegraphics[clip=true, trim = 5mm 2mm 10mm 2mm, width=0.95\linewidth]{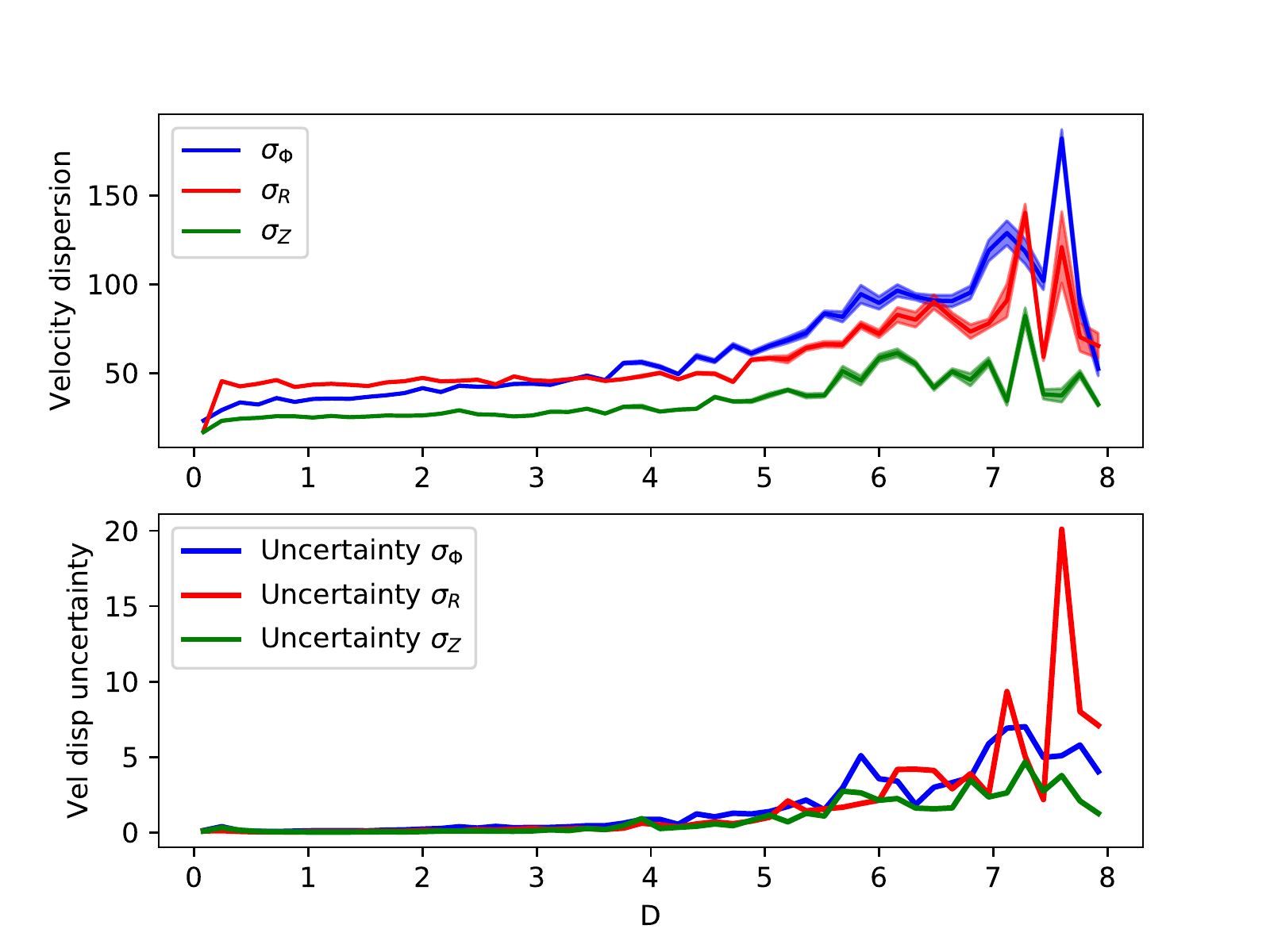}
\includegraphics[clip=true, trim = 5mm 2mm 10mm 2mm, width=0.95\linewidth]{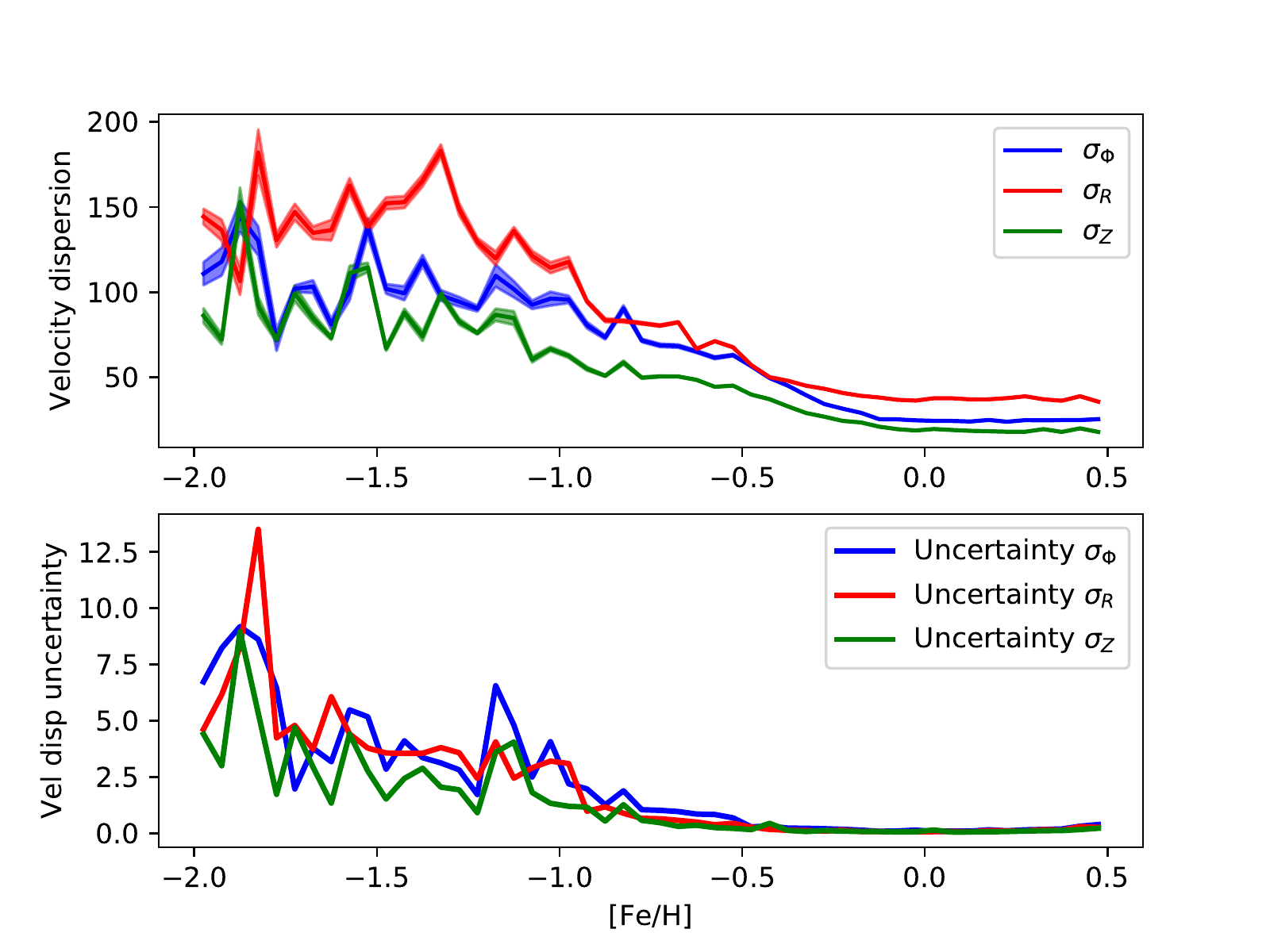}
\includegraphics[clip=true, trim = 5mm 2mm 10mm 2mm, width=0.95\linewidth]{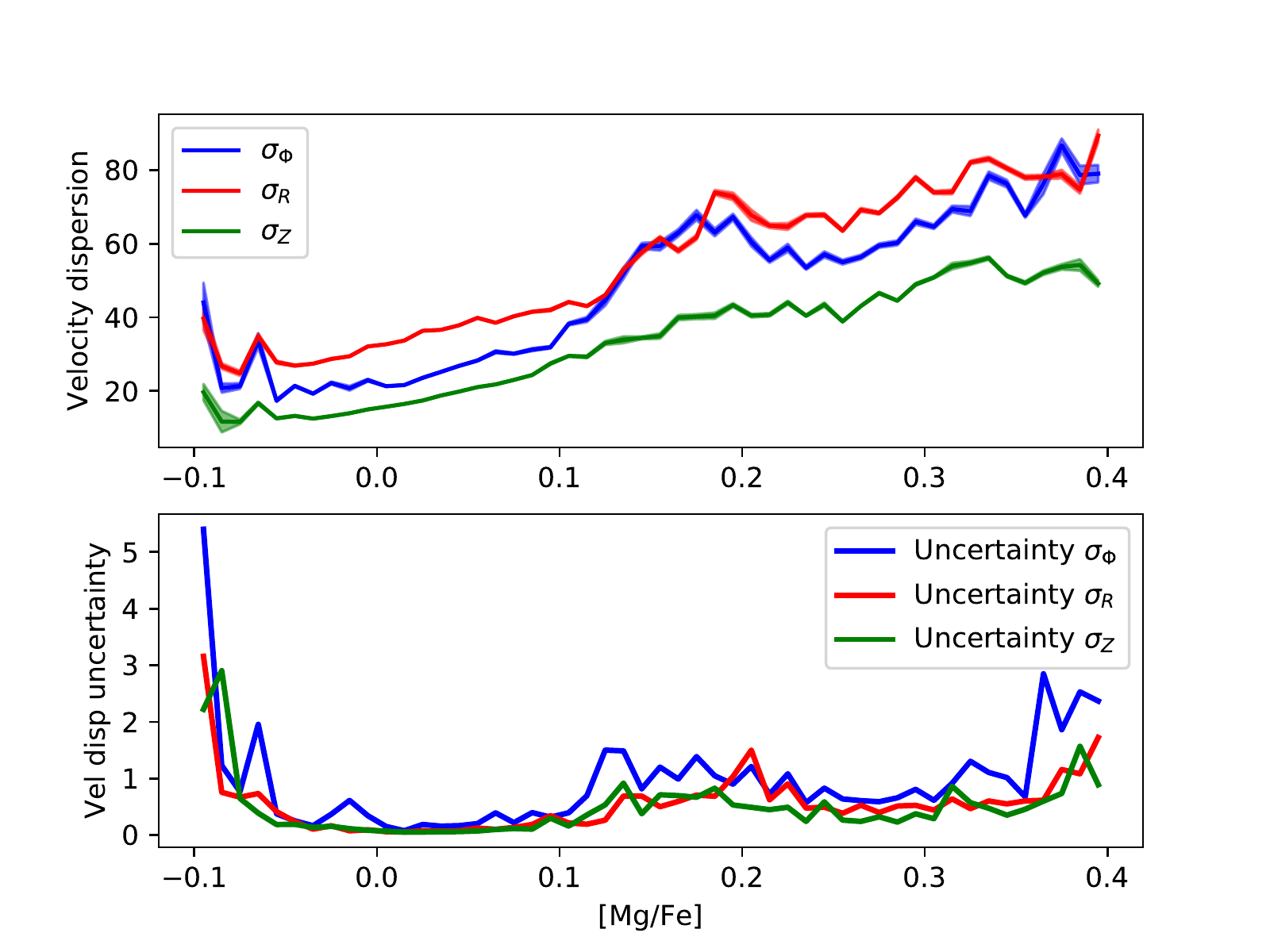}
\caption{\emph{Top figure:} Velocity dispersion (top panel) and velocity dispersion uncertainty (bottom panel) as a function the distance $D$ from the Sun. Blue, red and green curves correspond to the velocity dispersions in the aziumthal, radial and vertical direction,  as indicated in the legend. In the bottom panel, the solid lines indicate the mean relation, and the  color shaded areas indicate the 1$\sigma$ uncertainty, estimated through 100 random realizations, from the uncertainties on the observables (see text). These uncertainties are then reported in the bottom panel.  \emph{Middle and bottom figures:} As for the top, but now the velocity dispersions and corresponding uncertainties are shown as a function of  [Fe/H] and [Mg/Fe], respectively. In all plots, velocities are in units of km/s.}
\label{veldispuncert_vs_x}
\end{figure}

\begin{figure}
\centering
\includegraphics[clip=true, trim = 5mm 0mm 10mm 2mm, width=\linewidth]{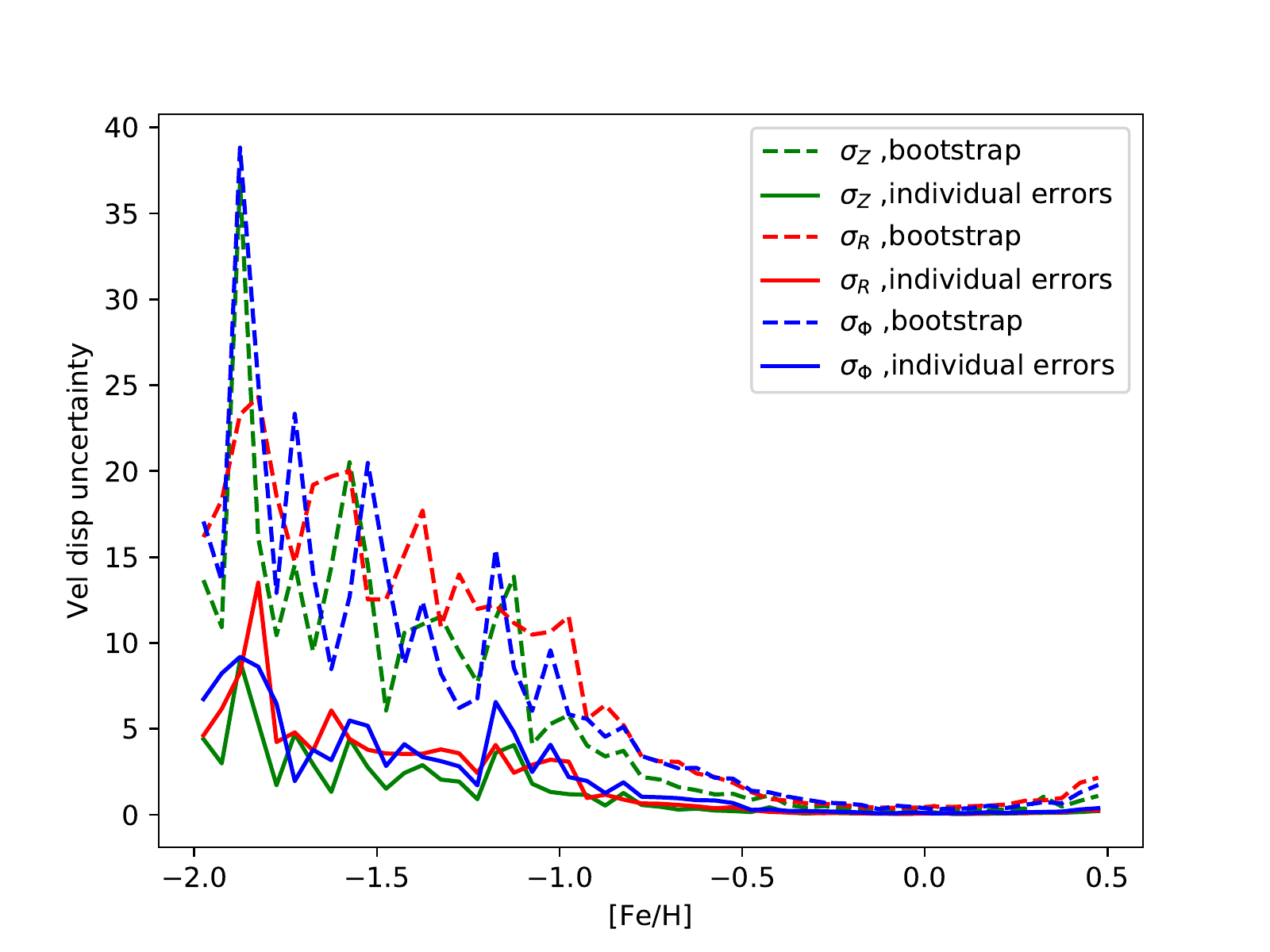}
\includegraphics[clip=true, trim = 5mm 0mm 10mm 2mm, width=\linewidth]{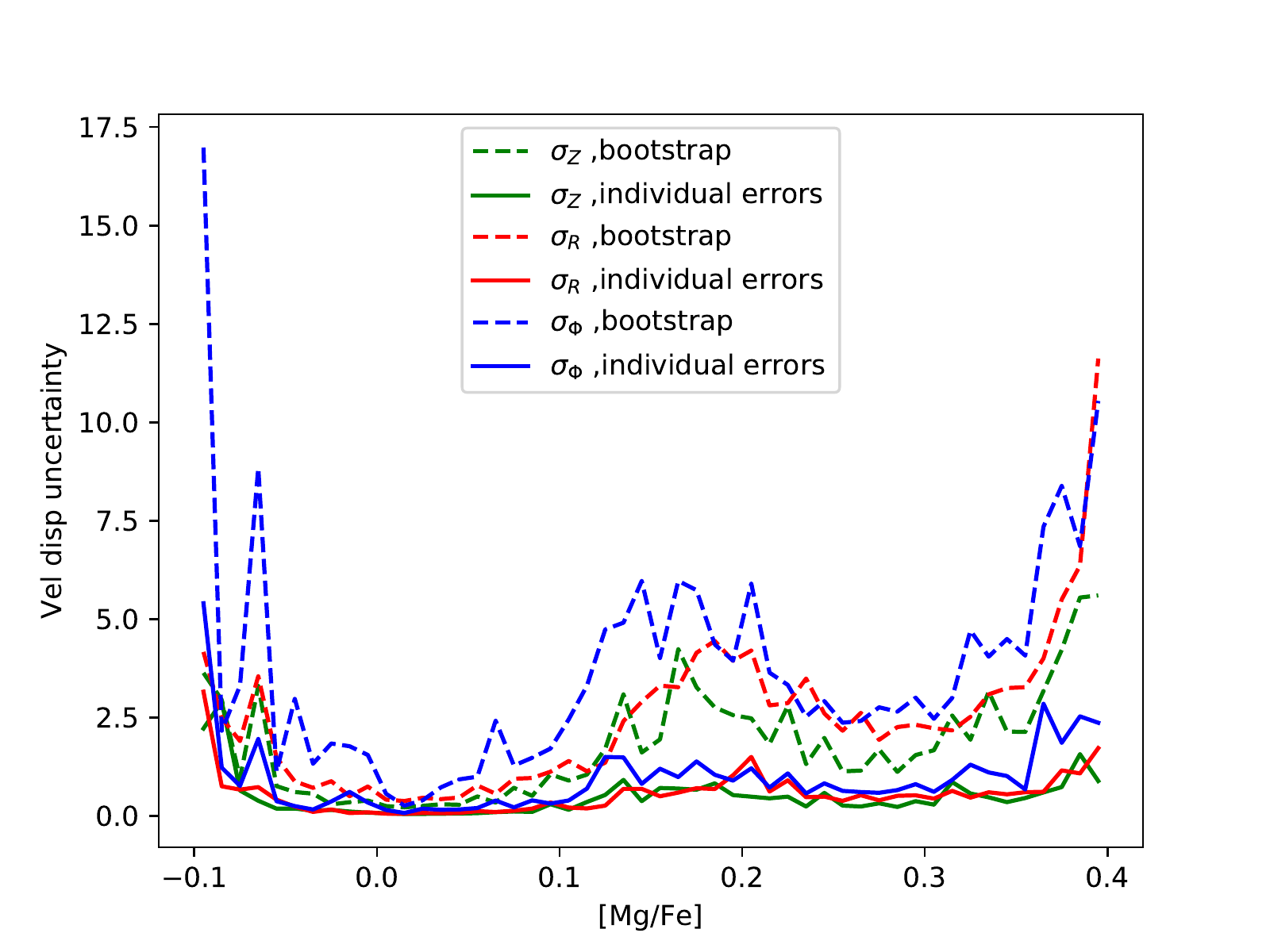}
\caption{Comparison between the uncertainties on velocity dispersions, due to the propagation of the uncertainties on the individual observables (solid lines, see Appendix~\ref{uncertainties} for details) and those estimated by bootstrapping technique (dashed lines, see Sect.~\ref{mean} for details). The uncertainties are shown as a function of [Fe/H] (\emph{top panel}) and as a function of [Mg/Fe] (\emph{bottotm panel}).}
\label{bootvsuncert}
\end{figure}

In this section, we, first, discuss the uncertainties in the velocities of stars in the sample, due to the propagation of the individual uncertainties in the observables (parallaxes, proper motions, and radial velocities). Then, we discuss how these uncertainties impact the velocities (and velocity dispersions)-abundances relation presented in  Sect.~\ref{mean}. We anticipate the conclusions of this section, by confirming that the statistical uncertainties estimated by the bootstrapping technique adopted in Sect.~\ref{mean} are always the dominant source of uncertainty, when uncertainties are significant.  

To estimate the uncertainties on the azimuthal, radial and vertical velocities $v_\Phi, v_R, v_z$, of each star, due to the propagation of the individual uncertainties on its parallax, proper motion and radial velocity, we have  assumed gaussian distributions of these errors, and generated 100 random realizations of these parameters per source. For each realization, we then make a transformation from the space of observables ($\alpha, \delta, \pi, {\mu_\alpha}^*, \mu_\delta$) to the Galactocentric  rest-frame \citep[see][]{johnson87, kepley07}, and calculate the corresponding values of $v_\Phi, v_R, v_z$.  The resulting uncertainties on  $v_\Phi, v_R, v_z$ are then finally estimated as the standard deviations of their values, over the 100 realizations. Their distributions are shown in Fig.~\ref{veluncert_histo}, and probe that for most of the stars uncertainties are low: the 75th percentile of the distribution is indeed equal to 3.56~km/s,  2.87~km/s and  2.22~km/s, for $v_\Phi, v_R$ and $v_z$, respectively, whereas the corresponding medians are 1.70~km/s,  1.42~km/s and 1.20~ km/s.

Uncertainties on $v_\Phi, v_R, v_z$ increase with the distance of stars in the sample from the Sun, as expected (see Fig.~\ref{veluncert_vs_x}, first column). Stars with halo kinematics ($v_\Phi \sim 0$~km/s,  high $v_R$ and $v_z$) have also, on average, larger uncertainties than stars with disc-like kinematics (see Fig.~\ref{veluncert_vs_x}, second column), and this finding can be explained because the fraction of halo stars increases with the distance from the Sun. Despite the larger uncertainties, the typical (median) uncertainty of stars with null $v_\Phi$, $|v_r| \ge 150$~km/s and  $|v_r| \ge 100$~km/s is equal or below 15~km/s.

Finally, because stars with halo-like kinematics have, on average, larger velocity uncertainties than stars with disc-like kinematics, a trend exists between the $v_\Phi, v_R$ and $v_z$ uncertainties and chemical abundances, with the uncertainties decreasing with [Fe/H] and increasing with [Mg/Fe] (see Fig.~\ref{veluncert_vs_x}, third and fourth columns).

The uncertainties on the velocity dispersions-abundance relations discussed in Sect.~\ref{mean} have been estimated by making use of the 100 random realizations described before. For each realization, we have computed the corresponding velocity dispersion-abundance relations (see Fig.~\ref{veldispuncert_vs_x}, middle and bottom plots, top panels), and estimated the corresponding standard deviations (see Fig.~\ref{veldispuncert_vs_x}, middle and bottom plots, bottom panels). As shown in this Figure, the uncertainties on the velocity dispersion-abundance relations are always very small, and most of time below 5~km/s (for the velocity dispersions--[Fe/H] relation), and below 1.5~ km/s   (for the velocity dispersions--[Mg/Fe] relation). For completeness, and for comparison with Fig.~\ref{veluncert_vs_x}, we also show in Fig.~\ref{veldispuncert_vs_x} the dependency of the uncertainties in the velocity dispersions on the distance $D$ from the Sun.
The comparison of the uncertainties on the velocity dispersions, as due to the propagation of individual errors on observable and the uncertainties calculated by bootstrapping technique (see Sect.~\ref{mean}) is shown in Fig.~\ref{bootvsuncert} and demonstrate that the statistical uncertainties calculated by bootstrapping are always larger (or equal) than those due to the individual errors. In particular, when uncertainties in the sample are significant (above 2-3~km/s), those calculated with the bootstrapping technique are always the dominant. As a consequence, the errors reported in Figs.~\ref{kins_vs_MgFe}, and \ref{kins_vs_FeH}, and estimated by bootstrapping the sample, constitute the dominant uncertainty among stars in our sample.

\section{Number of stars in the mean chemo-kinematic relations}\label{numbers_ck}

In Fig.~\ref{fig_numbers_ck}, we present the number distribution of stars, as a function of their [Mg/Fe](/[Fe/H]) ratio, for different [Fe/H] and [Mg/Fe] intervals. The adopted values for the [Fe/H] and [Mg/Fe] intervals are the same already used in Sect.~\ref{mean} and Figs.~\ref{kins_vs_MgFe}, \ref{kins_vs_FeH} and \ref{CR_map}, that is 10 metallicity intervals, ranging from [Fe/H]=-2.1 up to  [Fe/H]=0.25 (see Fig.~\ref{fig_numbers_ck}, top panel), and 6 [Mg/Fe] intervals, ranging from [Mg/Fe]=-0.1 up to [Mg/Fe]=0.4 (see Fig.~\ref{fig_numbers_ck}, bottom panel). The width of the intervals is the same as adopted in Sect.~\ref{mean}. 
We point out that in  Fig.~\ref{fig_numbers_ck}, in both panels, we have used a  number of bins along the $x$-axis greater than those used in Figs.~\ref{kins_vs_MgFe}, \ref{kins_vs_FeH} and \ref{CR_map} to make clear where the dip in the [Mg/Fe]-[Fe/H] sequence is and how it changes with [Mg/Fe] for different [Fe/H] intervals (see, for example, in the top panel, the values corresponding to [Mg/Fe]=[0.24, 0.27, 0.21, 0.18, 0.18, 0.18, 0.12] for [Fe/H]=[-1.5,  -1.25, -1.0, -0.75, -0.5, -0.25, 0.0]). 

\begin{figure}
\centering
\includegraphics[clip=true, trim = 0mm 0mm 0mm 0mm, width=0.7\linewidth]{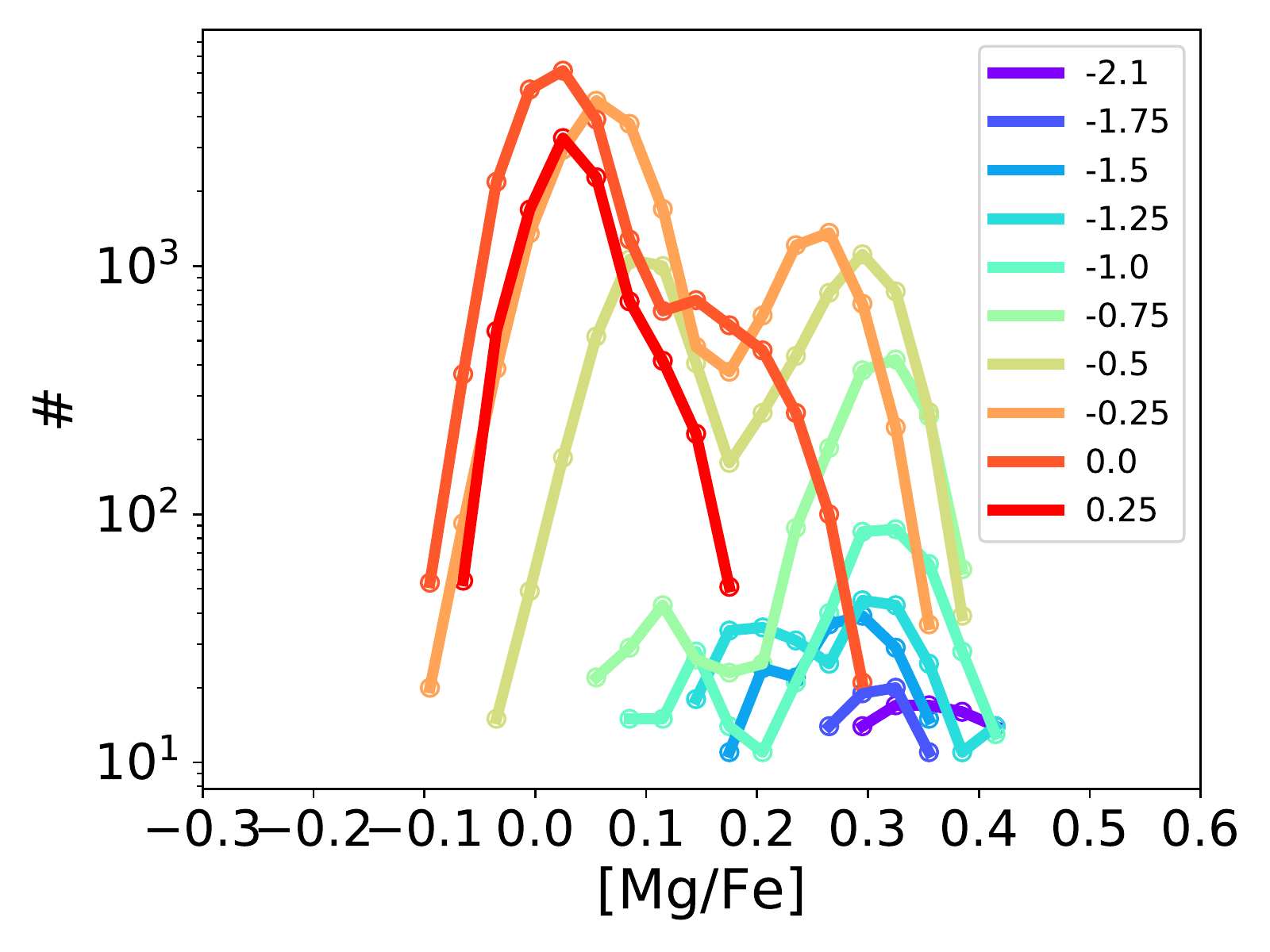}
\includegraphics[clip=true, trim = 0mm 0mm 0mm 0mm, width=0.7\linewidth]{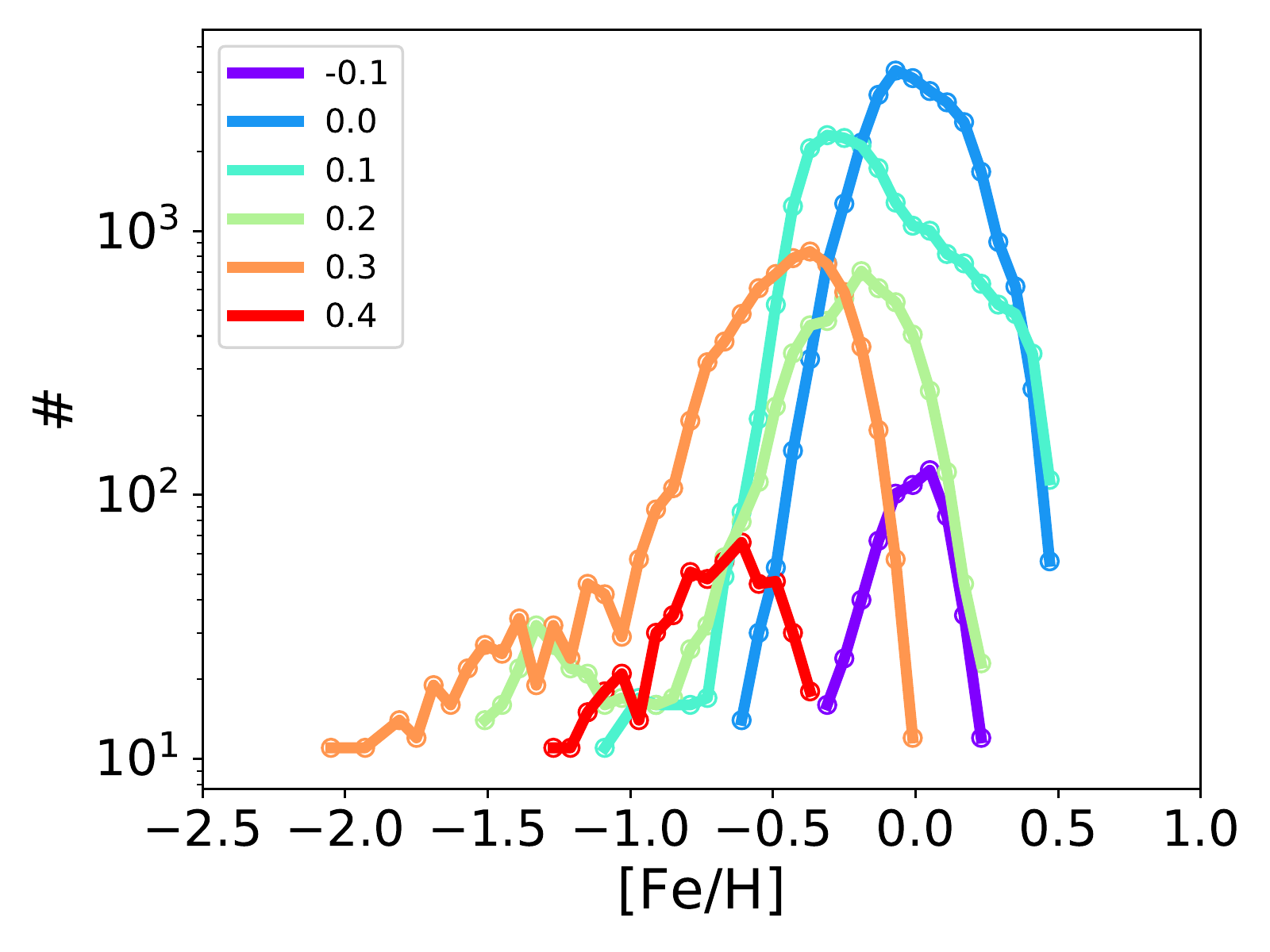}
\caption{\emph{Top panel:} Number of stars, as a function of their [Mg/Fe] ratio. Relations are given for bins in [Fe/H], as indicated in the legend. \emph{Bottom panel:} Number of stars, as a function of their [Fe/H] ratio. Relations are given for bins in [Mg/Fe], as indicated in the legend. }
\label{fig_numbers_ck}
\end{figure}

\end{appendix}

\end{document}